\documentclass[12pt,letterpaper]{JHEP3}

\usepackage{amscd,amsmath,amssymb,amsfonts,xspace,mathrsfs}
\usepackage[multiple]{footmisc}
\usepackage{url}

\numberwithin{equation}{section}

\def\varpi{t}

\def\det{\,{\rm det}\, }
\def\diag{{\rm diag}}
\def\sign{{\rm sign}}

\def\Im{\,{\rm Im}\,}
\def\Re{\,{\rm Re}\,}

\newcommand{\p}{\partial}

\def\({\left(}
\def\){\right)}
\def\[{\left[}
\def\]{\right]}
\def\<{\left\langle}
\def\>{\right\rangle}
\def\hf{{1\over 2}}
\def\haf{\textstyle{1\over 2}}
\newcommand{\half}{\frac{1}{2}}

\renewcommand{\d}{\mathrm{d}}
\newcommand{\de}{\mathrm{d}}

\newcommand{\I}{\mathrm{i}}

\newcommand{\cD}{\mathcal{D}}

\def\vrh{\varrho}

\newcommand{\cF}{\mathcal{F}}

\newcommand{\cS}{\mathcal{S}}

\newcommand{\cK}{\mathcal{K}}
\newcommand{\cM}{\mathcal{M}}
\newcommand{\cW}{\mathcal{W}}

\newcommand{\cX}{\mathcal{X}}

\newcommand{\cJ}{\mathcal{J}}

\newcommand{\ub}{\bar{u}}

\DeclareSymbolFont{AMSa}{U}{msa}{m}{n}
\DeclareSymbolFont{AMSb}{U}{msb}{m}{n}
\DeclareMathSymbol{\fieldR}{\mathalpha}{AMSb}{"52}

\newcommand{\Z}{{\mathbb Z}}
\newcommand{\cZ}{\mathcal{Z}}
\newcommand{\cI}{\mathcal{I}}

\newcommand{\cU}{\mathcal{U}}
\newcommand{\cA}{\mathcal{A}}

\newcommand{\pa}{\partial}
\newcommand{\nn}{\nonumber}

\newcommand{\eps}{\epsilon}
\newcommand{\veps}{\varepsilon}
\newcommand{\IR}{\mathbb{R}}
\newcommand{\IC}{\mathbb{C}}
\newcommand{\IZ}{\mathbb{Z}}

\newcommand{\sgn}{\mbox{sgn}}
\newcommand{\tzeta}{\tilde\zeta}

\newcommand{\talp}{\tilde\alpha}
\newcommand{\tc}{\tilde c}

\newcommand{\txi}{\tilde\xi}

\newcommand{\CP}{\IC P^1}

\def\bea{\begin{eqnarray}}
\def\eea{\end{eqnarray}}
\def\be{\begin{equation}}
\def\ee{\end{equation}}
\def\ba{\begin{align}}
\def\ea{\end{align}}
\def\bse{\begin{subequations}}
\def\ese{\end{subequations}}

\newcommand{\bfA}{{\boldsymbol A}}

\newcommand{\bfD}{{\boldsymbol D}}

\newcommand{\bfb}{{\boldsymbol b}}
\newcommand{\bfc}{{\boldsymbol c}}

\newcommand{\bfk}{{\boldsymbol k}}

\newcommand{\bfp}{{\boldsymbol p}}
\newcommand{\bfq}{{\boldsymbol q}}

\newcommand{\bft}{{\boldsymbol t}}
\newcommand{\bfz}{{\boldsymbol z}}
\newcommand{\bfx}{{\boldsymbol x}}
\newcommand{\bfy}{{\boldsymbol y}}

\def\ba{\bar a}

\def\by{\bar y}
\def\bz{\bar z}

\def\btau{\bar \tau}

\def\bZ{\bar Z}

\def\bF{\bar F}

\def\hE{\hat E}
\def\hM{\hat M}

\def\hV{\hat V}

\def\cbfq{\check {\boldsymbol q}}
\def\cbfp{\check {\boldsymbol p}}
\def\cbft{\check {\boldsymbol t}}

\def\cbfb{\check {\boldsymbol b}}
\def\cbfc{\check {\boldsymbol c}}
\def\cbfx{\check {\boldsymbol x}}
\def\cbfmu{\check {\boldsymbol \mu}}
\def\ckap{\check\kappa}
\def\cp{\check p}
\def\cx{\check x}

\def\ct{\check t}

\def\tq{\tilde q}

\def\cij#1{c}
\def\ci#1{c}

\newcommand{\kahler}{{K\"ahler}\xspace}

\newcommand{\qk}{{quaternion-K\"ahler}\xspace}

\def\txii#1{{\tilde\xi}^{[#1]}}

\def\ai#1{{\alpha}^{[#1]}}

\def\xii#1{\xi_{[#1]}}

\def\Hij#1{H^{[#1]}}

\def\Hij#1{H^{[#1]}}

\def\gfinv{n^{(0)}}

\def\ellg#1{\ell_{#1}}

\def\Xicl{\Xi_{\rm cl}}
\def\Fcl{F^{\rm cl}}
\def\Fws{F^{\rm w.s.}}
\def\bFcl{\bF^{\rm cl}}

\newcommand{\expe}[1]{{\bf E}\!\left( #1\right)}

\def\gamD#1{\tilde\gamma}
\def\GamD#1{\Gamma^{(#1)}}

\def\CYm{\mathfrak{Y}}
\def\CY{{\hat{\mathfrak{Y}}}}

\def\Nq{N}

\DeclareMathOperator{\Erf}{Erf}
\DeclareMathOperator{\Erfc}{Erfc}

\def\cla{\tilde c_a}
\def\cl0{\tilde c_0}

\newcommand{\bfeps}{{\boldsymbol \epsilon}}
\newcommand{\bfmu}{{\boldsymbol \mu}}

\newcommand{\bfch}{{\bf{\boldsymbol c}_2}}

\def\Dop#1{D^{(#1)}}
\def\bDop#1{\bar D^{(#1)}}

\def\Del#1{\mathscr{D}^{(#1)}}

\newcommand{\gammap}{\gamma}
\newcommand{\gammam}{-\gamma}

\def\trans{\vrh}

\def\gtrans{\rho_{h^g}}

\def\Erfc{\text{Erfc}}

\def\bOm{\bar\Omega}

\def\wh{\mathfrak{h}}
\def\bwh{\bar\wh}

\def\gH{h^g}

\def\cXsf{\cX^{\rm sf}}
\def\cXcl{\cX^{\rm cl}}
\def\cXclD{\cX^{\rm D1}}
\def\cXq{\cX^{(1)}}

\def\tcF{\tilde\cF}

\def\cFi#1{\cF^{(#1)}}
\def\cFq{\cF^{(1)}}

\def\tcJ{\tilde\cJ}

\def\cJi#1{\cJ^{(#1)}}
\def\cJq{\cJ^{(1)}}

\def\tgH{\tilde h^g}

\def\gHi#1{h^{g(#1)}}
\def\gHq{h^{g(1)}}

\def\hcJ{\widehat\cJ}

\def\hcJe{\widehat\cJ^{\rm ext}}

\def\whh{\widehat h}

\def\Dela{\Delta^{\bft}}
\def\Dg{\Delta^{(1)}}
\def\Dl{\Delta^{(2)}}
\def\hDl{\hat\Delta^{(2)}}

\def\xmn{x_{m,n}}
\def\xpmn{y_{m,n}}

\def\bOmMSW{{\bar \Omega}^{\rm MSW}}

\def\ff{\mathfrak{F}}
\def\fa{f}
\def\ww{w}

\def\wwx{\mathfrak{w}}

\def\nnf{\mathfrak{n}}

\def\Xxx{\mathfrak{X}}

\def\Ixx{\mathfrak{I}}

\def\Xv{\mathscr{X}}
\def\Cv{\mathscr{C}}

\def\altwo{\alpha_{(2)}}

\def\rpp#1#2{{\textstyle\sqrt{\frac{(p_{#1}t^2)}{(p_{#2}t^2)}}}}

\def\hKer{\hat\Phi^{(1)}}
\def\hKerr{\hat\Phi^{(2)}}
\def\sPhi{\Phi^{\rm (s)}}
\def\mPhi{\Phi_{\bfA,\bfA^\star}}

\def\PhiT{\Phi^E_{2,1}}
\def\hPhiT{\hat\Phi^E_{2,1}}
\def\hPhid{\hat\Phi^{(F)}}
\def\hPhip{\hat\Phi^{(E)}}

\def\Phian{\Phi^{{\rm an}}}
\def\Phianp{\Phi'^{\,{\rm an}}}

\def\aLam{\Lambda^{-}}

\def\cLam{\check\Lambda}

\def\vn{\vec n}

\def\under#1#2{\mathop{#1}\limits_{#2}}

\title{Multiple D3-instantons and mock modular forms II}

\preprint{L2C:17-011\\
CERN-TH-2017-040\\
IPhT-T17/020 \\
}

\author{Sergei Alexandrov$^{1,2}$, Sibasish Banerjee$^3$, Jan Manschot$^{4,5}$, Boris Pioline$^{2,6,7}$
\\
$^1$ {\it
Laboratoire Charles Coulomb (L2C), UMR 5221 CNRS-Universit\'e de
Montpellier, F-34095, Montpellier, France}\\

$^2$ {\it CERN, Theoretical Physics Department,
Case C01600, CERN, CH-1211 Geneva 23, Switzerland}\\

$^3$ {\it IPhT, CEA, Saclay, Gif-sur-Yvette, F-91191, France}\\

$^4$ {\it School of Mathematics, Trinity College, Dublin 2, Ireland}\\

$^5$ {\it Hamilton Mathematical Institute, Trinity College, Dublin 2, Ireland}\\

$^6$ {\it  Laboratoire de Physique Th\'eorique et Hautes
Energies, CNRS UMR 7589,
Universit\'e Pierre et Marie Curie,
4 place Jussieu, 75252 Paris cedex 05, France} \\

$^7$ {\it Sorbonne Universit\'es, UPMC Universit\'e Paris 6, UMR 7589, F-75005 Paris, France}\\

\vspace*{2mm} {\tt e-mail:
\email{sergey.alexandrov@umontpellier.fr},
\email{sibasish.banerjee@cea.fr},
\email{manschot@maths.tcd.ie},
\email{boris.pioline@cern.ch}
}

\vspace*{-3mm}

}

\abstract{
We analyze the modular properties of D3-brane instanton corrections to the hypermultiplet moduli
space in type IIB string theory compactified
on a Calabi-Yau threefold.
In Part I, we found a necessary condition for the existence of an isometric action of S-duality on
this moduli space: the generating function of DT invariants in the large volume attractor chamber
must be a vector-valued mock modular form with specified modular properties.
In this work, we prove that this condition is also sufficient at two-instanton order.
This is achieved by producing a holomorphic action of $SL(2,\IZ)$
on the twistor space which preserves the holomorphic contact structure.
The key step is to cancel the anomalous modular variation of the Darboux coordinates
by a local holomorphic contact transformation,
which is generated by a suitable indefinite theta series. For this purpose we introduce
a new family of theta series of signature $(2,n-2)$, find their modular completion,
and conjecture sufficient conditions for their convergence, which may be of independent mathematical interest.
}

\begin{document}

\section{Introduction}
\label{sec-intro}

Determining the exact metric on the hypermultiplet moduli space $\cM_{\rm H}$ in type II string theories
compactified on a Calabi-Yau (CY) threefold  $\CYm$ is a challenge whose importance can hardly be overestimated.
Indeed, this metric incorporates a large class of topological invariants of $\CYm$ such as its Euler number,
intersection numbers, Gromov-Witten (GW) and generalized Donaldson-Thomas (DT) invariants into
a tightly constrained structure.
In particular, the existence of an isometric action of the modular group $SL(2,\IZ)$  \cite{RoblesLlana:2006is}, 
which descends from the S-duality
symmetry of type IIB string theory in 10 dimensions, imposes strong constraints on generalized DT invariants,
which control  both D-brane instantons to the metric on $\cM_H$ and degeneracies of  
supersymmetric black hole micro-states. One may hope that this line of study will lead to new strategies
for determining these important invariants.

Major progress towards this goal was made in recent years by applying twistorial methods and exploiting dualities
(see e.g. \cite{Alexandrov:2011va, Alexandrov:2013yva} for a review). However, it is fair to say that this goal
has not been fully attained yet. The main missing piece are NS5-brane instantons,
which are currently understood only at linear order around the D-instanton corrected metric \cite{Alexandrov:2010ca}
(see \cite{Alexandrov:2014mfa,Alexandrov:2014rca} for an attempt to go beyond this approximation).
While D-instantons are in principle completely specified by the twistorial
construction of \cite{Alexandrov:2008gh,Alexandrov:2009zh} (which generalizes a similar construction
in rigid quantum field theories \cite{Gaiotto:2008cd}), their compatibility with S-duality is far from manifest.
Indeed, this construction is formulated using coordinates adapted to type IIA string theory, whereas the action of
S-duality on $\cM_H$ is fixed in terms of type IIB coordinates. Hence, understanding the modular properties of the metric requires
specifying the relation between these two sets of fields, known as the {\it mirror map}, which itself receives quantum corrections.
Furthermore, the existence of an isometric action of $SL(2,\IZ)$ on $\cM_{\rm H}$
is equivalent to the existence of a holomorphic $SL(2,\Z)$ action on the twistor space $\cZ$
which preserves the holomorphic contact structure.
Thus the Darboux coordinates used to trivialize this contact structure
must transform according to a {\it holomorphic contact transformation},
but neither holomorphy, nor the contact property are obvious
in the original formulation of \cite{Alexandrov:2008gh,Alexandrov:2009zh}.
This issue was investigated first in the case of D1-D(-1) instantons in \cite{Alexandrov:2009qq},
where S-duality in this particular sector was made manifest by introducing a different set of Darboux coordinates,
which are related to the original ones by a {\it local} contact transformation and
which transform under S-duality in the same way as the classical Darboux  coordinates.

In the case of D3-D1-D(-1) instantons, the issue of S-duality invariance is further complicated
by the fact that the generalized DT invariants which count these instantons have a non-trivial
dependence on the moduli of $\CYm$. In \cite{Alexandrov:2012au}, the consistency of D3-D1-D(-1) instantons
with S-duality was studied in the large volume limit and one-instanton approximation,
where these wall-crossing phenomena can be ignored. It was shown that the holomorphic contact structure on $\cZ$ is indeed invariant
under  $SL(2,\IZ)$, provided the relevant DT invariants are Fourier coefficients of a vector-valued
modular form, a property which follows from the  superconformal field theory  description of these invariants \cite{Maldacena:1997de}.
In particular, the contact potential (a function akin to a \kahler potential \cite{Alexandrov:2008nk}) 
was shown to transform as a modular form of the correct weight.
The Darboux coordinates used in the construction of  \cite{Alexandrov:2008gh,Alexandrov:2009zh}, 
on the other hand, do not transform covariantly
(like their classical, uncorrected counterparts), but have anomalous transformations.
The anomaly, however, like in the case of D1-D(-1) instantons,
can be cancelled by a local holomorphic contact transformation. The generating function
for this contact transformation turned out to be an indefinite theta series of signature $(1,b_2-1)$ of the type
considered by Zwegers \cite{Zwegers-thesis} (with an insertion of  the difference of two sign functions),
which is holomorphic but not modular, whereas the contact potential
was expressed in terms of an ordinary non-holomorphic Siegel theta series.

In Part I of this work \cite{Alexandrov:2016tnf}, we extended this analysis to second order in the
multi-instanton expansion, still in the large volume limit. In this case, to deal with the moduli dependence of DT invariants,
it is useful to trade them for the so called MSW invariants, which are defined as the DT invariants
in a particular chamber of the moduli space.
Then, at second order in the expansion in powers of these invariants,
we showed that the contact potential is again modular invariant, provided the MSW invariants
for a fixed D3-brane charge $\bfp=\bfp_1+\bfp_2$ are now Fourier coefficients
of a certain vector-valued {\it mock} modular form $h_{\bfp,\bfmu}$,
with a specific shadow constructed out of the MSW invariants of the
constituents with D3-brane charges $\bfp_1$ and $\bfp_2$. In the case where $\bfp$ cannot be decomposed
as a sum of two positive vectors, this statement reduces to the one already found at one-instanton level.
In contrast to that case, at two-instanton level, the contact potential involves an indefinite theta series of signature $(2,2b_2-2)$,
where one dangerous  direction is treated like for Siegel theta series,
while the other is regularized in Zwegers' fashion \cite{Manschot:2009ia}.

In the present work, we complete this investigation and show that under the same assumptions, the holomorphic
contact structure on $\cZ$ is indeed invariant under $SL(2,\IZ)$. This guarantees
that the quaternion-\kahler metric on $\cM_{\rm H}$ carries an isometric action of the same group
(a statement which goes much beyond the modular invariance of the contact potential).
As in the one-instanton case, we show that the modular anomalous variation of 
the Darboux coordinates at two-instanton order can be cancelled by a holomorphic contact transformation,
generated by a suitable holomorphic indefinite theta series of signature $(2,n-2)$.
In more detail, we proceed as follows:
\begin{enumerate}
\item
First, we evaluate the instanton corrected holomorphic Darboux coordinates $\Xi^I$
in the two-instanton, large volume approximation and express them, using
a suitably quantum corrected mirror map,
in terms of type IIB physical fields, which transform in a fixed way under S-duality.
The result turns out to be remarkably simple: the D3-instanton corrections 
are all expressed as certain non-linear functionals $\ff^I$
of a single function $\tcJ_\bfp$ on the twistor space,
\be
\delta\Xi^I=\ff^I[\tcJ_\bfp] .
\label{res-dc}
\ee
Furthermore, the functionals $\ff^I$ are such that, if $\tcJ_\bfp$ transformed as a modular form of weight $(-1,0)$,
then the S-duality transformation of all $\Xi^I$ would be manifestly holomorphic and would preserve
the contact structure, i.e. it would be a holomorphic contact transformation.

\item
The modular transformation of $\tcJ_\bfp$ can be deduced by expressing it
in terms of the partition function of MSW invariants $h_{\bfp,\bfmu}$.
Unfortunately, even under the assumption that $h_{\bfp,\bfmu}$ is a vector-valued mock modular form,
the function $\tcJ_\bfp$ is {\it not} modular, so additional steps are needed to achieve the proof.
Therefore, at the next step, we show that the action of a class of local contact transformations on the Darboux coordinates
is captured by a simple shift of the argument in \eqref{res-dc},
\be
\gtrans\ :\quad \delta\Xi^I\mapsto \ff^I[\hcJ_\bfp],
\qquad
\hcJ_{\bfp}=\tcJ_{\bfp}-2\pi\I \,\tgH_\bfp ,
\label{gtr-dc}
\ee
where $\tgH_\bfp$ is a function constructed in a simple way from the contact hamiltonian $\gH$ generating the transformation.

\item
Next, we look for a contact hamiltonian $\gH$ which cancels the modular anomaly of $\tcJ_\bfp$ via \eqref{gtr-dc},
i.e. such that the shifted function $\hcJ_\bfp$ transforms as a modular function of weight $(-1,0)$.
To this end, we aim at representing
$\hcJ_\bfp$ as (a sum of) product of the partition function $h_{\bfp,\mu}$  (or rather,
its non-holomorphic modular completion $\widehat{h}_{\bfp,\mu}=h_{\bfp,\mu}-\frac12 R_{\bfp,\bfmu}$)
times a suitable indefinite theta series. Although the requirements of convergence,
holomorphy and modularity turn out to be very non-trivial,
we manage to solve them, thereby proving the modular symmetry of the instanton corrected hypermultiplet moduli space.
\end{enumerate}

It is important to note that our construction  relies heavily on the existence of
a rational null vector belonging to the lattice used in the definition of theta series.
In our previous work on the one-instanton approximation \cite{Alexandrov:2012au}, we
implicitly assumed that the standard lattice of charges $\Lambda=H_4(\CYm,\IZ)$
admits such a rational
null vector. But in the course of the present project it became clear
that this is by no means guaranteed, e.g. CY threefolds with  $b_2=1$
admit no null vectors except zero, and it is easy to find examples of CY threefolds with $b_2=2$
which do not admit any rational null vector (such as the examples in \cite{Hosono:1993qy}).
Here we circumvent this problem by introducing an extended lattice $\cLam$ which contains
the original lattice $\Lambda$ but, thanks to the additional (convergent) directions,
is guaranteed to admit such a null vector.

Another important comment is that at two-instanton order,
the contact transformation removing the anomaly is generated by a holomorphic indefinite theta series of `conformal' signature $(2,n-2)$.
This theta series is similar to the one that we introduced in \cite{Alexandrov:2016enp}
(which was a spin-off of the present study), but it involves a more complicated kernel.
In Appendix \ref{sec-converge} we conjecture a set of sufficient conditions for the convergence of
indefinite theta series of signature $(2,n-2)$ whose kernel consists of a cyclic sum of products of sign functions,
which may be of independent mathematical interest. Our conjecture  is supported by extensive numerical
checks but we have not attempted to prove it. Of course, these
conditions are obeyed by the relevant theta series cancelling the modular anomaly at two-instanton order.

The outline of this work is as follows. In \S\ref{sec-review} we briefly review the twistorial
construction of the D-instanton corrected hypermultiplet moduli space in type IIB string vacua,
and the modular properties of the DT and MSW invariants which count D3-brane instantons.
In \S\ref{sec-dc} we compute the corrections to the Darboux coordinates at two-instanton
order, and express them in the form \eqref{res-dc}. In \S\ref{sec-gauge}, we determine
the gauge transformation which cancels the modular anomaly
of the Darboux coordinates, and show that the new Darboux coordinates have the correct
modular properties. We conclude in \S\ref{sec-disc} with a discussion of our main results.
Various details of the computation are relegated in Appendix A. In \S\ref{sec-theta},
we review the construction of indefinite theta series in Lorentzian and conformal signature, 
based on a theorem due to Vign\`eras \cite{Vigneras:1977}
which provides conditions for their modularity and plays a crucial role in our analysis.
We also introduce there several variants of modular theta series arising in the twistorial construction.
In \S\ref{sec-converge}, we spell out our conjecture for the convergence of a class of theta series
of signature $(2,n-2)$ whose kernel consists of a cyclic sum of products of sign functions,
and apply it to the gauge transformation arising in \S\ref{sec-gauge}. In \S\ref{sec-D1} we
discuss the modifications which are needed to incorporate worldsheet and D1-D(-1)-instantons
into our construction.

\section{Twistors, D-instantons and modularity}
\label{sec-review}

In this section we collect relevant facts about the hypermultiplet moduli space in type IIB
string vacua, the twistorial description of D-instanton corrections to its metric, the action of
S-duality and the modular properties of DT invariants which count D3-brane instantons.
A more extensive review of these issues can be found
in \cite{Alexandrov:2011va,Alexandrov:2013yva} or in our previous paper \cite{Alexandrov:2016tnf}.

\subsection{$\cM_H$ is the twistor formalism}
\label{subsec-MH}

In type IIB string theory compactified on a CY threefold $\CYm$, the hypermultiplet moduli space
$\cM_H$ is a \qk manifold of dimension $4b_2(\CYm)+4$,
which describes the dynamics of the ten-dimensional axio-dilaton $\tau=c^0+\I/g_s$,
the \kahler moduli $z^a=b^a+\I t^a$ (with $a=1\dots b_2$ running over a basis of
$H_2(\CYm,\IZ))$,
the  Ramond-Ramond (RR) scalars $c^a,\cla,\cl0$, corresponding to periods of the RR
2-form, 4-form and 6-form on a basis of $H^{\rm even}(\CYm,\IZ)$, and finally,
the NS axion $\psi$, dual to the Kalb-Ramond two-form $B$ in four dimensions. At tree-level,
the metric on $\cM_H$ is obtained from the moduli space $\cM_{\cS\cK}$ of
complexified K\"ahler deformations via the
$c$-map construction \cite{Cecotti:1988qn,Ferrara:1989ik}.
$\cM_{\cS\cK}$ is a special \kahler manifold characterized by the holomorphic prepotential $F(X)$
where $X^\Lambda$ are homogeneous complex coordinates on $\cM_{\cS\cK}$
such that $X^\Lambda/X^0=z^\Lambda$ (with $z^0=1$). Thus, the tree-level metric on $\cM_H$ is completely
determined by the prepotential.

Beyond the tree-level, the metric on $\cM_H$ receives quantum $g_s$-corrections.
At the perturbative level there is only a one-loop correction, proportional to the Euler characteristics $\chi_{\CYm}$.
The corresponding metric is a one-parameter deformation of the $c$-map found explicitly in a series of works
\cite{Antoniadis:1997eg,Gunther:1998sc,Antoniadis:2003sw,Robles-Llana:2006ez,Alexandrov:2007ec}.
At the non-perturbative level, there are corrections from D-branes
wrapped on complex cycles in $\CYm$ (described by coherent sheaves on $\CYm$),
and from NS5-branes wrapped on $\CYm$, which we consistently ignore throughout this paper.

The most efficient way to describe quantum corrections to $\cM_H$ is to work at the level of its twistor space $\cZ$,
a $\CP$-bundle over $\cM_H$ endowed with a canonical holomorphic contact structure \cite{MR664330}.
This contact structure is represented by a (twisted) holomorphic one-form $\cX$,
defined up to multiplication by a nowhere vanishing holomorphic function.
Locally it can always be expressed in terms of holomorphic Darboux coordinates 
$\{\Xi^I\}=\{\xi^\Lambda,\txi_\Lambda,\alpha\}$ as
\be
\cX^{[i]}=  \de\ai{i}+ \txii{i}_\Lambda \de \xii{i}^\Lambda\, ,
\label{contform}
\ee
where the index $\scriptsize [i]$ labels the
patches $\cU_i$ of an open covering of $\CP$. The global contact structure on $\cZ$
(hence, the metric on $\cM$) is then encoded in contact transformations between
Darboux coordinate systems on the overlaps $\cU_i \cap \cU_j$ \cite{Alexandrov:2008nk}.
In turn, these transformations can be parametrized by holomorphic functions $\Hij{ij}(\Xi)$,
which we call contact hamiltonians \cite{Alexandrov:2014mfa}. Such a function defines
the holomorphic vector field
\be
\begin{split}
X_H = \left( -\p_{\txi_\Lambda} H+\xi^\Lambda\p_\alpha H \right) \partial_{\xi^\Lambda}
+ \p_{\xi^\Lambda} H\, \partial_{\txi_\Lambda} +
\left( H-\xi^\Lambda\p_{\xi^\Lambda} H \right) \partial_{\alpha},
\end{split}
\label{contbr}
\ee
the exponentiation of which generates the contact transformation $\rho_H$,
\be
\rho_H\ :\quad \Xi^I \mapsto e^{X_H} \cdot \Xi^I.
\label{contact-eXH}
\ee
Thus, a set of such holomorphic functions associated to a covering of $\CP$
(satisfying obvious consistency conditions on triple overlaps)
uniquely defines a \qk manifold.

In particular, the D-instanton corrections to $\cM_H$ were found to be generated by
the contact hamiltonians \cite{Alexandrov:2008gh,Alexandrov:2009zh}
\be
H_\gamma(\xi,\txi)= \frac{\bOm_\gamma}{(2\pi)^2}\,\sigma_\gamma\cX_\gamma\, ,
\label{prepHnew}
\ee
where $\gamma=(p^\Lambda,q_\Lambda)$ is the D-brane charge (an element of $H^{\rm even}(\CYm)$),
$\bOm_\gamma$ is the rational DT invariant, and $\sigma_\gamma$
is a quadratic refinement of the intersection pairing on $H^{\rm even}(\CYm)$, a sign factor which we fix below.
Using notation $\expe{x}=e^{2\pi\I x}$, we also defined
\be
\cX_\gamma = \expe{p^\Lambda \txi_\Lambda-q_\Lambda \xi^\Lambda},
\label{defXg}
\ee
an exponentiated version of Darboux coordinates.
The contact hamiltonians $H_\gamma$ generate contact transformations connecting Darboux coordinates
on the two sides of the BPS rays $\ell_\gamma$ on $\CP$ which, in terms of the standard stereographic coordinate $t$,
extend from $t=0$ to $t=\infty$
\be
\ell_\gamma = \{ \varpi\in \CP\ :\ \ Z_\gamma/\varpi\in \I \IR^-\}
\label{BPSray}
\ee
along the direction fixed by the phase of the central charge function
\be
Z_\gamma(z)=q_\Lambda z^\Lambda -p^\Lambda F_\Lambda(z).
\label{centralch}
\ee

Combining the gluing conditions \eqref{contact-eXH} with certain regularity conditions on the Darboux coordinates,
one can derive integral equations which fix them in terms of coordinates on $\cM_H$ and the coordinate $t$ on $\CP$.
The equations on $\xi^\Lambda$ and $\txi_\Lambda$ can be written as
\be
\cX_\gamma(t) = \cXsf_\gamma(t)\, \expe{
\frac{1}{8\pi^2}
\sum_{\gamma'} \sigma_{\gamma'}\bOm_{\gamma'}\, \langle \gamma ,\gamma'\rangle
\int_{\ell_{\gamma'} }\frac{\de t'}{t'}\, \frac{t+t'}{t-t'}\,
\cX_{\gamma'}(t')},
\label{eqTBA}
\ee
where $\langle \gamma ,\gamma'\rangle=q_\Lambda p'^\Lambda-q'_\Lambda p^\Lambda$ and
\be
\cXsf_\gamma(t)=\expe{\frac{\tau_2}{2}\(\bZ_\gamma(\ub)\,t-\frac{Z_\gamma(u)}{t}\)
+p^\Lambda \tzeta_\Lambda- q_\Lambda \zeta^\Lambda}
\label{defXsf}
\ee
are the Fourier modes of the tree-level (or `semi-flat')
Darboux coordinates valid in the absence of D-instantons.
In the weak coupling limit, these equations can be solved iteratively, leading to a (formal) multi-instanton series
for $(\xi^\Lambda,\txi_\Lambda)$ in each angular sector.
This solution can then be used to compute the remaining
Darboux coordinate $\alpha$ \cite{Alexandrov:2009zh}
\bea
\alpha(t) &= &  -\hf(\sigma+\zeta^\Lambda \tzeta_\Lambda)+\frac{\tau_2^2}{8}\(\ub^\Lambda F_\Lambda+u^\Lambda\bF_\Lambda\)
-\frac{\tau_2^2}{4}\(\varpi^{-2} F+\varpi^2 \bF\)
-\frac{\tau_2\zeta^\Lambda}{2}\( \varpi^{-1}F_\Lambda-\varpi\bF_\Lambda\)
-\frac{\I\chi_\CY}{48\pi}\,\log\varpi\nn \\
&&
-\hf \sum_{\gamma} \int_{\ell_{\gamma}} \frac{\de \varpi'}{ \varpi'} \,
\biggl[ \frac{\varpi'+\varpi}{\varpi'-\varpi}
 \(\frac{1}{2 \pi \I}+q_{\Lambda}\xi^\Lambda(\varpi') -\pi\I q_\Lambda p^\Lambda H_{\gamma}(t') \)H_{\gamma}(t')
\biggr.
\label{alphaIIA}
\\
&&\biggl.
\quad
+\frac{\tau_2 p^\Lambda}{2}\Bigl( (\varpi^{-1}+\varpi'^{-1})F_\Lambda(u)+(\varpi+\varpi')  \bF_\Lambda(\ub) \Bigr)H_{\gamma}(t') \biggr]
-\frac{1}{8}\sum_{\gamma,\gamma'\in\Gamma}p^\Lambda q_\Lambda\int_{\ellg{\gamma}}\frac{\d \varpi'}{\varpi'}\,H_{\gamma}
\int_{\ellg{\gamma'}}\frac{\d \varpi''}{\varpi''}\,H_{\gamma'}.
\nn
\eea
Note that all instances of the prepotential $F(X)$ in \eqref{defXsf} and \eqref{alphaIIA} are evaluated at  $X^\Lambda=(1,u^a)$.
Together with  $(\tau_2, \zeta^\Lambda, \tzeta_\Lambda,\sigma)$ appearing in these equations, they
play the role of coordinates on $\cM_H$. They are adapted to the symmetries of the type IIA formulation\footnote{In particular,
$(\zeta^\Lambda,\tzeta_\Lambda)$ and $\sigma$ transform as symplectic vector and scalar, respectively.
This is consistent with that $(\xi^\Lambda,\txi_\Lambda)$ and $\talp=-2\alpha-\xi^\Lambda\txi_\Lambda$
also transform as vector and scalar. It is the latter coordinate $\talp$
that was usually used to present the D-instanton construction in previous papers.}
and therefore can be considered as natural coordinates on the moduli space of type IIA string theory compactified
on the mirror CY threefold $\CY$.
In particular, $u^a$ can be understood as the complex structure moduli of $\CY$.
The relation between the type IIA and type IIB fields will be explained in the next subsection.

\subsection{S-duality}
\label{subsec-S}

In the classical approximation (i.e. tree-level, large volume limit)
the metric on $\cM_H$ is known to be invariant under the action of $SL(2,\IR)$, which descends from the S-duality group
of type IIB supergravity in 10 dimensions. On the coordinates introduced in the beginning of the previous
subsection, an element
$\trans={\scriptsize \begin{pmatrix} a & b \\ c & d \end{pmatrix}} \in SL(2,\IR)$ acts by
\be\label{SL2Z}
\trans \ :\quad
\begin{array}{c}
\displaystyle{
\tau \mapsto \frac{a \tau +b}{c \tau + d} \, ,
\qquad
t^a \mapsto t^a |c\tau+d| \, ,
\qquad
\cla\mapsto \cla \,  ,}
\\
\displaystyle{
\begin{pmatrix} c^a \\ b^a \end{pmatrix} \mapsto
\begin{pmatrix} a & b \\ c & d  \end{pmatrix}
\begin{pmatrix} c^a \\ b^a \end{pmatrix} ,
\qquad
\begin{pmatrix} \cl0 \\ \psi \end{pmatrix} \mapsto
\begin{pmatrix} d & -c \\ -b & a  \end{pmatrix}
\begin{pmatrix} \cl0 \\ \psi \end{pmatrix}}.
\end{array}
\ee

To uncover this symmetry in the twistorial formulation of the previous subsection,
one should drop all integral contributions (tree level limit) and retain only the classical part of the prepotential
(large volume limit), determined by the triple intersection numbers $\kappa_{abc}$,
\be
\label{Flv}
\Fcl(X)=-\kappa_{abc}\frac{X^a X^b X^c}{6 X^0}\, .
\ee
Then, one should lift the symmetry to the twistor space, i.e.  supplement \eqref{SL2Z} with a transformation
of the fiber coordinate $t$, such that  the combined transformation acts  on $\cZ$
by a {\it holomorphic  contact transformation}, i.e. a holomorphic  transformation
which preserves the contact one-form $\cX$ up to a non-vanishing holomorphic factor.
The lift of the action of $SL(2,\IR)$ on $\cM_H$ was determined in \cite{Alexandrov:2008nk} and
is most easily formulated in terms of coordinate $z$ related to $t$ by
\be
\label{Cayley}
z=\frac{t+\I}{t-\I}\, .
\ee
Then it takes the simple form \cite{Alexandrov:2012au}
\be
\label{ztrans}
\trans \ :\quad
z\mapsto \frac{c\bar\tau+d}{|c\tau+d|}\, z\,  .
\ee

To check that this lift indeed generates a contact transformation, one has to relate the type IIA fields,
appearing in the expressions for the Darboux coordinates \eqref{defXsf} and \eqref{alphaIIA},
in terms of the type IIB ones. Such relation is known as {\it classical mirror map} and is given by \cite{Bohm:1999uk}
\be
\label{symptobd}
\begin{split}
u^a & =b^a+\I t^a\, ,
\qquad
\zeta^0=\tau_1\, ,
\\
\tzeta_a &=  \cla+ \frac{1}{2}\, \kappa_{abc} \,b^b (c^c - \tau_1 b^c)\, ,
\qquad
\tzeta_0 = \cl0-\frac{1}{6}\, \kappa_{abc} \,b^a b^b (c^c-\tau_1 b^c)\, ,
\\
\sigma &= -(2\psi+  \tau_1 \cl0) + \cla (c^a - \tau_1 b^a)
-\frac{1}{6}\,\kappa_{abc} \, b^a c^b (c^c - \tau_1 b^c)\, .
\end{split}
\ee
It is straightforward to check that the classical Darboux coordinates, after the substitution of \eqref{symptobd},
transform in the following way:
\be
\label{SL2Zxi}
\begin{split}
&
\xi^0 \mapsto \frac{a \xi^0 +b}{c \xi^0 + d} , \qquad
\xi^a \mapsto \frac{\xi^a}{c\xi^0+d} , \qquad
\txi_a \mapsto \txi_a +  \frac{ c}{2(c \xi^0+d)} \kappa_{abc} \xi^b \xi^c,
\\
&
\begin{pmatrix} \txi_0 \\ \alpha \end{pmatrix} \mapsto
\begin{pmatrix} d & -c \\ -b & a  \end{pmatrix}
\begin{pmatrix} \txi_0 \\  \alpha \end{pmatrix}
+ \frac{1}{6}\, \kappa_{abc} \xi^a\xi^b\xi^c
\begin{pmatrix}
c^2/(c \xi^0+d)\\
-[ c^2 (a\xi^0 + b)+2 c] / (c \xi^0+d)^2
\end{pmatrix} .
\end{split}
\ee
This shows that the transformation is holomorphic. Furthermore, applying it to the contact one-form \eqref{contform},
one finds that
\be
\cX\mapsto\frac{\cX}{c\xi^0+d},
\label{tr-contact}
\ee
which proves that it is indeed a contact transformation.

The continuous $SL(2,\IR)$ isometry is broken by quantum corrections. However,
a discrete subgroup $SL(2,\IZ)$ is expected to survive at the quantum level \cite{RoblesLlana:2006is}.
It is realized by the same transformation \eqref{SL2Z}, where all parameters are integer satisfying $ad-bc=1$,
except that the transformation law of $\cla$ is slightly modified and given now by \cite{Alexandrov:2010ca,Alexandrov:2012au}
\be
\trans \ :\quad \cla\mapsto \cla - c_{2,a} \varepsilon(\trans),
\label{trtca}
\ee
where $c_{2,a}$ are the components of the second Chern class on a basis of $H^{4}(\CYm,\IZ)$
and $\varepsilon(\trans)$ is the logarithm of the multiplier system of the Dedekind eta function.
In the absence of fivebrane instantons the lift to the twistor space \eqref{ztrans} also remains unchanged \cite{Alexandrov:2012bu}.
In particular, the claim is that in this limit, which includes
D3-D1-D(-1)-instantons as well as $\alpha'$-corrections,
a combination of \eqref{SL2Z}, \eqref{trtca} and \eqref{ztrans} acts by a contact transformation,
after a suitable modification of the mirror map \eqref{symptobd}.
It is this claim that we would like to verify in this paper, restricting ourselves to the two-instanton, large volume approximation.
In the main text, for simplicity, we ignore contributions from the one-loop $g_s$ and $\alpha'$-corrections as well as
from pure D1-D(-1)-instantons. In Appendix \ref{sec-D1},
generalizing the results of \cite{Alexandrov:2009qq,Alexandrov:2012bu,Alexandrov:2012au}, we show how
they can be included consistently with S-duality.

\subsection{D3-instantons in the large volume limit}
\label{subsec-D3}

In this subsection we specify the general construction of D-instantons outlined in section \ref{subsec-MH}
to the case of D3-instantons of the type IIB formulation.
We also review the modular properties of the relevant DT invariants and some of the results of
\cite{Alexandrov:2012au,Alexandrov:2016tnf} needed for the subsequent analysis.
For brevity, we shall extensively use the notation $(lkp)=\kappa_{abc}l^a k^b p^c$ and
$(kp)_a=\kappa_{abc}k ^b p^c$.

\subsubsection{DT and MSW invariants}
\label{subsubsec-DTMSW}

An instanton corresponding to a bound state of D3-D1-D(-1)-branes in type IIB is characterized by the charge vector
$\gamma=(0,p^a,q_a,q_0)\in\Gamma$ satisfying the following quantization conditions
\be
\label{fractionalshiftsD5}
p^a\in\IZ ,
\qquad
q_a \in \IZ  + \frac12 \,(p^2)_a ,
\qquad
q_0\in \IZ-\frac{1}{24}\, p^a c_{2,a} .
\ee
We assume that the charge $p^a$ corresponds to an effective divisor in Calabi-Yau and
belongs to the \kahler cone, i.e.
\be
\label{khcone}
p^3> 0,
\qquad
(r p^2)> 0,
\qquad
k_a p^a > 0,
\ee
for all effective divisors $r^a \gamma_a \in H_4^+(\CYm,\IZ)$ and
effective curves $k_a \gamma^a \in H_2^+(\CYm,\IZ)$, where $\gamma_a$ denotes an
integer basis of $ \Lambda=H_4(\CYm,\IZ)$, with $\gamma_a$ being irreducible divisors,
whereas $\gamma^a$ an integer basis of $\Lambda^*=H_2(\CYm,\IZ)$.
The set of charges with such properties will be denoted by $\Gamma_+$.
The charge $p^a$ gives rise to the quadratic form $\kappa_{ab}=\kappa_{abc} p^c$ on $\Lambda\otimes \IR\simeq \IR^{b_2}$
of signature $(1,b_2-1)$.
We use it to identify $\Lambda\otimes \IR$ and $\Lambda^*\otimes \IR$
and we use bold-case letters to denote the corresponding vectors. The lattice $\Lambda$ can also be
identified with its image in the dual lattice $\Lambda^*$. But since in general the map $\epsilon^a \mapsto \kappa_{ab} \epsilon^b$
is not surjective, the quotient $\Lambda^*/\Lambda$ is a finite group of order $|\det\kappa_{ab}|$.

The DT invariants $\bOm_\gamma$ associated to charges $\gamma\in\Gamma_+$
are invariant under the combination of the shift of the Kalb-Ramond field, $b^a\mapsto b^a +\epsilon^a$,
and the spectral flow transformation acting on the electric charges
\be
\label{flow}
q_a \mapsto q_a - (p\epsilon)_a,
\qquad
q_0 \mapsto q_0 - \epsilon^a q_a + \frac12\, (p\epsilon \epsilon).
\ee
It is important to take into account the shift of $b^a$ because the DT invariants are only
piecewise constant as functions of the complexified K\"ahler moduli $z^a=b^a+\I t^a$
and jump across walls of marginal stability $\cW_{\gamma_1,\gamma_2}$ where the phases of the central charge functions $Z_{\gamma_1}(z)$
and $Z_{\gamma_2}(z)$ become equal.
At these walls some BPS bound states of charge $\gamma=m_1\gamma_1+m_2\gamma_2$ with $m_1 m_2>0$ become unstable and decay into
multi-particle BPS states of charges $\gamma_1$, $\gamma_2$, hence explaining the jump in the BPS index.
For our purposes, it is enough to restrict to the case of primitive wall-crossing, where $(m_1,m_2)=\pm (1,1)$.

To deal with this wall-crossing phenomenon, it is convenient to introduce MSW invariants
$\bOmMSW_\gamma=\bOm_\gamma(\bfz_\infty(\gamma))$, defined as the DT invariants evaluated at the `large volume attractor point',
\be
\label{lvolatt}
\bfz_\infty(\gamma)=\lim_{\lambda\to +\infty}\(\bfb(\gamma)+\I\lambda \bft(\gamma)\)
= \lim_{\lambda\to +\infty}\(-\bfq+\I\lambda  \bfp\)\ .
\ee
The MSW invariants are therefore by definition independent of the moduli.
The wall-crossing behavior of the DT invariants is then captured by their expansion in terms of MSW invariants.
To second order in the $\bOmMSW$'s, this expansion reads \cite{Manschot:2009ia}
\bea
\bOm_\gamma(\bfz)&=&\bOmMSW_\gamma
+\frac{1}{2}\sum_{\gamma_1,\gamma_2\in \Gamma_+\atop \gamma_1+\gamma_2=\gamma}
(-1)^{\langle\gamma_1,\gamma_2\rangle} \langle\gamma_1,\gamma_2\rangle
\,\Dela_{\gamma_1\gamma_2}\,
\bOmMSW_{\gamma_1}\, \bOmMSW_{\gamma_2}
+\cdots,
\label{multiMSW}
\eea
where $\Dela_{\gamma_1\gamma_2}$ is a `sign factor'
\be
\Dela_{\gamma_1\gamma_2}=
\hf\,\Bigl[\sgn(\cI_{\gamma_1\gamma_2}(\bft))-\sgn\langle\gamma_1,\gamma_2 \rangle\Bigr],
\label{defDela}
\ee
which vanishes near the large volume attractor point, but becomes non-vanishing after crossing a wall of marginal stability where
a bound state of two constituents of charges $\gamma_1$ and $\gamma_2$ becomes stable.
Here $\cI_{\gamma_1\gamma_2}$ is defined by
\be
\cI_{\gamma_1\gamma_2}= -\frac{2\,{\rm Im}(Z_{\gamma_1} \bar Z_{\gamma_2})}{\sqrt{(p_1t^2)\, (p_2 t^2) \, (pt^2)}}
\simeq \frac{(p_2 t^2)\(q_{1,a}+(bp_1)_a\)t^a- (p_1 t^2)\(q_{2,a}+(bp_2)_a\)t^a}
{\sqrt{(p_1t^2)\, (p_2 t^2) \, (pt^2)}}\, ,
\label{defIgg}
\ee
where the second equality holds in the large volume limit.

An important property of the MSW invariants is that they are both independent of the moduli and invariant
under the spectral flow action \eqref{flow} on the charges.
As a result, they only depend on $p^a, \mu_a$ and $\hat q_{0}$, where we traded the electric charges $(q_a,q_0)$
for $(\epsilon^a, \mu_a,\hat q_0)$. The latter comprise the spectral flow parameter $\epsilon^a$, the residue class
$\mu_a\in \Lambda^*/\Lambda$ defined by the decomposition
\be
\label{defmu}
q_a = \mu_a + \frac12\, \kappa_{abc} p^b p^c + \kappa_{abc} p^b \epsilon^c,
\qquad
\bfeps\in\Lambda\, ,
\ee
and the invariant charge
\be
\label{defqhat}
\hat q_0 \equiv
q_0 -\frac12\, \kappa^{ab} q_a q_b\, ,
\ee
which is invariant under \eqref{flow}.
This allows to write $\bOmMSW_\gamma=\bOm_{\bfp,\bfmu}( \hat q_0)$.

Given that the invariant charge $\hat q_0$ is bounded from above by $\hat q_0^{\rm max}=\tfrac{1}{24}(p^3+c_{2,a}p^a)$,
it is natural to introduce a generating function of the MSW invariants
\be
\label{defchimu}
h_{\bfp,\bfmu}(\tau) = \sum_{\hat q_0 \leq \hat q_0^{\rm max}}
\bOm_{\bfp,\bfmu}(\hat q_0)\,
\expe{-\hat q_0 \tau }.
\ee
For an {\it irreducible} divisor $p^a \gamma_a \in H_4^+(\CYm,\IZ)$, this function
is known to be a vector-valued holomorphic modular form of negative weight $(-\frac{b_2}{2}-1,0)$
and multiplier system $M(\trans)=M_{\cZ}\times {M_\theta}^{-1}$
where $M_\theta$ is the multiplier system of the Siegel-Narain theta series
and $M_{\cZ}=\expe{\varepsilon(\trans)\,{\bfch}\cdot \bfp}$ with $\varepsilon(\trans)$ defined below \eqref{trtca}
\cite{Gaiotto:2006wm,deBoer:2006vg,Denef:2007vg}. However, in
\cite{Alexandrov:2016tnf} it was found that, if $\bfp$ is reducible
and can be written as a sum of two effective divisors,
$\bfp=\bfp_1+\bfp_2$, then $h_{\bfp,\bfmu}$ is a mock modular function whose
modular completion is provided by
\be
\whh_{\bfp,\bfmu}=h_{\bfp,\bfmu}-\hf\,R_{\bfp,\bfmu},
\label{defhh}
\ee
where $R_{\bfp,\bfmu}$ is a non-holomorphic function of $\tau$
constructed out of the MSW invariants.
Its explicit expression can be found in
\cite[Eq.(1.3)]{Alexandrov:2016tnf} in the case where $\bfp$ is reducible into at most two effective divisors. 
In this paper, we will only need the key property that  for any suitably decaying kernel $\Phi(\bfx):\IR^{b_2}\to \IC$, $R_{\bfp,\bfmu}$
satisfies the following identity\footnote{In \cite{Alexandrov:2016tnf}
the identity with kernel $\Phi=e^{-\pi x_+^2}$ was taken as the defining property of $R_{\bfp,\bfmu}$, but it is easy to see that
then it holds for a generic kernel. Note that we indicated the integer $\lambda$ controlling the weight of the theta series
since it is important for keeping track of powers of $\tau_2$.}
\bea
&&
\sum_{\bfmu\in\Lambda^\star/\Lambda}
R_{\bfp,\bfmu}\,\vartheta_{\bfp,\bfmu}[\Phi(\bfx),\lambda]
\label{hhtoRth}\\
&
=& -\frac{1}{4\pi\sqrt{2}} \,\sum_{\bfp_1+\bfp_2=\bfp}
\sum_{\bfmu_s\in \Lambda_s^\star/\Lambda_s}
h_{\bfp_1,\bfmu_1}\,h_{\bfp_2,\bfmu_2} \,
\vartheta_{\bfp_1,\bfp_2,\bfmu_1,\bfmu_2}\[|\Xxx|\beta_{\frac{3}{2}}\!\left({\textstyle\frac{\Xxx^2 }{(pp_1p_2)}}\right)
\Phi(\bfx_1+\bfx_2),\lambda+1\] ,
\nn
\eea
where $s\in\{1,2\}$, $\vartheta_{\bfp,\bfmu}[\Phi,\lambda]$ is the theta series  \eqref{Vignerasth}
for the lattice $\Lambda$, and $\vartheta_{\bfp_1,\bfp_2,\bfmu_1,\bfmu_2}[\tilde\Phi,\tilde\lambda]$
is the similar theta series for the doubled lattice $\Lambda_1\oplus\Lambda_2$ where $\Lambda_s$ is the lattice
with quadratic form defined by the charge $\bfp_s$.
On the r.h.s. we introduced the function $\beta_\nu(y)=\int_y^{+\infty} \de u\, u^{-\nu} e^{-\pi u}$, so that for $x\in\IR$
\be
\label{betahaf}
\beta_{\half}(x^2) = \Erfc(\sqrt{\pi} |x|),
\qquad
\beta_{\frac{3}{2}}(x^2)=2|x|^{-1}e^{-\pi x^2}-2\pi \beta_{\half}(x^2) ,
\ee
and used the notation $\Xxx=\bfx_1\cdot \bfp_2-\bfx_2\cdot \bfp_1$. 

\subsubsection{$\cX_\gamma$ at one-instanton order}
\label{subsubsec-Xg}

In the large volume limit, for $\gamma\in\Gamma_+$, 
the absolute value of the central charge function $Z_\gamma$ has the expansion
\be
\label{ZlargeV}
|Z_\gamma|=\half (p t^2)-q_0+(q+b)_+^2-(\bfq+\half\bfb)\cdot\bfb +\cdots,
\ee
where we used the notation $q_+ = \frac{q_a t^a}{\sqrt{(p t^2)}}$.
The twistorial integrals along BPS rays $\ell_\gamma$
are thus dominated by a saddle point at
\be
z'_\gamma\approx
-\I\frac{(q+b)_+}{\sqrt{(pt^2)}}.
\ee
For negative charge $\gamma\in-\Gamma_+$, the saddle point is at $z'_{-\gamma}=1/z'_\gamma$. This shows that in all integrands
we can send $z'$ either to zero or infinity keeping constant $t^a z' $ or $t^a/z'$, respectively.
This also motivates us to take the same limit for $z$ in all Darboux coordinates,
i.e. to evaluate them around the point $z=0$. This is sufficient to capture the metric on $\cM_H$
and, since the locus $z=0$ is invariant under S-duality, it also suffices to check
modular invariance  \cite{Alexandrov:2012au}.

In this combined limit, one finds the following result for the expansion of the Fourier modes $\cX_\gamma$
to the first order in the DT invariants \cite{Alexandrov:2012au}:
\be
\cX_\gamma=\cXcl_\gamma\(1+\cXq_\gamma+\cdots\),
\label{expcX}
\ee
where for $\gamma\in\Gamma_+$ one has
\be
\begin{split}
\vphantom{A\over A_A}
\cXcl_\gamma =&\, e^{-S^{\rm cl}_\bfp}\,
\expe{ - \hat q_0\tau - \frac{\tau}{2}\,(q+b)_-^2-\frac{\bar\tau}{2}\, (q+b)_+^2 +\bfc \cdot(\bfq +\haf \bfb)+\I \tau_2 (pt^2)
\left(z-z'_\gamma \right)^2},
\\
\cXq_{\gammap}=&\,
\frac{1}{2\pi}\sum_{\gamma'\in\Gamma_+}\sigma_{\gamma'}\bOm_{\gamma'}\int_{\ell_{\gamma'}}\de z'
\((tpp')-\frac{\I\langle\gamma,\gamma'\rangle}{z'-z}\)\cXcl_{\gamma'},
\end{split}
\label{cXzero}
\ee
and $S^{\rm cl}_\bfp$ is the leading part of the Euclidean D3-brane action
in the large volume limit
\be
S^{\rm cl}_\bfp = \pi\tau_2(pt^2) - 2\pi \I  p^a \tc_a .
\label{clactinst}
\ee
Note that $\cXcl_\gamma$ is the part of $\cXsf_\gamma$ \eqref{defXsf} obtained by using the classical mirror map \eqref{symptobd},
whereas $\cXq_{\gamma}$ has two contributions: one from the integral term in the equation \eqref{eqTBA}
and another from quantum corrections to the mirror map \eqref{inst-mp}.
For the opposite charge, the results can be obtained via the complex conjugation and the antipodal map,
$\cX_{-\gamma}(z)=\overline{\cX_\gamma(-\bz^{-1})}$.

Finally, we fix the quadratic refinement appearing in \eqref{prepHnew} and the above results for the Darboux coordinates.
We choose it to be $\sigma_\gamma=\expe{\hf\, p^a q_a}\sigma_\bfp$ where $\sigma_\bfp=\expe{\hf\, A_{ab}p^ap^b}$
and $A_{ab}$ is a matrix of half-integers satisfying \cite{Alexandrov:2010ca}
\be
\label{ALS}
A_{ab} p^p - \frac12\, \kappa_{abc} p^b p^c\in \IZ \quad \text{for}\ \forall p^a\in\IZ.
\ee
It is easy to check that such quadratic refinement satisfies the defining relation
$
\sigma_{\gamma_1}\sigma_{\gamma_2}=(-1)^{\langle\gamma_1,\gamma_2\rangle}\sigma_{\gamma_1+\gamma_2}.
$

\section{D-instanton corrected Darboux coordinates}
\label{sec-dc}

In this section we compute the expansion of Darboux coordinates at two-instanton order 
in the combined limit $t^a\to\infty$,$z\to 0$ keeping $zt^a$ fixed, and make manifest 
the restrictions imposed by modular symmetry on the instanton contributions.

\subsection{Quantum corrections as modular forms}
\label{subsec-qcmod}

Recall that our aim is to verify that S-duality acts on the twistor space by a holomorphic
contact transformation. To this end, one must first express all Darboux coordinates 
in terms of the type IIB fields using a suitably corrected version of the classical mirror map \eqref{symptobd},
and then apply the combination of \eqref{SL2Z}, \eqref{trtca} and \eqref{ztrans} to the result.
If the resulting transformation is indeed holomorphic, it will have to coincide with \eqref{SL2Zxi},
or be a small deformation thereof.
But the transformation \eqref{SL2Zxi} is already highly non-linear and mixes classical and quantum contributions
as well as various Darboux coordinates between each other. This makes the verification of the modular symmetry
very complicated.

To overcome this problem, we split all Darboux coordinates into classical and quantum pieces, $\Xi^I=\Xicl^I+\delta\Xi^I$,
and introduce combinations of the quantum contributions $\delta\Xi^I$ which
transform in a simple way under S-duality.
More precisely, we define\footnote{Recall that  in the absence of fivebrane instantons, the Darboux coordinate $\xi^0$ remains
equal to its classical value
$\xi^0=\tau+\frac{2\I\tau_2 z^2}{1-z^2}$ at the quantum level, so $\delta\xi^0=0$.
The result below generalizes \cite[Eq (4.5)]{Alexandrov:2012au}.}.
\be
\label{defdelta}
\begin{split}
\hat\delta\xi^a =&\, (1-z^2) \delta\xi^a ,
\\
\hat\delta\txi_a =&\,
\delta\txi_a+\kappa_{abc}\[b^b-\I z\(t^b  +\frac{\hat\delta\xi^b}{4\tau_2 z}\) \] \frac{\hat\delta\xi^c}{1-z^2},
\\
\hat \delta_+ \alpha =&\, \delta\alpha+\tau\, \delta \txi_0
+\kappa_{abc}\(b^a-\I z\(t^a  +\frac{\hat\delta\xi^a}{4\tau_2 z}\)\)
\Biggl[c^b-\tau b^b-\frac{\I\tau_2 z^2 b^b}{1-z^2}
\Biggr.
\\
&\, \Biggl.\qquad
-\frac{\tau_2 z(2-z^2)}{1-z^2}\( t^b +\frac{\hat\delta\xi^b}{4\tau_2 z} \)\Biggr]\frac{\hat\delta\xi^c}{1-z^2}
+\frac{\I(2-z^2)}{48\tau_2 (1-z^2)^2}\, \kappa_{abc}\hat\delta\xi^a\hat\delta\xi^b\hat \delta\xi^c,
\\
\hat\delta_-\alpha =&\, \delta\alpha+\bar\tau\, \delta\txi_0
+\kappa_{abc}\(b^a-\I z\(t^a  +\frac{\hat\delta\xi^a}{4\tau_2 z}\)\)
\Biggl[c^b-\tau_1 b^b-\frac{\I\tau_2 z^2 b^b}{1-z^2}
\Biggr.
\\
&\, \Biggl.\qquad
-\frac{\tau_2 z}{1-z^2}\( t^b +\frac{\hat\delta\xi^b}{4\tau_2 z} \)\Biggr]\frac{\hat\delta\xi^c}{1-z^2}
+\frac{\I}{48\tau_2 (1-z^2)^2}\, \kappa_{abc}\hat\delta\xi^a\hat\delta\xi^b\hat\delta\xi^c.
\end{split}
\ee
Then one can show that the Darboux coordinates transform under $SL(2,\IZ)$ as in \eqref{SL2Zxi}
if and only if the combinations \eqref{defdelta} undergo the following simple transformations
\be
\hat\delta\xi^a \mapsto  \frac{\hat\delta \xi^a}{c\tau+d},
\qquad
\hat\delta\txi_a \mapsto  \hat\delta \txi_a,
\qquad
\hat\delta_+ \alpha  \mapsto  \frac{\hat\delta_+\alpha}{c\tau+d},
\qquad
\hat\delta_- \alpha\mapsto  \frac{\hat\delta_-\alpha}{c\bar\tau+d}.
\label{SL2hatdel}
\ee
Thus, instead of the complicated transformation rules of quantum corrections to Darboux coordinates,
we need to verify that their combinations introduced above, which we denote collectively by $\hat\delta\Xi^I$,
transform as ordinary modular forms of suitable weight.

\subsection{Darboux coordinates at two-instanton order}
\label{subsec-dc2order}

In Appendix \ref{subsec-compDc} we compute the contributions $\hat\delta\Xi^I$
in the  following double approximation:
i) in the large volume limit combined with $z\to 0$ keeping $zt^a$ fixed,
ii) up to second order of the expansion in the DT invariants.
Remarkably, all $\hat\delta\Xi^I$ can be expressed only in terms of two functions, which we denote by $\tcF_\bfp$ and $\tcJ_\bfp$,
and a set of the modular-covariant derivative operators introduced in Appendix \ref{subsec-oper}.
Let us first define these functions.

To this end, we start from the two natural twistorial integrals
\be
\cF=\sum_{\gamma\in\Gamma_+} \int_{\ell_{\gammap}} \de z'H_{\gammap},
\qquad
\cJ(z)=\sum_{\gamma\in\Gamma_+} \int_{\ell_{\gammap}} \frac{\de z'}{z'-z}\,H_{\gammap} .
\label{defcJ}
\ee
Substituting \eqref{prepHnew} and the expansion of $\cX_\gamma$ \eqref{expcX},
one can represent these functions as a series in the DT invariants\footnote{$\sum\limits_\bfp$ is always understood
as a sum over charges satisfying \eqref{khcone}.}
\be
\cF=\sum_{\bfp}\cFi{1}_{\bfp}+\sum_{\bfp_1,\bfp_2}\cFi{2}_{\bfp_1\bfp_2}+\cdots\, ,
\qquad
\cJ(z)=\sum_{\bfp}\cJi{1}_\bfp(z)+\sum_{\bfp_1,\bfp_2}\cJi{2}_{\bfp_1\bfp_2}(z)+\cdots\, .
\label{expcJ}
\ee
The functions $\tcF_\bfp$ and $\tcJ_\bfp$
are obtained by {\it halving} the coefficient of the second order terms,
so that one defines
\be
\tcF_\bfp=\cFi{1}_{\bfp}+\hf\sum_{\bfp_1+\bfp_2=\bfp}\cFi{2}_{\bfp_1\bfp_2},
\qquad
\tcJ_\bfp(z)=\cJi{1}_{\bfp}(z)+\hf\sum_{\bfp_1+\bfp_2=\bfp}\cJi{2}_{\bfp_1\bfp_2}(z).
\label{deftcJp}
\ee
The first function $\tcF_\bfp$ was already introduced in \cite{Alexandrov:2016tnf}.
It was shown that in our approximation it transforms as a modular form of
weight $\(-\tfrac32,\tfrac12\)$.
The second function $\tcJ_\bfp$ is a natural extension of the first one to twistor space.
It will play a  central  role
in  determining the modular behavior of the Darboux coordinates in what follows.

Next, let us introduce non-linear functionals $\ff^I[\fa_\bfp]$ acting on a set of functions
on twistor space labeled by the charge vector $p^a$,
$\fa_\bfp\, : \, \cZ\to \IC$. They read as
\bea
\ff^a[\fa_\bfp] &=&
\sum_{\bfp} p^a \fa_\bfp
+\frac{\pi\I}{2}\sum_{\bfp_1,\bfp_2}(p_2^a-p_1^a)
\[ \(p_2^b\Dop{\bfp_1}_b\fa_{\bfp_1}\)\fa_{\bfp_2}-\fa_{\bfp_1}\(p_1^b\Dop{\bfp_2}_b\fa_{\bfp_2}\)\],
\nn\\
\vphantom{\rule{0pt}{0.8cm}}
\tilde\ff_a[\fa_\bfp] &=&
 \sum_{\bfp}\Dop{\bfp}_a\fa_\bfp
+\pi\I \sum_{\bfp_1,\bfp_2}\[ \(p_2^b\Dop{\bfp_1}_b\fa_{\bfp_1}\)\(\Dop{\bfp_2}_a\fa_{\bfp_2}\)
- \fa_{\bfp_1}\(p_1^b\Dop{\bfp_2}_b\Dop{\bfp_2}_a\fa_{\bfp_2}\)\],
\nn\\
\vphantom{\rule{0pt}{0.8cm}}
\ff_+[\fa_\bfp] &=&
\sum_{\bfp}\Del{\bfp}_+\fa_\bfp
+\pi\I \sum_{\bfp_1,\bfp_2}\[ \(p_2^a\Dop{\bfp_1}_a\fa_{\bfp_1}\)\(\Del{\bfp_2}_+\fa_{\bfp_2}\)
- \fa_{\bfp_1}\(p_1^a\Dop{\bfp_2}_a\Del{\bfp_2}_+\fa_{\bfp_2}\)
\right.
\label{deffunctional}\\
&&\left.
+2\tau_2\kappa^{ab}_{(\bfp_1+\bfp_2)}\kappa_{bcd}t^d\( p_1^c\( p_2^g \Dop{\bfp_1}_g\fa_{\bfp_1}\)\(\Dop{\bfp_2}_a\tcF_{\bfp_2}\)
- p_2^c\( p_2^g \Dop{\bfp_1}_g\Dop{\bfp_1}_a\fa_{\bfp_1}\)\tcF_{\bfp_2}
\right.\right.
\nn\\
&&\left.\left.\qquad
-p_1^c \fa_{\bfp_1}\(p_1^g \Dop{\bfp_2}_g\Dop{\bfp_2}_a\tcF_{\bfp_2}\)+p_2^c\(\Dop{\bfp_1}_a\fa_{\bfp_1}\)\(p_1^g\Dop{\bfp_1}_g\tcF_{\bfp_2}\)
\)
\],
\nn\\
\vphantom{\rule{0pt}{0.8cm}}
\ff_-[\fa_\bfp] &=&
\sum_{\bfp}\Del{\bfp}_- \fa_\bfp
+\pi\I \sum_{\bfp_1,\bfp_2}\[ \(p_2^a\Dop{\bfp_1}_a\fa_{\bfp_1}\)\(\Del{\bfp_2}_-\fa_{\bfp_2}\)
- \fa_{\bfp_1}\(p_1^a\Dop{\bfp_2}_a\Del{\bfp_2}_-\fa_{\bfp_2}\)\]
\nn\\
&&
-\I\tau_2 z \sum_{\bfp}(\bfp\cdot \bft^2)\overline{\tcF_\bfp}\, ,
\nn
\eea
where $\kappa^{ab}_{(\bfp)}$ is the inverse of the matrix $\kappa_{abc}p^c$.
Then  the calculation in Appendix \ref{subsec-compDc} shows that the combinations
\eqref{defdelta} are expressed as follows\footnote{The function $\tcJ_\bfp$ is discontinuous across BPS rays,
which in the large volume limit correspond to the straight lines $\Im z= -\frac{(\bfq+\bfb)\cdot\bft}{{(pt^2)}}$.
 Our manipulations are supposed to be valid
only away from these rays so that the delta functions on the twistor space, 
appearing formally from the action of the derivative operators,
can be safely ignored. \label{foot-deltas}}
\be
\hat\delta\Xi^I=\ff^I[\tcJ_\bfp].
\label{resdXi}
\ee

Given the modular properties of $\tcF_\bfp$ and all derivative operators entering \eqref{deffunctional},
it is immediate to see that $\ff^I[\tcJ_\bfp]$ would transform under $SL(2,\IZ)$ precisely as required by \eqref{SL2hatdel}
provided $\tcJ_\bfp$ were a modular form of weight $(-1,0)$.
However, as was found in \cite{Alexandrov:2012au}, this is not true even in the leading one-instanton approximation.
In the next section we explain how to cure this problem.

\section{Gauge transformation and modularity}
\label{sec-gauge}

While the QK metric on $\cM_H$ is uniquely specified by the holomorphic contact structure on the twistor space,
Darboux coordinates on $\cZ$ are only defined up to a local holomorphic contact transformation.
Upon performing such transformations in different patches of $\cZ$,
the contact hamiltonians relating these Darboux coordinates, and even the covering of $\cZ$, may change, but the
global holomorphic contact structure stays invariant. Nevertheless, this freedom plays an important role because
different choices of covering of $\cZ$ may help to make different symmetries manifest: e.g. the manifestly S-duality invariant
twistorial construction of D1-D(-1)-instantons
relevant in the type IIB formulation was related to the type IIA construction presented
in section \ref{subsec-MH} by such a change \cite{Alexandrov:2009qq}.
Therefore, one may expect that a similar trick will allow to remove the modular anomaly
of the D3-instanton contributions found in the previous section.

Keeping this idea in mind, let us study the effect of a class of local contact transformations,
which we call for brevity `gauge transformations', on the D-instanton corrected Darboux coordinates.
As was mentioned in section \ref{subsec-MH}, all contact transformations are generated by contact hamiltonians.
Then the class of transformations of interest is described by the following holomorphic functions\footnote{We will use
capital $H$ and small $h$ to denote contact hamiltonians generating transformations between Darboux coordinate systems
in two different patches or in the same patch (i.e., a gauge transformation), respectively.}
\be
\gH=\frac{1}{4\pi^2}\sum_{\bfp}e^{2\pi\I p^a\txi_a}g_{\bfp}(\xi),
\label{gengaugetr}
\ee
where $g_{\bfp}$ is any holomorphic function of $\xi^\Lambda$ regular in the neighborhood of $z=0$.
The motivation for considering this class of functions comes from the fact that it is consistent with the Heisenberg shift symmetry
$\txi_a\mapsto\txi_a+\tilde\eta_a$ with $\eta_a\in\IZ$. In particular, the simplest choice
$g_\bfp\sim \bOm_\gamma e^{-2\pi\I q_\Lambda\xi^\Lambda}$
produces a linear superposition of the contact hamiltonians $H_\gamma$ \eqref{prepHnew}
generating D-instanton corrections.

In Appendix \ref{subsec-effgauge} we show that the effect of the gauge transformation on the Darboux coordinates
is extremely simple: it affects only the instanton contributions and, when formulated in terms of the combinations \eqref{defdelta},
it  amounts to a shift in the argument of the functionals $\ff^I$,
\be
\gtrans\ :\quad \hat\delta\Xi^I\mapsto \hat\delta_g\Xi^I=\ff^I[\hcJ_\bfp]\, ,
\qquad
\hcJ_{\bfp}=\tcJ_{\bfp}-2\pi\I\tgH_{\bfp}\, ,
\label{gtr-dcinst}
\ee
where the function $\tgH_{\bfp}$ is constructed in the way similar to $\tcJ_{\bfp}$.
Namely, one first expands the contact Hamiltonian \eqref{gengaugetr} using \eqref{expcX}, thereby obtaining\footnote{Note that
this is {\it not} the expansion in powers of DT invariants
which may appear explicitly as coefficients in $g_{\bfp}$.
Rather, the first term corresponds to the evaluation of \eqref{gengaugetr} on the classical Darboux coordinates, whereas
the other terms arise by expanding in instanton corrections to these coordinates. See \eqref{defcGnew}. \label{foot-exphg}}
\be
\gH =\sum_{\bfp} \gHi{1}_{\bfp} + \sum_{\bfp_1,\bfp_2}\gHi{2}_{\bfp_1\bfp_2}+\cdots\, ,
\label{defgHpert}
\ee
and then halves the second order contribution
\be
\tgH_\bfp = \gHi{1}_{\bfp} + \hf\sum_{\bfp_1+\bfp_2=\bfp}\gHi{2}_{\bfp_1\bfp_2}.
\label{deftgH}
\ee
Thus, our problem is to find a {\it holomorphic} function $g_{\bfp}(\xi)$ such that the resulting
function $\hcJ_{\bfp}$ transforms as a modular form of weight $(-1,0)$. If it exists, then
according to the results of the previous section, S-duality will act on the gauge-transformed Darboux coordinates
by the classical map  \eqref{SL2Zxi}, which is a holomorphic contact transformation.

Hindsight from our earlier study of the one-instanton approximation\footnote{The
study in  \cite{Alexandrov:2012au} contained several errors that we shall rectify below.}
suggests the following ansatz
\be
\gH=\sum_{\gamma\in\Gamma_+} \Dg_{\gamma}
\, H_\gamma
+\hf\sum_{\gamma_1,\gamma_2\in\Gamma_+}\Dl_{\gamma_1\gamma_2}
\, H_{\gamma_1} \, H_{\gamma_2},
\label{gauge-full}
\ee
where $\Dg_{\gamma}$ and $\Dl_{\gamma_1\gamma_2}$ are two piecewise constant
functions of the moduli, which are further assumed to be independent of the zeroth
components of the electric charges. This last condition allows to factorize the sum over
electric charges and express the gauge transformation as a sum of indefinite theta series.
Indeed, using \eqref{multiMSW} to express
the DT invariants entering the contact hamiltonian $H_\gamma$ \eqref{prepHnew} in terms of
the MSW invariants, $\gH$ can be rewritten as
\be
\gH=\frac{1}{4\pi^2}\sum_{\gamma\in\Gamma_+}\bOmMSW_\gamma \Dg_{\gamma} \sigma_\gamma\cX_\gamma
+\frac{1}{32\pi^4}\sum_{\gamma_1,\gamma_2\in\Gamma_+}\bOmMSW_{\gamma_1}\bOmMSW_{\gamma_2}
\hDl_{\gamma_1\gamma_2}\sigma_{\gamma_1}\sigma_{\gamma_2}\cX_{\gamma_1}\cX_{\gamma_2},
\label{gauge-full-MSW}
\ee
where $\hDl_{\gamma_1\gamma_2}=\Dl_{\gamma_1\gamma_2}+4\Dela_{\gamma_1\gamma_2}\Dg_{\gamma}$,
with $\Dela_{\gamma_1\gamma_2}$ the sign factor defined in \eqref{defDela}.
Then due to the above property
and the spectral flow invariance of $\bOmMSW_\gamma$, both terms in \eqref{gauge-full-MSW}
can be represented in a factorized form
\be
\gH=\frac{1 }{4\pi^2}\sum_\bfp \sum_{\bfmu\in\Lambda^\star/\Lambda} h_{\bfp,\bfmu}\,
\hat\vartheta_{\bfp,\bfmu}(\Dg_{\gamma},0)
+ \frac{1 }{32\pi^4}\sum_{\bfp_1,\bfp_2}
\sum_{\bfmu_s\in\Lambda_s^\star/\Lambda_s} h_{\bfp_1,\bfmu_1}\,
h_{\bfp_2,\bfmu_2}\,
\hat\vartheta_{\bfp_1,\bfp_2,\bfmu_1,\bfmu_2}
(\sqrt{\tau_2}\hDl_{\gamma_1\gamma_2},1),
\label{factor-cG}
\ee
where $h_{\bfp,\bfmu}$ is the holomorphic generating function of the MSW invariants \eqref{defchimu} and the $\hat\vartheta$'s
are the $z$-dependent theta series \eqref{Vignerasth-z}
with a quadratic term in the exponential arising due to expressing the charge $q_0$
in terms of the invariant charge $\hat q_0$ \eqref{defqhat}.
In particular, $\hat\vartheta_{\bfp,\bfmu}(\Dg_{\gamma})$  is a theta series for the lattice $\Lambda$ with quadratic form
$\kappa_{abc} p^c$ of signature $(1,b_2-1)$,
whereas $\hat\vartheta_{\bfp_1,\bfp_2,\bfmu_1,\bfmu_2}(\sqrt{\tau_2}\hDl_{\gamma_1\gamma_2})$
is a theta series for the
doubled lattice $\Lambda_1\oplus\Lambda_2$ of signature $(2,2b_2-2)$.
We shall now fix the factors $\Dg_{\gamma}$  and $\Dl_{\gamma_1\gamma_2}$ in turn.

\subsection{Gauge transformation at one-instanton order, revisited}

At one instanton order, $\Dg_{\gamma}$ should be chosen so as to cancel the anomalous
variation of
\be
\cJi{1}_\bfp(z) = \frac{1}{4\pi^2}\sum_{q_\Lambda} \sigma_\gamma\, \bOmMSW_\gamma \int_{\ell_{\gammap}} \frac{\de z'}{z'-z}\,\cXcl_{\gammap}(z') .
\ee
It was recognized in \cite{Alexandrov:2012au} that, after replacing the charge $q_0$ in terms of
the invariant charge $\hat q_0$ and factorizing the sum over electric charges as in \eqref{factor-cG},
this object can be written as an Eichler integral
of a Siegel-type theta series for the lattice $\Lambda$. Since such an Eichler integral arises naturally in the modular
completion of Zwegers' indefinite theta series (see \S\ref{subsec-error1}),
it was  suggested to choose  $\Dg_{\gamma}$ such that $\gH$ produces an holomorphic theta
series whose modular anomaly cancels that of the Eichler integral.
This motivated the choice\footnote{This differs from  \cite[4.31]{Alexandrov:2012au} by the $z$-dependent shift in the
argument of the sign, which was overlooked.}
\be
\Dg_{\gamma} = \frac12 \Bigl[ \sign\bigl((\bfq+\bfb)\cdot \bft + (pt^2)\Im z\bigr)
- \sign\bigl((\bfq+\bfb)\cdot\bft' + 2(ptt')\Im z\bigr) \Bigr],
\label{oneinst-gt}
\ee
where $\bft'$ is an auxiliary null vector on the boundary of the \kahler cone,  such that $(pt'^2)=0$
and $(p t t')>0$.
In this expression, the first sign has a simple interpretation.
Indeed, the line on $\CP$ where its argument vanishes coincides with
the BPS ray $\ell_\gamma$ \eqref{BPSray} in the large volume limit. Therefore, the corresponding term
in \eqref{gauge-full} just removes the discontinuity in the Darboux coordinates along this BPS ray.
This is a typical effect of the gauge transformation mapping type IIA to type IIB twistorial construction
(cf. the gauge transformation \eqref{fungengauge} relevant for the description of
D1-D(-1)-instantons).\label{changepatches}

The second term in \eqref{oneinst-gt}, dependent on the null vector $\bft'$, is needed for convergence of the indefinite theta series.
The vector $\bft'$ must be null because otherwise modularity would require additional non-holomorphic terms which cannot be
produced by a holomorphic gauge transformation.
But, unfortunately, for CY threefolds with $b_2=1$, there are no non-trivial null vectors.
Moreover, as is explained in Appendix \ref{subsec-error1},
even when null vectors exist, the convergence of the indefinite theta series requires that $\bft'$ be a {\it rational} null vector,
since otherwise the theta series would diverge due to accumulation of lattice points near the light cone
leading to an infinite number of terms of the same strength.
But even when $b_2>1$, rational null vectors may not exist in general, as can be seen on the examples in \cite{Hosono:1993qy}.

In order to circumvent both problems, we propose to modify the previous construction
by {\it extending} the charge lattice.
First of all, let us choose an {\it integer vector} $\bft'\in H_4(\CYm,\IZ)$
inside the K\"ahler cone,
which implies in particular that  $(p_1p_2t')$, $(p_1tt')$ are positive for
all effective divisors $\bfp_1$, $\bfp_2$ and K\"ahler parameters $\bft$.
For example, $\bft'$ could be chosen to be one of these divisors. We now
supplement the original lattice $\Lambda$ with an auxiliary lattice of dimension $N_\bfp=(p t'^2)$,
parametrized by half-integers $n_i$, with
diagonal quadratic form $\kappa^-=\diag(-1,\dots,-1)$, which we denote by $\aLam_{N_\bfp}$.
The extended lattice $\cLam=\Lambda\oplus\aLam_{N_\bfp}$ then carries a quadratic form
\be
\ckap_{AB}=\(\begin{array}{cc}  \kappa_{ab} & 0\\ 0 &  -\delta_{ij}\end{array}\),
\ee
where $A=(a,i)$ and the indices $i,j$ run over $N_\bfp$ values. This quadratic form
may be obtained from an auxiliary
intersection form $\ckap_{ABC}$ with
\be
\ckap_{abc}=\kappa_{abc},
\qquad
\ckap_{abi}=\ckap_{aij}=0,
\qquad
\ckap_{ijk}=-\delta_{ij}\delta_{ik}
\ee
by contracting it with the extended charge vector $\cp^A=(p^a,1,\dots,1)$: $\ckap_{AB}=\ckap_{ABC}\cp^C$.
The lattice $\cLam$ possesses two crucial properties:
\begin{enumerate}
\item
$\Lambda$ and $\cLam$ have the same discriminant group,
so $\check\mu_A\in \cLam^*/\cLam$ can be chosen as $\check{\mu}_A=(\mu_a,0,\dots,0)$
with $\mu_a\in \Lambda^*/\Lambda$;
\item
the lattice $\cLam$ always admits  an integer null vector, namely $\ct'^A=(t'^a,1,\dots,1)$,
since $\ckap_{AB}\ct'^A \ct'^B=0$.
\end{enumerate}
Having introduced this extended lattice, we now take
\be
\Dg_{\gamma}  =
\frac12\, \(\sgn (\ww_\gamma (z))-
\frac{1}{\(\p_v\theta_1(\tau,0)\)^{N_\bfp}}\,
\sum_{n_i\in \IZ+\hf}\[ \prod_{i=1}^{N_\bfp}(-1)^{n_i} 2\pi\I n_i \,q^{\frac{n_i^2}{2}}\]
\sgn(\ww'_{\gamma,\vn}(z))\),
\label{deftDg}
\ee
where
\be
\begin{split}
\ww_\gamma(z) =&\,(\bfq+\bfb)\cdot \bft + (pt^2)\Im z,
\\
\ww'_{\gamma,\vn}(z) =&\, (\bfq+\bfb)\cdot\bft' +\sum_{i=1}^{N_\bfp} n_i + 2(ptt')\Im z,
\end{split}
\label{defww}
\ee
and $\theta_1(\tau,v)$ is the usual Jacobi theta series ($q=e^{2\pi\I \tau}$),
which is a holomorphic Jacobi form of weight 1/2,
\be
\theta_1(\tau,v)=\sum_{n\in \IZ+\hf}(-1)^{n}e^{2\pi\I n v}q^{\frac{n^2}{2}} .
\label{theta-div}
\ee

To interpret $\Dg_{\gamma}$ in terms of  the extended lattice, let us
also introduce the auxiliary components of the fields $\bfb$, $\bfc$ and the Darboux coordinates
\be
\xi^i_{\rm cl}=\tau b^i-c^i,
\qquad
\txi_i^{\rm cl}=-\hf\,b^i(c^i-\tau b^i).
\ee
Then it is easy to see that the theta series for the lattice $\Lambda$ with kernel $\Dg_{\gamma}$ can be expressed
in terms of a theta series on the extended lattice $\cLam$. The precise relation is given by
\be
\hat\vartheta_{\bfp,\bfmu}(\Dg_{\gamma};\tau,\bfb,\bfc,z)= \frac{1}{2\(\p_v\theta_1(\tau,0)\)^{N_\bfp}}
\left.\[\prod_{i=1}^{N_\bfp} \p_{c^i}\]\!\!
\(V^{-1}\hat\vartheta_{\cbfp,\cbfmu}(\sPhi_{\ct,\ct'};\tau,\cbfb,\cbfc,z)\)\right|_{b^i=c^i=0}.
\label{theta-ext}
\ee
Here $\,\check{ }\,$ indicates quantities living on $\cLam$, $\ct^A=(t^a,0,\dots,0)$, and we defined
\be
V =\expe{\sum_{i=1}^{N_\bfp}\txi_i^{\rm cl}}
\label{factorV}
\ee
as well as the kernel
\be
\sPhi_{\ct,\ct'}(\cbfx,y)=\sgn\(\wwx(y)\)-\sgn\(\wwx'(y)\),
\label{defPhis}
\ee
where $\wwx,\wwx'$ denote a rescaled version of the quantities \eqref{defww}, namely,\footnote{Note that
the definition of $\ct^A$ ensures that that for any extended vector $\cx_+=x_+$. \label{foot-cxx}}
\be
\begin{split}
\wwx(y) =&\,\cx_+ + \Im y,
\qquad
\wwx'(y) =\, \cbfx \cdot\cbft' + 2 \ct'_+\Im y.
\end{split}
\label{defwwx1}
\ee
Importantly, in the second sign function in \eqref{defPhis} the charge vector is contracted with the integer null vector $\cbft'$,
which ensures convergence of the theta series in \eqref{theta-ext}.
It is also worth noting that the derivatives with respect to the auxiliary fields $c^i$ in \eqref{theta-ext}
are needed because the sum over $n_i$ in all terms independent of the null vectors
would otherwise vanish, due to the vanishing of $\theta_1(\tau,v)$ at $v=0$.

\subsection{Gauge transformation at two-instanton order}

Similarly, at two-instanton order we introduce the extended version of the doubled lattice
$\cLam_{12}=\cLam_1\oplus\cLam_2=\Lambda_1\oplus\Lambda_2\oplus\aLam_{N_\bfp}$, and choose
\be
\begin{split}
\Dl_{\gamma_1\gamma_2} =&\, \frac{\pi^2\langle\gamma_1,\gamma_2\rangle}{\(\p_v\theta_1(\tau,0)\)^{N_\bfp}}
\sum_{n_i\in \IZ+\hf}\[\prod_{i=1}^{N_\bfp}(-1)^{n_i} 2\pi\I n_i \,q^{\frac{n_i^2}{2}}\]
\\
&\qquad \times \(\sgn(\cI_{\gamma_1\gamma_2})-\sgn (\ww'_{\gamma_1,\vn_1})\)
\(\sgn(\ww'_{\gamma,\vn})-\sgn(\ww_{\gamma_2})\),
\end{split}
\label{deftDl}
\ee
where the vector $\vn=(\vn_1,\vn_2)$ has $N_{\bfp_1}+N_{\bfp_2}=N_{\bfp}$
components. From \eqref{factor-cG} it follows that it leads to the two-instanton contribution which can
be written in terms of a theta series
with the following kernel
\bea
\label{conv-doubletheta}
\hDl_{\gamma_1\gamma_2}
&=& \frac{\pi^2\langle\gamma_1,\gamma_2\rangle}{\(\p_v\theta_1(\tau,0)\)^{N_\bfp}}
\sum_{n_i\in \IZ+\hf}\[\prod_{i=1}^{N_\bfp}(-1)^{n_i} 2\pi\I n_i \,q^{\frac{n_i^2}{2}}\]
\Bigl[\sgn(\cI_{\gamma_1\gamma_2})\(\sgn(\ww_\gamma)-\sgn(\ww_{\gamma_2})\)
\Bigr.
\nn\\
&& \Bigl.
+\sgn (\ww'_{\gamma_1,\vn_1})\(\sgn(\ww_{\gamma_2})-\sgn(\ww'_{\gamma,\vn})\)
+\sgn\langle\gamma_1,\gamma_2\rangle \, \(\sgn(\ww'_{\gamma,\vn}) - \sgn(\ww_\gamma)\)\Bigr].
\eea
Proceeding as above, this theta series can then be recast as a theta series
for the doubled extended lattice $\cLam_{12}$ with signature
$(2,2b_2-2+N_\bfp)$, with kernel given by $\Xxx$ (see \eqref{defXxx}) times
the factor in \eqref{conv-doubletheta} involving sign functions. Although the resulting theta series
is formally the sum of two theta series of the type  considered in \cite{Alexandrov:2016enp},
neither of the two is separately convergent. Instead, it is an example of a more
general type of indefinite theta series which we introduce
in Appendix \ref{sec-converge}. In this appendix, relying on a conjectural set of sufficient
conditions for convergence, we argue
that $\hDl_{\gamma_1\gamma_2}$ does lead to a convergent theta series.

\bigskip

Upon making these choices, it is clear that the contact hamiltonian \eqref{gauge-full} fits in
the class of contact transformations described by \eqref{gengaugetr}.
Thus, we can apply the result \eqref{gtr-dcinst} for the effect of the gauge transformation
on the Darboux coordinates, which is captured by the function $\hcJ_{\bfp}$.
In the next section we compute this function explicitly and prove that it transforms as a modular form of weight $(-1,0)$.

\subsection{Modularity}
\label{sec-modularity}

Starting from \eqref{defcJ} and \eqref{gauge-full}, substituting the expansion \eqref{expcX}, and
following the procedure leading to the definition of $\hcJ_{\bfp}$ \eqref{gtr-dcinst}, one arrives at the following
expression for this function
\bea
\hcJ_{\bfp}&=&
\frac{1}{4\pi^2}\sum_{q_\Lambda} \sigma_\gamma \bOm_\gamma
\[\int_{\ell_{\gammap}} \frac{\de z'}{z'-z}\,\cXcl_{\gammap}(z')-2\pi\I \Dg_{\gamma}  \cXcl_\gamma(z)\]
\nn\\
&+&\frac{1}{16\pi^3}\sum_{\gamma_1,\gamma_2\in\Gamma_+\atop \bfp_1+\bfp_2=\bfp}
\sigma_{\gamma_1}\sigma_{\gamma_2} \bOm_{\gamma_1}\bOm_{\gamma_2}\left\{(tp_1p_2)
\[\int_{\ell_{\gammap_1}}  \frac{\de z_1}{z_1-z}\,\cXcl_{\gammap_1}(z_1)-2\pi\I \Dg_{\gamma_1}  \cXcl_{\gamma_1}(z)\]
\int_{\ell_{\gamma_2}} \de z_2 \cXcl_{\gamma_2}(z_2)
\right.
\nn\\
&&\qquad
-\I\[\langle\gamma_1,\gamma_2\rangle\int_{\ell_{\gamma_1}}\frac{\de z_1}{z_1-z}\int_{\ell_{\gamma_2}}\frac{\de z_2}{z_2-z_1}
\cXcl_{\gamma_1}(z_1) \cXcl_{\gamma_2}(z_2)
\right.
\nn
\\ && \qquad \qquad\qquad \left.\left.
-2\pi\I \langle\gamma_1,\gamma_2\rangle \Dg_{\gamma_1}  \cXcl_{\gamma_1}(z)\int_{\ell_{\gamma_2}}\frac{\de z_2}{z_2-z}\cXcl_{\gamma_2}(z_2)
+\Dl_{\gamma_1\gamma_2}\cXcl_{\gamma_1}(z)\cXcl_{\gamma_2}(z)\]\right\}.
\label{tcJres}
\eea

Our aim here is to understand its modular properties. To this end, we rewrite $\hcJ_{\bfp}$ in terms of the MSW invariants by means of
the expansion \eqref{multiMSW}. Then, using the spectral flow invariance of $\bOmMSW_\gamma$,
one can represent all sums over charges in a factorized form, in a similar way as was done in \eqref{factor-cG}.
Introducing further the extended lattice and auxiliary components of the $\bfb$ and $\bfc$ fields as was explained in the previous section,
one finds that (c.f. \eqref{theta-ext})
\be
\hcJ_{\bfp}= \frac{1}{\(\p_v\theta_1(\tau,0)\)^{N_\bfp}}
\left.\[\prod_{i=1}^{N_\bfp} \p_{c^i}\]
\(V^{-1}\hcJe_{\bfp}\)\right|_{b^i=c^i=0},
\label{fullJp}
\ee
where
\be
\begin{split}
\hcJe_{\bfp}=&\, \frac{1}{4\pi\I}\[\sum_{\bfmu\in\Lambda^*/\Lambda}
\whh_{\bfp,\bfmu}\, {\hat\vartheta_{\cbfp,\cbfmu}(\hKer_{\cbfp},0)}
 +\frac{1}{8\sqrt{2}}\sum_{\bfp_1+\bfp_2=\bfp}
\sum_{\bfmu_s\in\Lambda_s^*/\Lambda_s}h_{\bfp_1,\bfmu_1}\,h_{\bfp_2,\bfmu_2}\,
{\hat\vartheta_{\cbfp_1,\cbfp_2,\cbfmu_1,\cbfmu_2}( \hKerr_{\cbfp_1\cbfp_2},1)}\].
\end{split}
\label{fullJpext}
\ee
Here the theta series $\hat\vartheta_{\cbfp,\cbfmu}$ \eqref{Vignerasth-z} is defined on the extended lattice $\cLam$,
$\hat\vartheta_{\cbfp_1,\cbfp_2,\cbfmu_1,\cbfmu_2}$  is its version for the doubled extended lattice $\cLam_{12}$,
and the two kernels are given by
\bea
\hKer_{\cbfp}(\cbfx,y)&=& \hE_1(\cx_+,y)-
\sgn\(\wwx'\),
\label{Phi-first}\\
\hKerr_{\cbfp_1\cbfp_2}(\cbfx_1,\cbfx_2,y) &=&\Biggl\{\frac{(tp_1p_2)}{\pi\sqrt{(p_2t^2)}}\,
e^{-\pi \(\cx_{2+}-\I y_2\)^2}\hKer_{\cbfp_1}(\cbfx_1,y_1)
-\Xxx \hM_2\(\rpp{2}{1},\cx_{1+},\cx_{2+},y\)
\Biggr.
\nn\\
&&
-\Xxx\Bigl[\sgn (\wwx_1)- \sgn\(\wwx'_1\)\Bigr]\hM_1(\cx_{2+},y_2)
\nn\\
&&+\Xxx \Bigl[\sgn(\Ixx)-\sgn (\wwx'_1)\Bigr]
\Bigl[\sgn(\wwx')-\sgn(\wwx_2)\Bigr]
\label{Phi-second}\\
&&\Biggl.
+\Xxx\[\sgn(\Ixx )-\sgn(\Xxx )\(1+\frac{1}{2\pi}\,\beta_{\frac{3}{2}}\!\left({\textstyle\frac{\Xxx^2 }{(pp_1p_2)}}\right)\)\]
\hKer_{\cbfp}(\cbfx_1+\cbfx_2,y)\Biggr\}
+(1\leftrightarrow 2),
\nn
\eea
where $\hE_1$, $\hM_1$, $\hM_2$ are the $y$-dependent versions of the usual and double error functions introduced
in appendices \ref{subsec-error1} and \ref{subsec-error2}.
Besides, we used rescaled versions of various quantities, which after rescaling are all
expressed in terms of $\cbfx_s=\sqrt{2\tau_2}(\cbfq_s+\cbfb)$, $y=\sqrt{2\tau_2(pt^2)}z$
and $y_s=\sqrt{2\tau_2(p_st^2)}z=\rpp{s}{}\, y$ (cf. \eqref{Vignerasth-z}).
In particular, $\Xxx$ and $\Ixx$ are rescaled versions of $\langle\gamma_1,\gamma_2\rangle$ and $\cI_{\gamma_1\gamma_2}$, respectively,
defined in \eqref{defXxx}, $\wwx,\wwx'$ have been already introduced in \eqref{defwwx1}, and
\be
\wwx_s(y)= \cx_{s+}+\Im y_s,
\qquad
\wwx'_s(y) = \cbfx_s \cdot\cbft'_s + 2\ct'_{s+}\Im y_s,
\label{defwwx}
\ee
where $\cbft'_s$ is the null vector for the extended lattice $\cLam_s$.

Note that the first contribution in \eqref{fullJpext} is written in terms of the modular completion \eqref{defhh} of the generating function
of the MSW invariants. The latter has a modular anomaly at the two-instanton order and therefore the difference
between $h_{\bfp,\bfmu}$ and $\whh_{\bfp,\bfmu}$ must be taken into account here, leading
to the very last
contribution in \eqref{Phi-second} via the property \eqref{hhtoRth}.

The representation \eqref{fullJp} shows that $\hcJ_{\bfp}$ is a modular function of weight $(-1,0)$ provided
its extended version $\hcJe_{\bfp}$ is modular with weight $(\half N_\bfp-1,0)$. Indeed, let us assume that $\hcJe_{\bfp}$ is
such a modular form. Then, we observe that the factor $V$
affects only the exponential part of the automorphy factor dependent on $b^i$, $c^i$ and does not affect the modular weight.
Therefore, it can be safely cancelled without spoiling the modularity.\footnote{In fact,
this factor is absent in the standard  definition of theta series found e.g. in \cite{MR781735}.} Moreover, it can be checked that
the remaining function depends on the auxiliary components $b^i$, $c^i$ only in the combinations $v^i=c^i-\tau b^i$
(except the dependence on $b^i$ through the sign functions which is anyway piecewise constant) so that
taking the derivatives $\p_{c^i}$ and subsequently setting $b^i=c^i=0$ is equivalent to extracting the first coefficient
of the Fourier expansion in all $v^i$. Since $v^i$ transform under $SL(2,\IZ)$ homogenously with modular weight $(-1,0)$,
the corresponding Fourier coefficient is also modular with weight $(\tfrac32 N_\bfp-1,0)$. Finally,
due to the same reason (or using the identity $\p_v\theta_1(\tau,0)=2\pi(\eta(\tau))^3$),
one concludes that $\p_v\theta_1(\tau,0)$ has modular weight $(\tfrac32,0)$ which brings the total weight of \eqref{fullJp}
to $(-1,0)$.

Thus, it remains to understand the modular properties of $\hcJe_{\bfp}$.
First, we concentrate on the one-instanton term in \eqref{fullJpext}.
Since $\cbft'$ is a null vector on the extended lattice,
as is shown in Appendix \ref{subsec-error1}, the kernel $\hKer_{\cbfp}$
is annihilated by the generalized Vign\'eras' operator $\hV_0$ \eqref{zdep-Vignop-mod}.
Thus, $\hat\vartheta_{\cbfp,\cbfmu}(\hKer_{\cbfp})$ is a modular theta series of weight $(\half(b_2+N_\bfp),0)$.
Taking into account that $\whh_{\bfp,\bfmu}$ is a modular form of weight $(-\frac{b_2}{2}-1,0)$, we conclude that the first term
in \eqref{fullJp} transforms as a modular form of the requisite weight $(\half N_\bfp-1,0)$.

To understand the modular properties of the second order contribution, we split the kernel \eqref{Phi-second}
into several pieces
\be
\hKerr_{\cbfp_1\cbfp_2}=-\hPhiT+\hPhid_{\cbfp_1\cbfp_2}+\hPhip_{\cbfp_1\cbfp_2}+(1\leftrightarrow 2).
\label{splitker}
\ee
The first term $\hPhiT$ is defined in \eqref{defhPhiT2}. Although it is defined on the usual doubled lattice $\Lambda_{12}$, it can
be trivially extended to $\cLam_{12}$ due to the property described in footnote \ref{foot-cxx}.
The other two terms are given by
\bea
\hPhid_{\cbfp_1\cbfp_2}&=& - F\({\textstyle\frac{\Xxx}{\sqrt{(pp_1p_2)}}}\)\hKer_{\cbfp}(\cbfx_1+\cbfx_2,y),
\label{diagPhi}
\nn\\
\hPhip_{\cbfp_1\cbfp_2} &=&   \sgn(\wwx'_1)
\[\Xxx\(\hE_1(\cx_{2+},y_2)-\sgn(\wwx')\)
-\frac{e^{-\pi \(\cx_{2+}-\I y_2\)^2}}{\pi\sqrt{(p_2t^2)}}\],
\label{sanomH}
\eea
where we introduced the function
\be
F(x)=|x|\(1+\frac{1}{2\pi}\,\beta_{\frac32}(x^2)\)
=x \, {\rm Erf}(\sqrt{\pi} x) + \frac{1}{\pi} \,e^{-\pi x^2},
\ee
and used the following sum rule afforded by the linear dependence of the  three hyperplanes $\Ixx=0$, $\wwx=0$,
$\wwx_2=0$,
\be
\sgn(\Ixx)\(\sgn(\wwx)-\sgn(\wwx_2)\) +\sgn(\wwx)\sgn(\wwx_2)=1 .
\ee
The contribution from $1$ on the r.h.s. vanishes after multiplication by $\Xxx$ and symmetrization.

If the above kernels are all annihilated
by the generalized Vign\'eras' operator $\hV_1$ for the extended doubled lattice,
obtained by replacing all vectors by their extended counterparts in
\eqref{zdep-Vignop-two},
 then the theta series based on them will be modular forms of weight $(b_2+\half N_\bfp+1,0)$.
Then, after multiplication by the two generating functions of MSW invariants appearing in the second term in \eqref{fullJpext}, they will
give rise to modular forms of weight $(\half N_\bfp-1,0)$, as required.
In Appendix \ref{subsec-error2} it is shown that $\hPhiT$ is indeed annihilated by $\hV_1$,
whereas for the other kernels one finds
\be
\begin{split}
\hV_1\cdot \hPhid_{\cbfp_1\cbfp_2} =&\, 0,
\\
\hV_1\cdot \hPhip_{\cbfp_1\cbfp_2} =&\, 4(p_1p_2t')\, \delta(\wwx'_1)\, \hKer_{\cbfp_2}(\cbfx_2,y_2).
\end{split}
 \label{anomVign}
 \ee
Thus, the last kernel $\hPhip_{\cbfp_1\cbfp_2}$ leads to an anomaly in the  Vign\'eras'
equation, which is proportional to a delta function
localized at $\wwx'_1=0$.

As we argue in Appendix \ref{subsec-thetaanom},  such delta function anomalies do not always
signify the breakdown of  modular symmetry.
Some of them lead to modular anomalies in the theta series which are themselves
localized on the codimension 1 loci in twistor space
where the delta functions are supported, which in our context correspond to the boundaries between different patches.
For such localization to happen, it is necessary that the anomaly in Vin\'eras' equation be itself an eigenfunction
of Vign\'eras' operator with a different eigenvalue.
Using the property $x\delta'(x)=-\delta(x)$, it is immediate to see that in our case this requirement does hold,
\be
\hV_{-1}\cdot \(\hV_1\cdot \hPhip_{\cbfp_1\cbfp_2}\)=0.
\ee
This should come as no surprise because from the discussion below Eq. \eqref{oneinst-gt} it follows that
the equation $\wwx'=0$ is expected to describe the position of discontinuities
of the Darboux coordinates after the gauge transformation,
i.e. a boundary of the new patches. Alternatively, we can view the kernel $\hPhip_{\cbfp_1\cbfp_2}$
as the limit $\bft'^2\to 0$ of a continuous kernel which does satisfy Vign\'eras
equation. In this limit, error functions become discontinuous sign functions, but
Gaussian terms similar to the last term in \eqref{diagPhi} lead to delta functions. The kernel
$\hPhip_{\cbfp_1\cbfp_2}$ is then obtained after dropping  such delta
function terms, which do not contribute to the theta series away from the singular loci,
at the expense of producing  an anomaly in Vign\'eras' equation.

Thus, we conclude that the anomaly can be dropped and the full kernel \eqref{splitker}
is annihilated by $\hV_1$, which implies that both functions $\hcJe_{\bfp}$ and $\hcJ_{\bfp}$
possess correct modular properties.
This proves that S-duality acts isometrically on the twistor space $\cZ$ and its quaternion-K\"ahler base, which is
the instanton corrected hypermultiplet moduli space $\cM_H$.

\section{Discussion}
\label{sec-disc}

Our main result is a proof that, at two-instanton order and in the large volume approximation,
D3-instanton corrections to the hypermultiplet moduli space $\cM_H$ are consistent
with the S-duality of type IIB string theory, provided the  MSW invariants are Fourier coefficients
of a vector-valued mock modular form with specified modular properties. To show that $\cM_H$ admits
an isometric action of $SL(2,\IZ)$, we produced a set of Darboux coordinates for the holomorphic
contact structure on the twistor space $\cZ$, which transform under $SL(2,\IZ)$ in the same way \eqref{SL2Zxi}
as the Darboux coordinates for the classical, uncorrected moduli space. These new
Darboux coordinates were obtained by finding appropriate quantum  corrections  \eqref{inst-mp} to
the mirror map and by applying a local holomorphic contact transformation \eqref{gauge-full}
to the Darboux coordinates used in the original type IIA formulation of the D-instanton corrected $\cM_H$.

Our twistorial construction makes it clear that at $k-$th order in the instanton expansion,
the local contact transformation, which is needed to cancel the modular anomaly, is generated
by an indefinite theta series of signature $(k,n-k)$ for some $n\geq k b_2$; and  moreover,
that the modular completion of such indefinite theta series involves a $k$-th iterated integral over the twistor fiber.
This physical input was key to our previous work
\cite{Alexandrov:2016enp}, where we developped a general machinery for dealing with such indefinite theta series,
with particular emphasis on the `conformal' case  ($k=2$). It is worth noting that
the twistorial construction of D-instantons automatically produces the modular completion without any guesswork.

While in  \cite{Alexandrov:2016enp} we focused on indefinite theta series
whose kernel is given by product of difference of sign functions, in the course of this work
we have encountered a more general class of theta series,  whose
kernel is a cyclic sum of products of signs. Although these  theta series may be formally
decomposed as a  sum of series of the type considered in \cite{Alexandrov:2016enp},
each term in this formal decomposition need not be separately convergent.
We have put forward a conjecture providing a set of sufficient conditions ensuring the convergence
of this more general class of theta series, which has passed extensive numerical tests.
It would be  interesting to prove our conjecture, possibly  along the lines of \cite{kudla2016theta},
or by relating it to the theta series for polyhedral cones considered in \cite{westerholt2016indefinite}.

To achieve convergence of the theta series, we had to extend the lattice of electric charges
by adding a number  of auxiliary directions. A similar trick is often used in the mathematics
literature, where the modular properties of Ramanujan's mock theta functions are exposed by expressing them
as the ratio of an indefinite theta series by a Jacobi theta series \cite{Zwegers-thesis,MR2605321}.
Such extended lattices also arise geometrically from blowing-up a two cycle
in a four-manifold \cite{MR1623706, 0961.14022, Manschot:2014cca} or performing
a flop transition in a Calabi-Yau \cite{Toda:2013}. Note that the presence of auxiliary directions,
and the fact that the number $N_\bfp$ scales
linearly with the magnetic charge, was key for the cancellation of anomalies.\footnote{For instance,
if we chose  a generic vector $\bft'$ such that $(pt'^2)=0$  as in \eqref{oneinst-gt},
ignoring the divergence due to its non-rationality, then $\bft'$ would depend on the charge $\bfp$ in a non-linear way.
As a result, we would lose the orthogonality of $\Cv_4$ and $\Cv_5$ (see \eqref{defvecC})
appearing in the theta series for the gauge transformation, which in turn would spoil both modularity and convergence conditions.}
Nevertheless, we admit that this construction leads to a complicated gauge transformation and, moreover, brings in an ambiguity
in choice of the integer vector $\bft'$. The physical meaning of this ambiguity and the geometric interpretation
of the resulting gauge transformation are not clear to us.

It is worth noting that, at the level of the one-instanton approximation,
the extended lattice construction is not strictly necessary.
Indeed, while the indefinite theta series defined by the kernel \eqref{oneinst-gt}
diverges if the null vector $\bft'$ is not rational, its variation under a modular transformation
is still convergent, given by a period integral of a convergent Siegel-type theta series, and it is still holomorphic.
While this fact is not enough for defining Darboux coordinates which transform covariantly under modular symmetry,
it suffices to show that the modular anomaly at one-instanton level can be absorbed
by a complex contact transformation.
It may be possible to use a similar strategy at the
two-instanton level, but we found it more convenient to use the extended lattice trick.
While the resulting construction might look somewhat {\it ad hoc}, as was emphasized above,
it has the advantage of producing a set of Darboux coordinates which transform like their
classical counterparts, with no modular anomaly. We hope that this feature will help in finally
determining the exact, fully corrected hypermultiplet moduli space.

\bigskip

{\noindent \bf Acknowledgments:}
S.B. thanks CEFIPRA for financial support.

\medskip

\appendix

\section{Details of computations}
\label{sec-details}

In this section, we collect technical details about the computation of instanton corrections to
the Darboux coordinates, and about the effect of local contact transformations.

\subsection{Modular operators}
\label{subsec-oper}

Let us introduce the following derivative operators
\begin{subequations}
\bea
\label{defDw-new}
\cD_{\wh} = \frac{1}{2\pi \I}\(\partial_\tau+\frac{\wh}{2\I\tau_2}+ \frac{\I t^a}{4\tau_2}\, \partial_{t^a}\),&&
\\
\Dop{\bfp}_a =  \frac{1}{4\pi\tau_2} \left[
\pa_{b^a} + \bar\tau \pa_{c^a} - \I\pi\kappa_{abc}p^c (c^b - \bar\tau b^b) \right],&&
\qquad
\bDop{\bfp}_a = (\Dop{\bfp}_a)^\star.
\eea
\end{subequations}
They respect the modularity in the sense that, acting on theta functions of weight $(\wh,\bwh)$, the operators
$\cD_{\wh}$, $\Dop{\bfp}_a$, $\bDop{\bfp}_a$ preserve their form and raise the modular weight by (2,0), (1,0), (0,1), respectively.
Furthermore, we allow the modular derivative \eqref{defDw-new} to act on theta functions living on the twistor space $\cZ$
in which case it acquires an additional term
\be
\hat\cD_{-1}=\cD_{-1}-\frac{z}{8\pi\tau_2}\,\p_z.
\label{defDw-Z}
\ee
Using these operators, we also define
\begin{subequations}
\bea
\Del{\bfp}_+&=&-2\I \tau_2 \kappa^{ab}_{(\bfp)}\Dop{\bfp}_a \bDop{\bfp}_b-\frac{b_2+1}{2\pi\I},
\\
\Del{\bfp}_-&=& 2\I \tau_2 \hat\cD_{-1}+(c^a-\bar\tau b^a) \Dop{\bfp}_a.
\eea
\end{subequations}

A straightforward calculation shows that the action of the above operators
on the Fourier modes of the classical Darboux coordinates \eqref{cXzero} is given by
\bea
\cD_{\wh}\cXcl_{\gamma} &=&-\(\hat q_0+\hf\,(\bfq+\bfb)^2
-\frac{\I z}{2}\(q_a t^a+(btp)\)+\frac{\wh}{4\pi\tau_2}\)\cXcl_{\gamma},
\nn\\
\hat\cD_{\wh}\cXcl_{\gamma} &=&-\(\hat q_0+\hf\,(\bfq+\bfb)^2
-\I z\(q_a t^a+(btp)\)-\frac{z^2}{2}\,(pt^2)+\frac{\wh}{4\pi\tau_2}\)\cXcl_{\gamma},
\nn\\
\Dop{\bfp}_a \cXcl_{\gamma} &=& \(q_a+ \kappa_{abc} p^c (b^b -\I z t^b)\) \cXcl_{\gamma},
\nn\\
\bDop{\bfp}_a \cXcl_{\gamma} &=& \frac{\I}{2\tau_2} \,\kappa_{abc} p^c (c^b -\tau b^b-2\tau_2 z t^b) \cXcl_{\gamma},
\label{actder-Xg}\\
\Del{\bfp}_+\cXcl_{\gamma} &=& \(\frac{\I}{2\pi}+\(c^a -\tau b^a-2\tau_2 z t^a\)\(q_a+ \kappa_{abc} p^c (b^b -\I z t^b)\)\) \cXcl_{\gamma},
\nn\\
\Del{\bfp}_-\cXcl_{\gamma} &=& \(\frac{\I}{2\pi}-2\I\tau_2 q_0+q_a\(c^a -\tau b^a-2\tau_2 z t^a\)+
\kappa_{abc} p^c (b^b -\I z t^b)\(c^a-\tau_1 b^a-\tau_2 z t^a\)\) \cXcl_{\gamma}.
\nn
\eea
In the case when the operator and the Darboux coordinates in the last relation depend on different fiber coordinates,
one finds instead
\be
\Del{\bfp}_-\cXcl_{\gamma}(z') = \(\frac{\I}{2\pi}-2\I\tau_2 q_0+q_a\(c^a -\tau b^a-\tau_2 z' t^a\)+
\kappa_{abc} p^c (b^b -\I z' t^b)\(c^a-\tau_1 b^a\)\) \cXcl_{\gamma}(z').
\ee
Finally, it is useful to note the following two relations
\bea
p_2^a\(\Dop{\bfp_1}_a\cXcl_{\gamma_1}\)\cXcl_{\gamma_2}-p_1^a\(\Dop{\bfp_2}_a\cXcl_{\gamma_2}\)\cXcl_{\gamma_1}&=&
\Bigl(\langle\gamma_1,\gamma_2\rangle+\I(z_2-z_1)(tp_1p_2)\Bigr)\cXcl_{\gamma_1}\cXcl_{\gamma_2},
\eea
\be
\Del{\bfp}_-\int \de z' \, K(z,z')X(z')=\int \de z'\[K(z,z'){\Del{\bfp}_-}'X(z')
+\frac{X(z')}{4\pi\I}\(z\p_z+\p_{z'}z'\)K(z,z')\],
\label{propDm}
\ee
where prime on $\Del{\bfp}_-$ indicates that it depends on $z'$. Note that $\(z\p_z+\p_{z'}z'\)\frac{1}{z'-z}=0$.

\subsection{Instanton contributions to Darboux coordinates}
\label{subsec-compDc}

To evaluate the D3-instanton corrections to Darboux coordinates in terms of the type IIB fields, we must first
define the quantum corrected mirror map. We borrow the results from \cite{Alexandrov:2012au}
where the leading order in the large volume expansion of the one-instanton corrections to the classical mirror map \eqref{symptobd}
was found. The corrections are given by the following twistorial integrals
\be
\begin{split}
\delta u^a=&\,-\frac{\I}{2\tau_2}{\sum_{\gamma\in\Gamma_+}}p^a\[
\int_{\ell_{\gammap}}\de z\,(1-z)\,H_{\gammap}
+\int_{\ell_{\gammam}}\frac{\de z}{z^3} (1-z)\,H_{\gammam}\]
\\
\delta\zeta^a=&\, -3\sum_{\gamma\in\Gamma_+} p^a\Re\(\int_{\ell_{\gammap}} \de z\, z\,H_{\gammap}\),
\\
\delta\tzeta_a =&\,
\kappa_{abc}t^b \sum_{\gamma\in\Gamma_+} p^c\Im\(\int_{\ell_{\gammap}} \de z\,H_{\gammap}\),
\\
\delta\tzeta_0=&\,
-\kappa_{abc}t^b\sum_{\gamma\in\Gamma_+} p^c\Im \int_{\ell_{\gammap}}\de z
\( b^a -\frac{\I}{2}\, t^a z \)H_{\gammap},
\\
\delta\sigma=&\,
-\kappa_{abc}t^b\sum_{\gamma\in\Gamma_+}p^c \Im\int_{\ell_{\gammap}}\de z
\( c^a -\( \frac{\I}{2}\,\tau_1-\tau_2\) t^a z \) H_{\gammap}.
\end{split}
\label{inst-mp}
\ee
We assume that these relations continue to hold beyond the one-instanton approximation and use them
in our two-instanton calculations.
Substituting them into \eqref{eqTBA} and \eqref{alphaIIA}, and taking the combined limit $t^a\to\infty$, $z\to 0$ and $zt^a$ fixed,
one finds that D3-instanton contributions to the Darboux coordinates are given by
\bea
\delta\xi^a &=&
\sum_{\gamma\in\Gamma_+}p^a \int_{\ell_{\gammap}} \frac{\de z'}{z'-z} \, H_{\gammap},
\nn\\
\delta\txi_a &=&
\sum_{\gamma\in\Gamma_+} \int_{\ell_{\gammap}} \de z'\(\frac{q_a}{z'-z} -\I\kappa_{abc} t^b p^c \)H_{\gammap},
\nn\\
\delta\txi_0 &=&
\sum_{\gamma\in\Gamma_+}\[ \int_{\ell_{\gammap}} \de z'\(\frac{q_0}{z'-z}
+\kappa_{abc} t^b p^c\(\I b^a+ \frac{z'+z}{2}\,t^a\)\)H_{\gammap}
-\frac{z}{2}\, (pt^2)
\int_{\ell_{\gammam}}\frac{\de z'}{(z')^2} H_{\gammam}\],
\nn\\
\delta\alpha&=&
\sum_{\gamma\in\Gamma_+}\[ \int_{\ell_{\gammap}} \de z'\(-\frac{\frac{1}{2\pi\I}
+ q_0\tau+q_a(\tau b^a-c^a+2 \tau_2 z't^a)}{z'-z}
\right.\right.
\label{DClimit}\\
&&\left.\qquad\qquad
-\hf\,\kappa_{abc}p^c\(2\I c^a t^b
+4\tau_2 b^a t^b+(\tau_1-3\I\tau_2)\(z+z'\)t^at^b\)  \)
H_{\gammap}
\nn\\
&&
\left.\qquad
+\frac{\tau z}{2}\,(pt^2) \int_{\ell_{\gammam}} \frac{\de z'}{(z')^2}\,H_{\gammam}\]-\altwo.
\nn
\eea
where
\be
\altwo(z)
= \frac{1}{32\pi^4}\sum_{\gamma_1,\gamma_2\in\Gamma_+ }\sigma_{\gamma_1}\sigma_{\gamma_2}\bar\Omega(\gamma_1)\bar\Omega(\gamma_2)
\int_{\ell_{\gammap_1}}\frac{\de z_1}{z_1-z}\int_{\ell_{\gammap_2}} \de z_2
\(\frac{\langle\gamma_1,\gamma_2\rangle}{z_2-z_1}+\frac{p_1^a q_{2,a}}{z_2-z}\)\cXcl_{\gammap_1}\cXcl_{\gammap_2}.
\label{defUp}
\ee

In section \ref{subsec-qcmod}, it was argued that the Darboux coordinates are not by
themselves well-suited to study the fate of S-duality.
Instead, it is useful to consider the combinations \eqref{defdelta}
which are expected to have very simple modular transformation properties.
In our combined large volume limit, they reduce to
\bea
\hat\delta\xi^a & \approx & \delta\xi^a ,
\nn\\
\hat\delta\txi_a &\approx &
\delta\txi_a+\kappa_{abc}\(b^b-\I z t^b  -\frac{\I \delta\xi^b}{4\tau_2 } \)\delta\xi^c,
\nn\\
\hat \delta_+ \alpha &\approx &
\delta\alpha+\tau\, \delta \txi_0
+\kappa_{abc}\(b^a-\I z t^a  -\frac{\I \delta\xi^a}{4\tau_2}\)
\[c^b-\tau b^b-2\tau_2 z t^b -\hf\,\delta\xi^b\]\delta\xi^c
\nn\\
&&
+\frac{\I}{24\tau_2}\, \kappa_{abc}\delta\xi^a\delta\xi^b\delta\xi^c,
\label{defdelta-exp}\\
\hat\delta_-\alpha &\approx &
\delta\alpha+\bar\tau\, \delta\txi_0
+\kappa_{abc}\(b^a-\I z t^a  -\frac{\I \delta\xi^a}{4\tau_2}\)
\[c^b-\tau_1 b^b-\tau_2 z t^b -\frac{1}{4}\,\delta\xi^b\]\delta\xi^c
\nn\\
&&
+\frac{\I }{48\tau_2}\, \kappa_{abc}\delta\xi^a\delta\xi^b\delta\xi^c.
\nn
\eea
Substituting \eqref{DClimit} into these expressions and restricting to the two-instanton order, one obtains
\bea
\hat\delta\xi^a &=&
\frac{1}{4\pi^2}\sum_{\gamma\in \Gamma_+}p^a \sigma_\gamma \bar\Omega(\gamma) \int_{\ell_{\gammap}} \frac{\de z'}{z'-z}\,\cX_{\gammap},
\label{hatdelta-res}\\
\hat\delta\txi_a &=&
\frac{1}{4\pi^2}\sum_{\gamma\in \Gamma_+} \sigma_\gamma \bar\Omega(\gamma) \int_{\ell_{\gammap}}
\frac{\de z'}{z'-z}\(q_a+\kappa_{abc}(b^b-\I z't^b)p^c\)\cX_{\gammap}
\nn\\
&&
-\frac{\I}{64\pi^4\tau_2}\sum_{\gamma_1,\gamma_2\in\Gamma_+ }
\kappa_{abc}p_1^bp_2^c\,\sigma_{\gamma_1}\sigma_{\gamma_2}\bar\Omega(\gamma_1)\bar\Omega(\gamma_2)
\int_{\ell_{\gammap_1}} \frac{\de z_1}{z_1-z}\,\cX_{\gammap_1}\int_{\ell_{\gammap_2}} \frac{\de z_2}{z_2-z}\,\cX_{\gammap_2}
\nn\\
\hat\delta_+\alpha &=&
\frac{1}{4\pi^2}\sum_{\gamma\in \Gamma_+} \sigma_\gamma \bar\Omega(\gamma) \int_{\ell_{\gammap}}
\frac{\de z'}{z'-z}
\left[-\frac{1}{2\pi\I}+(c^a-\tau b^a-2\tau_2 z' t^a)\(q_a+\kappa_{abc}p^c(b^b-\I z' t^b)\) \right]\cX_{\gammap}
-\altwo
\nn\\
&&
-\frac{\I\(c^a-(\tau_1+3\I\tau_2) b^a-4\tau_2 z t^a\)}{64\pi^4\tau_2}
\sum_{\gamma_1,\gamma_2\in\Gamma_+ }
\kappa_{abc}p_1^bp_2^c\,\sigma_{\gamma_1}\sigma_{\gamma_2}\bar\Omega(\gamma_1)\bar\Omega(\gamma_2)
\int_{\ell_{\gammap_1}} \frac{\cX_{\gammap_1}\de z_1}{z_1-z}\int_{\ell_{\gammap_2}} \frac{\cX_{\gammap_2}\de z_2}{z_2-z}
\nn\\
\hat\delta_-\alpha&=&
\frac{1}{4\pi^2}\sum_{\gamma\in \Gamma_+} \sigma_\gamma \bar\Omega(\gamma)\left\{ \int_{\ell_{\gammap}}
\frac{\de z'}{z'-z}
\biggl[-\frac{1}{2\pi\I}-2\I\tau_2 q_0+q_a(c^a-\tau b^a-2\tau_2 z' t^a)
\biggr.\right.
\nn\\
&& \left.\biggl.
+\kappa_{abc}p^c(b^b-\I z' t^b)(c^a-\tau_1 b^a-\tau_2 z' t^a) \biggr]\cX_{\gammap}
+\I\tau_2 z (pt^2)\int_{\ell_{-\gammap}} \frac{\de z'}{(z')^2}\,\cX_{-\gammap} \right\}
-\altwo
\nn\\
&&
-\frac{\I\(c^a-\tau b^a-2\tau_2 z t^a\)}{64\pi^4\tau_2}
\sum_{\gamma_1,\gamma_2\in\Gamma_+ }
\kappa_{abc}p_1^bp_2^c\,\sigma_{\gamma_1}\sigma_{\gamma_2}\bar\Omega(\gamma_1)\bar\Omega(\gamma_2)
\int_{\ell_{\gammap_1}} \frac{\cX_{\gammap_1}\de z_1}{z_1-z}\int_{\ell_{\gammap_2}} \frac{\cX_{\gammap_2}\de z_2}{z_2-z}.
\nn
\eea
Finally,  substituting the instanton expansion of $\cX_\gamma$ \eqref{expcX},  one can verify that
the resulting expressions can be generated as in \eqref{resdXi} by applying the modular operators, introduced in Appendix \ref{subsec-oper},
to the function $\tcJ_{\bfp}$ \eqref{deftcJp}.
To perform this check, the relations \eqref{actder-Xg}-\eqref{propDm} are very helpful.

\subsection{The effect of gauge transformation}
\label{subsec-effgauge}

Let us now find how the gauge transformation generated by the contact hamiltonian \eqref{gengaugetr} affects
the instanton corrected Darboux coordinates calculated in the previous subsection.
To this end, first, we obtain the exponentiated action of the vector field \eqref{contbr}, up to second order in $g_{\bfp}$.
It reads as
\bea
e^{X_{\gH}}\cdot  \xi^a &\approx & \xi^a
-\frac{\I}{2\pi}\sum_{\bfp}e^{2\pi\I p^a\txi_a}\[p^a g_{\bfp}
+\frac{\I}{4\pi}\sum_{\bfp_1+\bfp_2=\bfp}p_1^a\(p_1^b g_{\bfp_1}\p_{\xi^b}g_{\bfp_2}-p_2^b g_{\bfp_2}\p_{\xi^b}g_{\bfp_1}\)\],
\nn\\
e^{X_{\gH}}\cdot \txi_\Lambda &\approx & \txi_\Lambda
+\frac{1}{4\pi^2}\sum_{\bfp}e^{2\pi\I p^a\txi_a}\[\p_{\xi^\Lambda} g_{\bfp}
+\frac{\I}{4\pi}\sum_{\bfp_1+\bfp_2=\bfp}\(p_1^a\p_{\xi^\Lambda} g_{\bfp_1}\p_{\xi^a}g_{\bfp_2}-p_2^a g_{\bfp_2}\p_{\xi^a}\p_{\xi^\Lambda}g_{\bfp_1}\)\],
\nn\\
e^{X_{\gH}}\cdot  \alpha &\approx & \alpha
+\frac{1}{4\pi^2}\sum_{\bfp}e^{2\pi\I p^a\txi_a}\Biggl[\(1-\xi^\Lambda\p_{\xi^\Lambda}\) g_{\bfp}
\Biggr.
\nn\\
&& \Biggl. \qquad
+\frac{\I}{4\pi}\sum_{\bfp_1+\bfp_2=\bfp}\(p_1^a\(1-\xi^\Lambda\p_{\xi^\Lambda}\) g_{\bfp_1}\p_{\xi^a}g_{\bfp_2}
+p_2^a g_{\bfp_2}\,\xi^\Lambda\p_{\xi^a}\p_{\xi^\Lambda}g_{\bfp_1}\)\Biggr].
\label{gaugetr-Dc}
\eea
Next, we expand the contact hamiltonian as in \eqref{defgHpert}. This gives (see footnote \ref{foot-exphg})
\be
\begin{split}
\gHi{1}_{\bfp} =&\,\frac{1}{4\pi^2} \,e^{2\pi\I p^a\txi^{\rm cl}_a}g_{\bfp}(\xi_{\rm cl}),
\\
\gHi{2}_{\bfp_1\bfp_2}=&\,
2\pi\I\(\gHq_{\bfp_1}\(p_1^a\Dop{\bfp_2}_a \cJq_{\bfp_2}\)-\(p_2^a\Dop{\bfp_1}_a \gHq_{\bfp_1}\)\cJq_{\bfp_2}\),
\end{split}
\label{defcGnew}
\ee
where we have used that, to the first order in instantons,
\be
2\pi\I p_1^a\,\delta\txi_a\, g_{\bfp_1}(\xi_{\rm cl})+\delta\xi^a \p_{\xi^a}g_{\bfp_1}(\xi_{\rm cl})=
\sum_{\bfp_2}\[2\pi\I p_1^a \(\Dop{\bfp_2}_a -\kappa_{abc}(b^b-\I zt^b)p_2^c\)\cJq_{\bfp_2}
+p_2^a\cJq_{\bfp_2}\p_{\xi^a}\]g_{\bfp_1}(\xi_{\rm cl})
\ee
and
\be
\Dop{\bfp}_a\gHq_{\bfp}= \(\frac{\I}{2\pi}\,\p_{\xi^a}+\kappa_{abc}p^c(b^b-\I z t^b)\)\gHq_{\bfp}.
\label{actDop-hg}
\ee
It is also useful to note similar identities for the other derivative operators
\be
\begin{split}
\Del{\bfp}_+\gHq_{\bfp}=&\,\(\frac{\I}{2\pi}-\xi_{\rm cl}^a\Dop{\bfp}_a\)\gHq_{\bfp},
\\
\Del{\bfp}_-\gHq_{\bfp}=&\,\(\frac{\I}{2\pi}\(1-2\I\tau_2\p_{\xi^0}-\xi_{\rm cl}^a\p_{\xi^a}\)+
\kappa_{abc}p^c(b^b-\I z t^b)(c^a-\tau_1 b^a-\tau_2 z t^a)\)\gHq_{\bfp}.
\end{split}
\label{actDel-hg}
\ee
Then plugging the expansion \eqref{defgHpert} into \eqref{gaugetr-Dc} and using the above identities,
by a lengthy but straightforward calculation,
one can show that all terms produced by the gauged transformation result from a simple shift of the function $\tcJ_{\bfp}$
determining the original Darboux coordinates, as described in \eqref{gtr-dcinst}.

\section{Indefinite theta series and their modular properties}
\label{sec-theta}

In this section we introduce several variants of indefinite theta series which appear in the main text.
These constructions are all based on Vign\'eras' theorem \cite{Vigneras:1977}, which we recall for completeness.

\subsection{Vign\'eras' theorem}
\label{subsec-Vign}

Let $\Lambda$ be an $n$-dimensional lattice equipped with a bilinear form
$(\bfx,\bfy)\equiv \bfx\cdot\bfy$, where $\bfx,\bfy\in \Lambda \otimes \mathbb{R}$, such that its associated quadratic form
has signature $(n_+,n_-)$ and is integer valued, i.e. $\bfk^2\equiv \bfk\cdot \bfk\in\IZ$ for $\bfk\in\Lambda$.
Furthermore, let $\bfp\in\Lambda$  be a characteristic vector
(such that $\bfk\cdot(\bfk+ \bfp)\in 2\mathbb{Z}$, $\forall \,\bfk \in \Lambda$),
$\bfmu\in\Lambda^*/\Lambda$ a glue vector, and $\lambda$ an arbitrary integer.
With the usual notation $q=\expe{\tau}$, we consider the following family of theta series
\be
\label{Vignerasth}
\vartheta_{\bfp,\bfmu}(\Phi,\lambda;\tau, \bfb, \bfc)=\tau_2^{-\lambda/2} \sum_{{\bfk}\in \Lambda+\bfmu+\hf\bfp}
(-1)^{\bfk\cdot\bfp}\,\Phi(\sqrt{2\tau_2}(\bfk+\bfb))\, q^{-\frac12 (\bfk+\bfb)^2}\,\expe{\bfc\cdot (\bfk+\haf\bfb)}
\ee
defined by a kernel $\Phi(\bfx)$ such that the function $f(\bfx)\equiv \Phi(\bfx)\, e^{\tfrac{\pi}{2}\,\bfx^2}
\in L^1(\Lambda\otimes\IR)$ so that the sum is absolutely convergent.
Irrespective of the choice of this kernel and of the parameter $\lambda$, any such theta series satisfies the
following elliptic properties
\be
\begin{split}
\label{Vigell}
\vartheta_{\bfp,\bfmu}\left({\Phi} ,\lambda; \tau, \bfb+\bfk,\bfc\right) =&(-1)^{\bfk\cdot\bfp}\,
\expe{-\haf\, \bfc\cdot \bfk} \vartheta_{\bfp,\bfmu}\left({\Phi} ,\lambda; \tau, \bfb,\bfc\right),
\\
\vphantom{A^A \over A_A}
\vartheta_{\bfp,\bfmu}\left({\Phi}, \lambda; \tau, \bfb,\bfc+\bfk \right)=&(-1)^{\bfk\cdot\bfp}\,
\expe{\haf\, \bfb\cdot \bfk} \vartheta_{\bfp,\bfmu}\left({\Phi} ,\lambda; \tau, \bfb,\bfc\right).
\end{split}
\ee

Now let us require that in addition the kernel satisfies the following two conditions:
\begin{enumerate}
\item
Let $D(\bfx)$ be any differential operator of order $\leq 2$, and
$R(\bfx)$ any polynomial of degree $\leq 2$. Then $f(\bfx)$ defined above must
be such that $f(\bfx)$, $D(\bfx)f(\bfx)$ and $R(\bfx)f(\bfx)\in
L^2(\Lambda\otimes\mathbb{R})\bigcap L^1(\Lambda\otimes\mathbb{R})$.
\item
$\Phi(\bfx)$ must satisfy
\be
\label{Vigdif}
V_\lambda\cdot \Phi(\bfx)=0,
\qquad
V_\lambda= \partial_{\bfx}^2   + 2\pi \( \bfx\cdot \pa_{\bfx}  - \lambda\)  .
\ee
\end{enumerate}
Then in \cite{Vigneras:1977} it was proven that
the theta series \eqref{Vignerasth} transforms as a vector-valued modular form of
weight $(\lambda+n/2,0)$ (see Theorem 2.1 in \cite{Alexandrov:2016enp} for the
detailed transformation under $\tau\to-1/\tau$). We refer to $V_\lambda$ as Vign\'eras' operator.
The simplest example is the Siegel
theta series, which arises by choosing $\Phi(\bfx)=e^{-\pi \bfx_+^2}$ where $\bfx_+$
is the projection of $\bfx$ on a fixed positive plane of dimension $n_+$, which is
annihilated by $V_{-n_+}$.

\subsection{Generalizations}
\label{subsec-shift}

In the main text we need various generalizations of Vign\'eras' theorem,
which allow to describe theta series of a more general form than \eqref{Vignerasth}.

First, we lift the theta series \eqref{Vignerasth} to twistor space, i.e. we allow the kernel to depend
on the $\CP$ variable $z$ and its complex conjugate.
To state the conditions for modularity, note that the combinations $\sqrt{\tau_2}z$ and $\bz/\sqrt{\tau_2}$
transform with modular weight $(-1,0)$ and $(1,0)$, respectively.
Therefore, expanding the kernel $\Phi(\bfx,z,\bz)$, generating a modular theta series of weight $(\wh,0)$,
in a Laurent series around $z=0$,
\be
\Phi(\bfx,z,\bz)=\sum_{k,l}\Phi_{k,l}(\bfx)(\sqrt{\tau_2}z)^k (\bz/\sqrt{\tau_2})^l
\ee
the coefficients $\Phi_{k,l}$ must define themselves modular theta series of weight $(\wh+k-l,0)$
and hence satisfy Vign\'eras' equation with $\lambda_{k,l}=\lambda+k-l$ and $\lambda=\wh-n/2$.
This immediately implies that the full kernel $\Phi(\bfx,z,\bz)$ should be annihilated
by the following twistorial extension of Vign\'eras' operator,
\be
V_\lambda\cdot \Phi(\bfx,z,\bz)=0,
\qquad
V_\lambda=\p_{\bfx}^2+2\pi\(\bfx\cdot \p_{\bfx}-z\p_z+\bz\p_{\bz}-\lambda\).
\label{zdep-Vignop}
\ee

Second, let us introduce a shifted version of the operators $\bar\bfD$ defined in \eqref{defDw-new},
\be
\begin{split}
\bfD^\star=&\, \bar\bfD-\frac{\I}{2\tau_2}\(\bfc-\tau \bfb\)
\\
=&\,\frac{1}{4\pi\tau_2}\(\p_{\bfb}+\tau\p_{\bfc}-\I\pi \(\bfc-\tau \bfb\)\).
\end{split}
\ee
The motivation for this definition is that, upon acting on theta series \eqref{Vignerasth},
this new operator, like $\bfD$ but in contrast to $\bar\bfD$, preserves their form, up to a factor of $\tau_2$.
To avoid this mismatch, let us introduce two  vectors $\bfA$ and $\bfA^\star$, which
depend on $\tau$ and other moduli so as to transform as modular forms of weight $(-1,0)$ and $(0,-1)$,
respectively (see \eqref{choiceA-z} below for a relevant example).
Then the two operators, $\bfA\cdot \bfD$ and $\bfA^\star\cdot \bfD^{\star}$,
act on the theta series \eqref{Vignerasth} via
\be
\begin{split}
\bfA\cdot\bfD\ :\quad & \Phi(\bfx)\mapsto \frac{1}{2\pi\sqrt{2\tau_2}}\,\bfA\cdot (\p_{\bfx}+2\pi\bfx) \Phi(\bfx),
\\
\bfA^\star\cdot \bfD^{\star}\ :\quad & \Phi(\bfx)\mapsto\frac{1}{2\pi\sqrt{2\tau_2}} \,\bfA^\star\cdot\p_{\bfx} \Phi(\bfx).
\end{split}
\label{mapkernel}
\ee
In particular, if $\Phi$ satisfies \eqref{Vigdif} so that \eqref{Vignerasth} is a modular form
of weight $(\lambda+n/2,0)$, then $(\bfA\cdot\bfD)\vartheta_{\bfp,\bfmu}(\Phi ,\lambda)$
and $(\bfA^\star\cdot \bfD^{\star})\vartheta_{\bfp,\bfmu}(\Phi ,\lambda)$ are also modular
with the same weight.

These operators can be used to obtain yet another generalization of Vign\'eras' theorem.
To this end, we introduce the following operator
\be
D[\bfA,\bfA^\star]= \sum_{n=0}^\infty\frac{(2\pi)^n}{n!}\Bigl(\bfA\cdot \bfD+\bfA^\star\cdot \bfD^{\star}\Bigr)^n ,
\ee
which maps modular forms to modular forms of the same weight.
Acting on theta series \eqref{Vignerasth}, it changes the kernel into
\bea
D[\bfA,\bfA^\star]\ :\quad \Phi(\bfx)&\mapsto & \mPhi(\bfx)=\sum_{n=0}^\infty\frac{1}{n!(2\tau_2)^{n/2}}
\Bigl(\bfA\cdot (\p_{\bfx}+2\pi\bfx)+\bfA^\star\cdot \p_{\bfx}\Bigr)^n\Phi(\bfx)
\nn\\
&=& \exp\[\frac{\pi}{2\tau_2}\,\bfA\cdot\(\bfA+\bfA^\star\)+\frac{2\pi}{\sqrt{2\tau_2}}\,\bfA\cdot \bfx\]
\exp\[\frac{1}{\sqrt{2\tau_2}}\(\bfA+\bfA^\star\)\cdot \p_{\bfx}\]\Phi(\bfx)
\nn\\
&=&\exp\[\frac{\pi}{2\tau_2}\,\bfA\cdot\(\bfA+\bfA^\star\)+\frac{2\pi}{\sqrt{2\tau_2}}\,\bfA\cdot \bfx\]
\Phi\(\bfx+\frac{\bfA+\bfA^\star}{\sqrt{2\tau_2}}\),
\label{modifiedker-l}
\eea
where we summed the series and applied the Baker-Campbell-Hausdorff formula.
This result shows that the theta series with the kernel \eqref{modifiedker-l} is also modular with weight $(\lambda+n/2,0)$,
provided the vectors $\bfA$ and $\bfA^\star$
transform with modular weight $(-1,0)$ and $(0,-1)$, respectively.\footnote{In the derivation of \eqref{modifiedker-l}
it was implicitly assumed that $\bfA$ and $\bfA^\star$ do not depend on $\bfb$ and $\bfc$. But it is easy to see that
the above statement holds also in the presence of such dependence.}

As an important application, let us choose
\be
\bfA=-2\I\tau_2 z \bft,
\qquad
\bfA^\star=2\I\tau_2\bz\bft_\perp,
\label{choiceA-z}
\ee
where $\bft$ is a timelike vector, $|\bft|=\sqrt{\bft\cdot\bft}$ and
\be
\bft_\perp=\bft-\frac{\bft\cdot\bft}{\bft\cdot \bft'}\,\bft'
\ee
such that $\bft\cdot\bft_\perp=0$.
Substituting \eqref{choiceA-z} into \eqref{modifiedker-l}, one concludes that the kernel
\be
\mPhi(\bfx,z)=e(x_+,\sqrt{2\tau_2}\,|\bft| z)\,
\Phi\(\bfx-\I\sqrt{2\tau_2}\(z\bft-\bz\bft_\perp\)\),
\label{modker-z}
\ee
where
\be
e(x,y)=e^{-\pi y^2-2\pi\I x y},
\label{expfac}
\ee
generates a modular theta series, consistently with the fact that it
satisfies the generalized Vign\'eras' equation \eqref{zdep-Vignop}.

Since the exponential factor \eqref{expfac} arises repeatedly in our analysis,
it is convenient to absorb it into the definition of the theta series on twistor space.
Further multiplying by the modular invariant factor $\sigma_{\bfp}\, e^{- S^{\rm cl}_{\bfp}}$,
the theta series then take the form\footnote{Quantities with (without) a hat correspond to the new (old) frame, respectively.}
\be
\hat\vartheta_{\bfp,\bfmu}(\tilde\Phi,\lambda;\tau, \bfb, \bfc,z)
= \tau_2^{-\lambda/2} \!\!
\sum_{{\bfk}\in \Lambda+\bfmu+\hf\bfp}\!\!
\sigma_\gamma \hat\Phi\(\sqrt{2\tau_2}(\bfk+\bfb),\sqrt{2\tau_2}\,|\bft|z\)\,
\expe{p^a\txi_a^{\rm cl}-k_a\xi^a_{\rm cl}-\frac{\tau}{2}\, \bfk^2},
\label{Vignerasth-z}
\ee
where $\txi_a^{\rm cl}$, $\xi^a_{\rm cl}$ denote the classical Darboux coordinates in the double limit $t^a\to\infty$,
$z\to 0$ with $zt^a$ fixed. Eq. \eqref{Vignerasth-z} transforms as a vector-valued
modular form of weight $(\lambda+\tfrac{n}{2},0)$ provided the new kernel $\hat\Phi$ satisfies a modified version of  \eqref{zdep-Vignop},
\be
\hV_\lambda\cdot\hat\Phi(\bfx,y)=0,
\qquad
\hV_\lambda= \p_{\bfx}^2+2\pi\(\(\bfx-{\textstyle\frac{2\I y}{|\bft|}}\,\bft\)\cdot \p_{\bfx}-y\p_y+\by\p_{\by}-\lambda\),
\label{zdep-Vignop-mod}
\ee
where $y=\sqrt{2\tau_2}\,|\bft|z$.

\subsection{Indefinite theta series of Lorentzian signature}
\label{subsec-error1}

We now discuss several important examples of indefinite theta series of signature $(1,n-1)$,
referring  to \cite{Alexandrov:2016enp} for more details on this construction.

Let us first introduce the error and complementary error functions
\be
\label{M1Erfc}
E_1(u) = \Erf(\sqrt{\pi}u)\, ,
\qquad
M_1(u)=-\sgn(u)\, \Erfc\(\sqrt{\pi}|u|\) ,
\ee
such that $E_1(u)=M_1(u)+\sgn(u)$. It is straightforward to see that $E_1(u)$ is
a smooth solution of Vign\'eras' equation on $\IR$ with $\lambda=0$ which asymptotes
to $\sgn(u)$ as $|u|\to\infty$, while $M_1(u)$ is a solution of the same equation which
is exponentially suppressed as $|u|\to\infty$ and smooth except at $u=0$. The latter
statements are most easily seen using  the  integral representation
\begin{equation}
\label{defM1}
M_1(u)=\frac{\I}{\pi}\int_{\ell}\,\frac{\de y}{y}\,
e^{-\pi y^2-2\pi\I u y},
\end{equation}
where the contour $\ell=\IR-\I u$ runs parallel to the real axis through the saddle point.
Indeed,  the second equality in \eqref{M1Erfc} follows by
changing the integration variable, $y=y'-\I u$, and using the identity
\be
\int_{\IR} \frac{\de y'}{y'-\I\alpha}\, e^{-\beta^2 y'^2} =
\I \pi \,\sgn (\Re(\alpha))\, e^{\alpha^2\beta^2} \Erfc(\sgn(\Re(\alpha\beta)) \alpha\beta).
\label{ident-int}
\ee

In order to construct convergent theta series with signature $(1,n-1)$, let us introduce two vectors
$\bft, \bft'\in \IR^{1,n-1}$ with positive norm, $\bft^2, \bft'^2>0$, such that $\bft\cdot\bft'\geq 0$. Defining
\be
x_+ = \frac{\bfx\cdot \bft}{\sqrt{\bft\cdot\bft}},
\qquad
x'_+ = \frac{\bfx\cdot \bft'}{\sqrt{\bft'\cdot\bft'}} ,
\ee
the kernel
\be
\label{PhiE1ttp}
\widehat\Phi_{\bft,\bft'}(\bfx) = E_1( x_+) - E_1(x'_+)
= \bigl[ \sgn(x_+) - \sgn(x'_+) \bigr] + M_1(x_+) - M_1(x'_+)
\ee
satisfies the assumptions of Vign\'eras' theorem. Indeed, it is a smooth solution of
\eqref{Vigdif} with $\lambda=0$, and
the square bracket in \eqref{PhiE1ttp} vanishes whenever $x_+$ lies
in the dangerous region $\bfx^2>0$, while the last two terms (multiplied by $e^{\tfrac{\pi}{2}\,\bfx^2}$)
decay exponentially. Thus, the theta series $\vartheta_{\bfp,\bfmu}(\widehat\Phi_{\bft,\bft'})$
is a vector-valued modular form of weight $(n/2,0)$  \cite{Zwegers-thesis}.
However, it is non-holomorphic due to the last two terms in \eqref{PhiE1ttp}. In contrast, the
theta series with kernel
\be
\Phi_{\bft,\bft'}(\bfx) =  \sgn(x_+) - \sgn(x'_+)
\ee
is holomorphic but not modular. Since the theta series with kernels $M_1(x_+)$ and $M_1(x'_+)$
can be written as Eichler integrals of a Siegel-type theta series (see e.g. Remark 2.4 in  \cite{Alexandrov:2016enp}),
the modular anomaly of $\vartheta_{\bfp,\bfmu}\left(\Phi_{\bft,\bft'}\right)$ involves
period integrals of these Siegel-type theta series.

An important observation is that either of the last two  terms in \eqref{PhiE1ttp} can be eliminated by
letting $\bft'$ (or $\bft$) approach a rational
null vector $\bft'_r$, i.e. a vector $\bft'_r\in \Lambda/N$ for some integer $N$ with $(\bft'_r)^2=0$.
For generic values of $\bfb$, the limit is smooth, and leads to a theta series  with kernel
\be
\label{PhiE1ttr}
\widehat\Phi_{\bft,\bft'_r}(\bfx) = \left[ \sgn(\bfx\cdot\bft) - \sgn(\bfx\cdot\bft'_r) \right] + M_1(x_+) .
\ee
It is worth noting that \eqref{PhiE1ttr} is annihilated by Vign\'eras' operator $V_0$,
despite the fact that it is  discontinuous at $\bfx\cdot\bft'_r=0$.
The series $\vartheta_{\bfp,\bfmu}(\widehat\Phi_{\bft,\bft'_r})$ however diverges whenever
$\bft'_r\cdot (\bfk+\bfb)=0$ for some $\bfk\in\Lambda+\bfmu+\tfrac12 \bfp$ \cite{Zwegers-thesis}.

It is also crucial that the null vector $\bft'_r$ be rational since, otherwise, 
$\vartheta_{\bfp,\bfmu}\left(\Phi_{\bft,\bft'}\right)$ will
diverge in the limit $\bft'\to \bft'_r$. This point is best illustrated
on an example, e.g.
\be
\label{th11div}
\sum_{(k_1,k_2)\in \IZ^2} \tfrac{1}{2} k_1 \(\sgn(k_1)-\sgn(2k_1-\sqrt{2}k_2) \) q^{-k^2/2}
\ee
corresponding to an indefinite theta series with quadratic form
$\diag(2,-1)$ and with $t=(1,0)$ and $t'=(1,\sqrt{2})$.
For \eqref{th11div} to converge, the following ``half'' theta series
\be
\sum_{k_1>0,\,2k_1-\sqrt{2}k_2<0} q^{-(2k_1^2-k_2^2)/2}.
\ee
should certainly converge.
However, the subset of integers $(k_1,k_2)=(\ell, \lceil \ell \sqrt{2} \rceil)$ in this sum has norm
$2k_1^2-k_2^2=2\ell \sqrt{2}\{-\ell\sqrt{2}\}-\{-\ell\sqrt{2}\}^2\equiv -2n$, where
$0\leq \{x\}<1$ is the rational part of $x$, and it appears that  an infinite number of $\ell$
contribute for any given exponent $n$. For example, all $\ell$ of the form
$\ell(r)=\tfrac{1}{2}\left( (1+\sqrt{2})^{2r+1}+(1-\sqrt{2})^{2r+1}\right)$
for any $r>0$, contribute $q^1$ to the sum leading to the divergence
of the theta series. The series $\ell(r)$ is known as the
Newman-Shanks-Williams series of primes.

Next, let us apply the result \eqref{modker-z} from the previous subsection to the kernel \eqref{PhiE1ttr}.
It allows to conclude that
\be
\Phi_1(\bfx,y)=e(x_+,y)\,
\Bigl[E_1\(x_{+}-\I y\)- \sgn\(\bfx\cdot \bft'+2t'_+\Im y\)\Bigr]
\label{modPhi-z}
\ee
is annihilated by $V_0$ and generates a modular theta series of weight $(n/2,0)$.
Note that the argument of the sign function is nothing but
$\wwx'(y)$ introduced in \eqref{defwwx1}, whereas
the first term can be split as in \eqref{PhiE1ttr},
\be
\hE_1(x_+,y)\equiv \Erf\[\sqrt{\pi}\(x_+ - \I y\)\]=\hM_1(x_+,y) + \sgn \(x_+ + \Im y\),
\label{defE1z}
\ee
where the argument of the sign coincides with $\wwx(y)$ and we defined
\be
\hM_1(u,y)=-\sgn(\Re v)\, \Erfc\(\sgn(\Re v)\sqrt{\pi}v\),
\qquad
v=u-\I y.
\ee
Using again the identity \eqref{ident-int}, this function can be rewritten in the following integral form
\be
\hM_1(u,y) = \frac{\I}{\pi} \int_{\ell} \,\frac{\de y'}{y'-y}\, e^{\pi(y^2-y'^2)+2\pi\I (y-y') u},
\label{defM1-z}
\ee
which generalizes \eqref{defM1}.
After the coordinate change $y=\sqrt{2\tau_2}\,|\bft|z$, this is exactly the integral which arises in our twistorial construction.
Finally, note that the factor $e(x_+,y)$ in \eqref{modPhi-z} is the same as the one which was absorbed to define
the rescaled theta series \eqref{Vignerasth-z}. In this new frame, for the quadratic form $\kappa_{ab}=\kappa_{abc}p^c$,
the kernel \eqref{modPhi-z} corresponds precisely to $\hKer_\bfp$ \eqref{Phi-first},
which shows that $\hat\vartheta_{\bfp,\bfmu}(\hKer_\bfp)$
is a modular theta series of weight $(\frac{b_2}{2},0)$.

\subsection{Indefinite theta series of conformal signature}
\label{subsec-error2}

We now turn to the case of signature $(2,2n-2)$.
Analogues $E_2(\alpha;u_1,u_2)$
and $M_2(\alpha;u_1,u_2)$ of the error functions \eqref{M1Erfc}
satisfying Vign\'eras' equation on $\IR^{2}$ were constructed in \cite{Alexandrov:2016enp}.
For the purposes of this paper, it will be useful to introduce a one-parameter
generalization of these functions, namely\footnote{These functions
can be further generalized by making the coefficient in front of $y$ arbitrary.
However, this coefficient can be reabsorbed into $y$ since Vign\'eras' equation is invariant under  this rescaling.
We used this freedom to fix the coefficient to the value appropriate to our applications.
}

\be
M_2(\alpha;u_1,u_2,y) = -  \frac{1}{\pi^2}\int_{\ell_1}\frac{\de y_1}{y_1-\frac{y}{\sqrt{1+\alpha^2}}}
\int_{\ell_2}\frac{\de y_2}{y_2-\alpha y_1}\,
e^{-\pi(y_1^2+y_2^2)-2\pi\I\(u_1 y_1+u_2 y_2\)} ,
\label{defM2-z}
\ee
and
\bea
\label{defE2-z}
&&E_2(\alpha;u_1, u_2,y) = M_2(\alpha;u_1, u_2,y)
+e_{\alpha}(y)\biggl[\sgn\(u_1+\frac{\Im y}{\sqrt{1+\alpha^2}}\) \hM_1\(u_2,\frac{\alpha y}{\sqrt{1+\alpha^2}}\)
\biggr.\\
&&\qquad
+\sgn(u_2-\alpha u_1) \,
\hM_1\(\frac{u_1+\alpha u_2}{\sqrt{1+\alpha^2}}\,,y\)
\biggl.
+ \sgn\(u_2+\frac{\alpha\Im y}{\sqrt{1+\alpha^2}}\)\,  \sgn\(\frac{u_1+ \alpha u_2}{\sqrt{1+\alpha^2}}+\Im y\)\biggr],
\nn
\eea
where $e_{\alpha}(y)=e\(\frac{u_1+\alpha u_2}{\sqrt{1+\alpha^2}}, y\)$
is the exponential factor \eqref{expfac} with a suitable choice of the first argument.
By acting with the operator $V_\lambda$ and integrating by parts, it is straightforward to check that both of these functions
are solutions of Vign\'eras' equation  with $\lambda=0$. Analogously to $M_1(u)$,
$M_2(\alpha;u_1, u_2,y)$ is exponentially suppressed away from the origin in the $(u_1,u_2)$ plane,
and is discontinuous across the lines $u_1=-\frac{\Im y}{\sqrt{1+\alpha^2}}$ and $u_2=\alpha u_1$,
where the zero of one of the denominators sits on the integration contour. Across these loci,
it behaves as
\bea
M_2(\alpha_1,\alpha_2;u_1,u_2,y) &\sim &
-\sgn\(u_1+\frac{\Im y}{\sqrt{1+\alpha^2}}\) e_{\alpha}(y)\, \hM_1\(u_2,\frac{\alpha y}{\sqrt{1+\alpha^2}}\)
\qquad \mbox{near } u_1=-\frac{\Im y}{\sqrt{1+\alpha^2}},
\nn\\
M_2(\alpha_1,\alpha_2;u_1,u_2,y) &\sim & - \sgn(u_2-\alpha u_1) \, e_{\alpha}(y)\,
\hM_1\(\frac{u_1+\alpha u_2}{\sqrt{1+\alpha^2}}\,, y\)
\quad\qquad\qquad\  \mbox{near } u_2=\alpha u_1.
\nn
\eea
The additional terms in
\eqref{defE2-z} ensure that $E_2(\alpha;u_1, u_2,y)$ is a smooth function of $(u_1,u_2)$,
which asymptotes to
\be
E_2(\alpha;u_1, u_2,y) \to  e_{\alpha}(y)\, \sgn\(u_2+\frac{\alpha\Im y}{\sqrt{1+\alpha^2}}\)
\sgn\(\frac{u_1+ \alpha u_2}{\sqrt{1+\alpha^2}}+\Im y\)
\ee
as $u_1^2+u_2^2\to\infty$.

Starting from these functions, we now construct solutions of Vign\'eras' equation on $\IR^{2,n-2}$
parametrized by two linearly independent timelike vectors\footnote{We reserve bold letters for the vectors
in a Lorentzian space and use calligraphic fonts such as $\Xv, \Cv$ for vectors in
$\IR^{2,n-2}$.}  $\Cv_1$, $\Cv_2$ such that $\Delta_{12}\equiv \Cv_1^2 \Cv_2^2 -(\Cv_1,\Cv_2)^2>0$,
namely
\be
\Phi^M_2(\Cv_1,\Cv_2;\Xv,y) =
M_2\( \frac{(\Cv_1,\Cv_2)}{\sqrt{\Delta_{12}}};\frac{(\Cv_{1\perp 2},\Xv)}{|\Cv_{1\perp 2}|},\frac{(\Cv_2,\Xv)}{|\Cv_2|},y \),
\label{defM2-Cz}
\ee
\be
\Phi^E_2(\Cv_1,\Cv_2;\Xv,y) =
E_2\( \frac{(\Cv_1,\Cv_2)}{\sqrt{\Delta_{12}}};\frac{(\Cv_{1\perp 2},\Xv)}{|\Cv_{1\perp 2}|},\frac{(\Cv_2,\Xv)}{|\Cv_2|},y \),
\label{defE2-Cz}
\ee
where $\Cv_{k\perp l}$ is the projection of $\Cv_k$ on the subspace orthogonal to $\Cv_l$. Both
of these functions are annihilated by $V_0$, everywhere in the case of $\Phi^E_2$,
away from the loci $(\Cv_{1\perp 2},\Xv)=-\frac{\Cv_{1\perp 2}^2}{|\Cv_1|}\, \Im y$
and $(\Cv_{2\perp 1},\Xv)=0$ in the case of $\Phi^M_2$.
At $y=0$, they reduce to the ``boosted double  error functions" introduced in \cite{Alexandrov:2016enp}.

Finally, one can upgrade the function \eqref{defE2-Cz} to a smooth solution of the Vign\'eras' equation with $\lambda=1$ parametrized by
an additional vector $\Cv_3$. To this end, one can act on $\Phi^E_2$ by the operator $(\Cv_3, (\Xv +  \frac{1}{2\pi} \p_\Xv))$,
which, as can be seen from \eqref{mapkernel}, realizes the action of the covariant derivative rising the holomorphic weight by 1.
A straightforward evaluation gives
\bea
&&\PhiT(\Cv_1,\Cv_2,\Cv_3; \Xv,y)= (\Cv_3, (\Xv +  \frac{1}{2\pi} \p_\Xv)) \, \Phi^E_2(\Cv_1,\Cv_2; \Xv,y)
\nn\\
&&\quad
= \(\Cv_3,\Xv-\I y \frac{\Cv_1}{|\Cv_1|}\) \Phi^E_2(\Cv_1,\Cv_2; \Xv,y)
\label{defT2-Cz}\\
&&\qquad
+ \frac{(\Cv_2,\Cv_3)}{\pi|\Cv_2|} \,e^{-\frac{\pi\(\Cv_2, \Xv\)^2}{\Cv_2^2}} E_1\(\tfrac{(\Cv_{1\perp2}, \Xv)}{|\Cv_{1\perp2}|};
\tfrac{\sqrt{\Delta_{12}}}{|\Cv_1||\Cv_2|}\, y\)
+ \frac{(\Cv_1,\Cv_3)}{\pi|\Cv_1|} \, e^{-\frac{\pi\(\Cv_1, \Xv \)^2}{\Cv_1^2}} E_1\(\tfrac{(\Cv_{2\perp1},\Xv)}{|\Cv_{2\perp1}|}\).
\nn
\eea

In order to compare the results with the main text, it will be convenient to introduce
an analogue of the theta series \eqref{Vignerasth-z} with rescaled kernels.
In this case the rescaling is given by $e_\alpha(y)$ so that we define
\be
\hM_2=e_\alpha^{-1} M_2,
\qquad
\hE_2=e_\alpha^{-1} E_2,
\qquad
\hat\Phi^{M,E}_2=e_\alpha^{-1} \Phi^{M,E}_2,
\qquad
\hPhiT=e_\alpha^{-1} \PhiT,
\label{def-hfun2}
\ee
where in the last two relations the parameters $\alpha$ and $u_i$ 
of the exponential factor are chosen as in \eqref{defM2-Cz}, \eqref{defE2-Cz}.

We are interested in the particular case of this general construction where the
signature $(2,n-2)$ lattice is a direct sum $\Lambda_{1}\oplus \Lambda_{2}$,
with quadratic form $\Xv^2=(p_1x_1^2)+(p_2x_2^2)$ where  $\Xv=(\bfx_1,\bfx_2)$.
Moreover, the vectors determining the function $\hPhiT$ are given by
\be
\Cv_1 = (\bft,\bft),
\qquad
\Cv_2 = (0,\bft),
\qquad
\Cv_3 =  (\bfp_2,-\bfp_1).
\ee
The norms of these vectors are
\be
\Cv_1^2=(pt^2),
\qquad
\Cv_2^2=(p_2t^2),
\qquad
\Cv_3^2=(pp_1p_2).
\ee
In addition, one finds
\be
\Cv_{1\perp 2} = (\bft,0),
\qquad
\Cv_{2\perp1} = \frac{1}{(pt^2)} \, \bigl(-(p_2t^2) \bft, (p_1t^2) \bft\bigr),
\ee
\be
(\Cv_1,\Cv_2)=(p_2t^2)\, ,
\qquad
(\Cv_2,\Cv_3)=-(p_1p_2t)\, ,
\qquad
(\Cv_1,\Cv_3)=0,
\qquad
\Delta_{12}=(p_1t^2)(p_2t^2)\, .
\label{CCC-scpr}
\ee
Comparing \eqref{defE2-Cz} with \eqref{CCC-scpr}, we see that in our case the arguments of the $E_2$ function
are taken to be $u_i=x_{i+}$ and $\alpha=\sqrt{\frac{(p_2t^2)}{(p_1t^2)}}$, which implies that
$e_\alpha(y)=e(x_+,y)$ where $\bfx=\bfx_1 +\bfx_2$ so that
$x_+= \sqrt{\frac{(p_1t^2)}{(pt^2)}}\,x_{1+}+\sqrt{\frac{(p_2t^2)}{(pt^2)}}\,x_{2+}$.
Substituting these results into \eqref{defE2-z} and \eqref{defT2-Cz} and
writing down the result using the rescaled kernels \eqref{def-hfun2}, one finds
\be
\begin{split}
\hat\Phi_2^E (\bfx_1,\bfx_2,y) =&\,
\hM_2\(\rpp{2}{1}, x_{1+},x_{2+}, y\)
+\sgn(\wwx_1(y)) \hM_1(x_{2+},y_2)
\\
&\,
-\sgn(\Ixx) \,
\hM_1\(x_+,y\)
+ \sgn(\wwx_2(y))\,  \sgn(\wwx(y)),
\label{defhPhiE2}
\end{split}
\ee
\be
\hPhiT(\bfx_1,\bfx_2,y) =
\Xxx\, \hat\Phi_2^E (\bfx_1,\bfx_2,y)
-{\textstyle\frac{(p_1p_2t)}{\pi\sqrt{(p_2t^2)}}}\,
e^{-\pi \(x_{2+}-\I y_2\)^2}
\hE_1(x_{1+}, y_1),
\label{defhPhiT2}
\ee
where we used notations from \eqref{defwwx1}, \eqref{defwwx}, \eqref{defXxx} and $y_s=\rpp{s}{}\,y$.
By construction these kernels satisfy $\hV_0\cdot \hat\Phi_2^E=0$ and $\hV_1\cdot \hPhiT =0$
where $\hV_\lambda$ is the double lattice version of \eqref{zdep-Vignop-mod}
where $\bfx$ should be replaced by $\Xv=(\bfx_1,\bfx_2)$ and the vector $\bft$ by $(\bft,\bft)$.
Explicitly, the operator reads
\be
\hV_\lambda= \p_{\bfx_1}^2+\p_{\bfx_2}^2+2\pi\(\(\bfx_1-\textstyle{\frac{2\I y\,\bft}{\sqrt{(pt^2)}}}\)\cdot \p_{\bfx_1}
+\(\bfx_2-\textstyle{\frac{2\I y\,\bft}{\sqrt{(pt^2)}}}\)\cdot \p_{\bfx_2}-y\p_y+\by\p_{\by}-\lambda\).
\label{zdep-Vignop-two}
\ee

\subsection{Safe and dangerous anomalies}
\label{subsec-thetaanom}

Now we would like to address the following question: Let $\Phian$ is a kernel satisfying
Vign\'eras' equation up to terms proportional to delta functions and derivatives thereof.
Does it imply that the corresponding theta series $\vartheta_{\bfp,\bfmu}(\Phian)$ has an anomalous modular transformation?
And, if so, is the kernel governing the modular anomaly also proportional to a delta function,
or can it have a more general support?

This question is important for our analysis of modularity because in our set-up,
delta function anomalies in modular transformations can be safely ignored.
Indeed, discontinuities in the Darboux coordinates and the gauge transformation, which lead eventually to delta functions
resulting from the action of Vign\'eras' operator, always correspond to boundaries between two patches on the twistor space.
A typical example is the discontinuity across a BPS ray corresponding to $\ww_\gamma(z)=0$.
Since Darboux coordinates are only defined away from these loci, such modular anomalies
supported on delta functions do not play any role
(cf. footnote \ref{foot-deltas}).

To discuss this issue, in this section we allow the kernels defining theta series to be
not necessarily functions, but in general distributions with delta-function support on some codimension one loci.
To understand when a localized anomaly in the action of Vign\'eras' operator leads to a localized anomaly in the
modular transformation of the corresponding theta series, let us assume that
\be
V_\lambda\cdot\Phian=\sum_{m}\cA_m ,
\ee
where $\cA_m$ are given by a combination of delta functions and its derivatives, i.e.
\be
\cA_m(\bfx)=\sum_{l}\[ a_{m,l}\,\delta(\wwx_{m,l}(\bfx))+b_{m,l}\,\delta'(\wwx_{m,l}(\bfx))\],
\ee
where $\wwx_{m,l}(\bfx)$ is a set of real linear forms. We claim that,
{\it if each of the $\cA_m(\bfx)$'s is annihilated by Vign\'eras' operator $V_{\lambda_m}$ for
some $\lambda_m\ne \lambda$,
then the anomaly appearing in the modular transformation of $\vartheta_{\bfp,\bfmu}(\Phian)$
is localized at the zeros of $\wwx_{m,l}(\sqrt{2\tau_2}(\bfk+\bfb))$ for some $\bfk\in\Lambda+\bfmu+\tfrac12\bfp$. }
Indeed, it is easy to see that
\be
\widehat\Phi=\Phian+\sum_m\frac{\cA_m}{2\pi(\lambda-\lambda_m)}
\label{Phian-complet}
\ee
is annihilated by $V_\lambda$ and thus generates a modular theta series of weight $(\lambda+n/2,0)$.
Furthermore, the condition $V_{\lambda_m}\cdot \cA_m=0$ implies that
the theta series constructed from $\cA_m$ are also modular,
but with different weights given by $(\lambda_m+n/2,0)$. Combining this information, we can obtain the modular transformation of
$\vartheta_{\bfp,\bfmu}(\Phian)$ which is found to be\footnote{We ignored the possibility of having
a non-trivial multiplier system which is irrelevant for the present discussion.}
\be
\vartheta_{\bfp,\bfmu}(\Phian,\lambda) \mapsto (c\tau+d)^{\lambda+\frac{n}{2}}
\[\vartheta_{\bfp,\bfmu}(\Phian,\lambda)+\sum_m
\tfrac{\tau_2^{\hf(\lambda_m-\lambda)}}{2\pi(\lambda-\lambda_m)}\,
\(1-\(\tfrac{c\tau+d}{c\bar\tau+d}\)^{\hf(\lambda_m-\lambda)}\)
\vartheta_{\bfp,\bfmu}(\cA_m,\lambda_m)\].
\label{anomtr}
\ee
Since the anomaly is a linear combination of $\vartheta_{\bfp,\bfmu}(\cA_m)$, it is localized
at the zeros of $\wwx_{m,l}(\sqrt{2\tau_2}(\bfk+\bfb))$.
Let us illustrate this situation in two examples:

\smallskip

{\bf Example 1:} First, we consider the theta series defined by the kernel \eqref{PhiE1ttr}
where {\it both} vectors $\bft$ and $\bft'$ are taken to be timelike, which we denote by $\Phian_{\bft,\bft'}$.
Then the action of Vign\'eras' operator gives
\be
\cA\equiv V_0\cdot \Phian_{\bft,\bft'}=-2(\bft')^2\,\delta'(\bfx\cdot \bft').
\ee
In this case the anomaly $\cA$ does not satisfy Vign\'eras' equation for any $\lambda$. Instead, one has
\be
V_{-2}\cdot \cA=-2(\bft')^4\,\delta^{(3)}(\bfx\cdot \bft'),
\ee
where we used the property $x\,\delta^{(n)}(x)=-n\,\delta^{(n-1)}(x)$. Continuing in this way, one generates an infinite set
of derivatives of $\delta(\bfx\cdot \bft')$. This infinite series can in fact be
resummed into a smooth solution of Vign\'eras' equation.
Indeed, let us consider
\be
\widehat\Phi_{\bft,\bft'}=\sum_{n=0}^\infty \frac{1}{(4\pi)^n n!}\[\prod_{k=0}^{n-1} V_{-2k}\]\cdot\Phian_{\bft,\bft'}.
\ee
Proceeding as above, it is easy to show that
\be
\begin{split}
\widehat\Phi_{\bft,\bft'}=&\, E_1(x_+)-\sum_{n=0}^\infty \frac{1}{n!}\,\frac{(\bft')^{2n}}{(4\pi)^n}\,\sgn^{(2n)}(\bfx\cdot \bft')
= E_1(x_+)-e^{\frac{(\bft')^{2}}{4\pi}\, \p_{x}^2}\left.\sgn(x)\right|_{x=\bfx\cdot \bft'}.
\end{split}
\ee
Computing the action of  the heat kernel operator acting on the sign function in Fourier space,
one sees that the second term is equal to $E_1(x'_+)$, so that $\widehat\Phi_{\bft,\bft'}$ coincides with the kernel \eqref{PhiE1ttp}
which is smooth and annihilated by $V_0$. Thus, despite the fact that the anomaly $\cA$ is proportional to the derivative of delta function,
it becomes `delocalized' by the iterated action of Vign\'eras' operator, and the
modular anomaly of the theta series based on $\Phian_{\bft,\bft'}$ is a theta series with
a smooth kernel (which is itself a period integral of a non-anomalous theta series).

\smallskip

{\bf Example 2:}
Now let us consider the following kernel
\be
\Phianp_{\bft,\bft'}=x_+ \widehat\Phi_{\bft,\bft'}+\frac{1}{\pi}\, e^{-\pi x_+^2},
\ee
where $\bft'$ is now taken to be null. This kernel satisfies
\be
\cA'\equiv V_1\cdot \Phianp_{\bft,\bft'}=-4t'_+\,\delta(\bfx\cdot \bft').
\ee
In contrast to the previous case, due to the condition $\bft'^2=0$, the anomaly $\cA'$ is annihilated by $V_{-1}$.
Then according to the general results \eqref{Phian-complet} and \eqref{anomtr},
the modular completion of $\Phianp_{\bft,\bft'}$ is given by
\be
\widehat\Phi'_{\bft,\bft'}=\Phianp_{\bft,\bft'}+\frac{1}{4\pi}\,\cA'.
\label{Phian-compl}
\ee
The anomaly in the transformation of $\vartheta_{\bfp,\bfmu}(\Phianp_{\bft,\bft'})$ is localized
at solutions of $(\bfk+\bfb)\cdot\bft'=0$ and can be safely dropped. This  should come as no surprise because
the kernel \eqref{Phian-compl} is obtained by acting with the modular covariant derivative
on the modular theta series $\vartheta_{\bfp,\bfmu}(\widehat\Phi_{\bft,\bft'})$
(see \eqref{mapkernel})
\be
D_+=\frac{\bft\cdot\bfD}{\sqrt{\bft^2}}\ :\quad
\widehat\Phi_{\bft,\bft'}\mapsto \widehat\Phi'_{\bft,\bft'}.
\ee
Dropping the delta function term encoded by $\cA'$ is in fact equivalent 
to ignoring the action of the derivative operator on the sign functions
in this theta series. This is precisely what we did in our analysis of 
instanton corrected Darboux coordinates, see footnote \ref{foot-deltas}.

\smallskip

The lesson from these examples is that delta-function anomalies in Vign\'eras' equation
can be ignored if they are themselves eigenfunctions of Vign\'eras' operator with a
different eigenvalue.

\section{Convergence of indefinite theta series of signature $(2,n-2)$}
\label{sec-converge}

Extended numerical tests suggest the validity of the following
\medskip

{\bf Conjecture:} {\it
Let $\Xv\in\IR^n$ equipped with a bilinear form $(\,\cdot\,,\,\cdot\,)$ of signature $(2,n-2)$ as in Appendix \ref{subsec-Vign},
$\{\Cv_k\}$ is a set of $N$ vectors in $\IR^n$, $\{\eps_k=\pm 1\}$ is a set of $N$ signs, and
\be
\Phi(\Xv)=P(\Xv)\times\left\{
\begin{array}{ll}
\sum\limits_{k=1}^N \eps_k\, \sgn(\Cv_k,\Xv)\,\sgn(\Cv_{k+1},\Xv) &\quad \mbox{for $N$ even},
\\
\eps+\sum\limits_{k=1}^N \eps_k \, \sgn(\Cv_k,\Xv)\, \sgn(\Cv_{k+1},\Xv)& \quad \mbox{for $N$ odd},
\end{array}
\right.
\label{convker}
\ee
where $\Cv_{N+1}\equiv \Cv_1$, $P(\Xv)$ is a function of at most polynomial growth, and
the signs are required to satisfy
\be
\left\{
\begin{array}{ll}
\prod\limits_{k=1}^N\eps_k=(-1)^{N/2} &\quad \mbox{for $N$ even},
\\
\prod\limits_{k=1}^N\eps_k=(-1)^{(N+1)/2}\eps & \quad \mbox{for $N$ odd}.
\end{array}
\right.
\label{condsigns}
\ee
Then provided the following three conditions are satisfied, the theta series
defined in \eqref{Vignerasth} using the kernel \eqref{convker}
is convergent:
\be
\begin{array}{ll}
1) & \Cv_k^2\ge 0,
\\
2) &\Delta_{k,k+1}>0\quad \mbox{if \, $\Cv_k^2,\, \Cv_{k+1}^2>0$, \ or}\quad \Delta_{k,k+1}=0 \quad \mbox{if one of them is null},
\\
3) & \eps_k \eps_{k+1}(\Cv_{k\perp k+1},\Cv_{k+2\perp k+1})<0,
\end{array}
\label{condconv}
\ee
where $\Delta_{kl}=\Cv_k^2 \Cv_l^2 -(\Cv_k,\Cv_l)^2$ and $\Cv_{k\perp l}$ 
is the projection of $\Cv_k$ on the subspace orthogonal to $\Cv_l$.
}

Some comments are in order:
\begin{itemize}
\item
The statement for odd $N$ follows from the one for even $N$.
Let us assume for simplicity that there are at least two consecutive timelike vectors (not null), which we label by 1 and $N$.
Then we start from the kernel \eqref{convker} for $N+1$ where we specialize to the case $\Cv_{N+1}=\Cv_N$ and rename
$\eps_N\to \eps$, $\eps_{N+1}\to \eps_N$. This produces the kernel for $N$ odd with the signs satisfying \eqref{condsigns}.
Thus, it remains only to check the conditions \eqref{condconv}. They all follow immediately except
\be
\eps_{N-1}\eps_N(\Cv_{N-1\perp N},\Cv_{1\perp N})<0.
\ee
To prove this condition, we start from the two original conditions for $N+1$, which after relabeling the signs $\epsilon_k$, read
\be
\begin{split}
\eps_{N-1}\eps(\Cv_{N-1\perp N},\Cv_{N+1\perp N})< &\, 0,
\\
\eps\eps_N(\Cv_{N\perp N+1},\Cv_{1\perp N+1})< &\, 0.
\end{split}
\label{twocondN}
\ee
Taking their product, we see that we need to show that in the limit $\Cv_{N+1}=\Cv_N$, the scalar product
$(\Cv_{N-1\perp N},\Cv_{1\perp N})$ is of different sign than
$(\Cv_{N-1\perp N},\Cv_{N+1\perp N})(\Cv_{N\perp N+1},\Cv_{1\perp N+1})$.
The problem is that both $\Cv_{N\perp N+1}$ and $\Cv_{N+1\perp N}$ vanish in this limit.
To avoid this, let us take $\Cv_{N+1}=\Cv_N+\delta \Cv$ where $\delta \Cv$ is infinitesimally small.
A straightforward calculation shows that
\be
\begin{split}
(\Cv_{N-1\perp N},\Cv_{N+1\perp N})(\Cv_{N\perp N+1},\Cv_{1\perp N+1})\approx &\,
- (\Cv_{N-1\perp N},\delta \Cv)(\delta \Cv,\Cv_{1\perp N})
\\
=&\,
- (\Cv_{N-1\perp N},\delta \Cv_{\perp N})(\delta \Cv_{\perp N},\Cv_{1\perp N}).
\end{split}
\label{scalprodrel}
\ee
Note that the condition $\Delta_{N,N+1}>0$ implies
\be
\Delta(\Cv_N,\delta \Cv)>0 \ \Rightarrow\ \delta \Cv_{\perp N}^2>0.
\label{cond-delC}
\ee
Similarly, the positivity of $\Delta_{N+1,1}$ leads to the positivity of $\Cv_{1\perp N}^2$.
Finally, we can take $\delta \Cv$ such that $\delta \Cv_{\perp N}=\veps\, \Cv_{1\perp N}$
since both vectors are timelike and orthogonal to $\Cv_N$.
Here $|\veps|\ll 1$ and $\sgn(\veps)=-\eps\eps_N$ so that the second condition in \eqref{twocondN}
is fulfilled automatically.
Then the r.h.s. of \eqref{scalprodrel} becomes $- \veps^2 \Cv_{1\perp N}^2 (\Cv_{N-1\perp N},\Cv_{1\perp N})$,
which proves the desired property.

\item
In the case of even $N$, the signs $\epsilon_k$ can be brought to a standard form
by changing signs of the vectors $\Cv_k$.
To this end, let us redefine
\be
\Cv_k\ \mapsto\ (-1)^{\[\frac{k-1}{2}\]}\(\prod_{l=1}^{k-1} \eps_k \)\Cv_k.
\ee
As a result, the kernel becomes
\be
\Phi(\Xv)=P(\Xv)\sum\limits_{k=1}^N (-1)^{k-1}\sgn(\Cv_k,\Xv)\,\sgn(\Cv_{k+1},\Xv),
\label{convker-red}
\ee
whereas the third condition in \eqref{condconv} takes the form
\be
(\Cv_{k\perp k+1},\Cv_{k+2\perp k+1})>0.
\label{condconv-red}
\ee
For $N=4$, this reduces to the conformal theta series first considered in \cite{Alexandrov:2016enp},
whose convergence conditions were clarified in \cite{kudla2016theta}. The convergence
conditions \eqref{condconv} are natural generalizations of those in \cite{kudla2016theta},
and presumably allow to construct a compact two-dimensional surface $S$ in the space $D$
of oriented positive 2-planes, bounded by an $N$-sided polygon, such that $S$ intersects the
codimension 2 subspace $\Xv^\perp \subset D$ if and only if $\Phi(\Xv)\neq 0$. The compactness
of $S$ would then ensure the convergence of the theta series. We have not tried to
prove this rigorously, but we conducted extensive numerical checks which give
us confidence that these conditions are sufficient.

\item
For $P=1$, the modular completion of the holomorphic theta series $\vartheta_{\bfp,\bfmu}(\Phi)$ is
the theta series $\vartheta_{\bfp,\bfmu}(\widehat\Phi)$ whose kernel is obtained
by replacing each product of two signs by the corresponding generalized error function \eqref{defE2-Cz},
\be
\widehat\Phi(\Xv)=\left\{
\begin{array}{ll}
\sum\limits_{k=1}^N \eps_k\, \Phi^E_2(\Cv_k,\Cv_{k+1};\Xv,0) &\quad \mbox{for $N$ even},
\\
\eps+\sum\limits_{k=1}^N \eps_k \, \Phi^E_2(\Cv_k,\Cv_{k+1};\Xv,0)& \quad \mbox{for $N$ odd}.
\end{array}
\right.
\label{convkerc}
\ee
When $P$ is a non-trivial homogeneous polynomial, the modular completion can be obtained
by applying the heat kernel operator as in Theorem 3.11 in \cite{Alexandrov:2016enp}.

\end{itemize}

\subsection{Application: convergence of the gauge transformation}
\label{subsec-conv-app}

Let us apply the above conjecture for analyzing the convergence of the second term in the contact hamiltonian \eqref{factor-cG}.
Its convergence would follow from the convergence  of a theta series with quadratic form
$(\Xv,\Xv)=(p_1x_1^2)+(p_2x_2^2)-\sum_{i=1}^{N_\bfp}n_i^2$, where $\Xv=(\bfx_1,\bfx_2,n_i)$,
and kernel
\bea
\Phi(\Xv) &= & \Xxx \Bigl[\sgn(\Ixx )\(\sgn((\bfx_1+\bfx_2)\cdot \bft)-\sgn(\bfx_2\cdot \bft)\)
\Bigr.
\nn\\
&&
+\sgn \(\bfx_1\cdot \bft'+\nnf_1\)\(\sgn(\bfx_2\cdot \bft)-\sgn((\bfx_1+\bfx_2)\cdot \bft'+\nnf)\)
\label{conv-doubletheta2}\\
&& \Bigl.
+\sgn(\Xxx) \, \(\sgn((\bfx_1+\bfx_2)\cdot \bft'+\nnf) - \sgn((\bfx_1+\bfx_2)\cdot \bft)\)\Bigr],
\nn
\eea
where
\be
\begin{split}
\Xxx=&\,  \bfx_1\cdot \bfp_2-\bfx_2\cdot \bfp_1,
\\
\Ixx=&\, (p_2t^2)\bfx_1\cdot \bft-(p_1t^2)\bfx_2\cdot\bft,
\\
\nnf_1=\sum_{i=1}^{N_{\bfp_1}}n_i,
\qquad &
\nnf_2=\sum_{i=N_{\bfp_1}+1}^{N_{\bfp}}n_i,
\qquad
\nnf=\nnf_1+\nnf_2.
\end{split}
\label{defXxx}
\ee
The kernel \eqref{conv-doubletheta2} is nothing but the factor appearing in
\eqref{conv-doubletheta} with the charges replaced by $\bfx_s$
and taken to be large together with $n_i$, which allows to drop $z$-dependent terms.

The expression \eqref{conv-doubletheta2} is of the form \eqref{convker} 
for $n=2b_2+N_\bfp$ and $N=6$ with the vectors $\Cv_k$ given by
\bea
&
\Cv_1=(\bft,\bft,\under{\underbrace{0,\dots,0}}{N_\bfp}),
\qquad
\Cv_2=((p_2t^2)\bft,-(p_1t^2)\bft,\under{\underbrace{0,\dots,0}}{N_\bfp}),
&
\nn\\
&
\Cv_3=(0,\bft,\under{\underbrace{0,\dots,0}}{N_\bfp}),
\qquad
\Cv_4=(\bft',0,\under{\underbrace{1,\dots,1}}{N_{\bfp_1}},\under{\underbrace{0,\dots,0}}{N_{\bfp_2}}),
&
\label{defvecC}\\
&
\Cv_5=(\bft',\bft',\under{\underbrace{1,\dots,1}}{N_\bfp}),
\qquad
\Cv_6=(\bfp_2,-\bfp_1,\under{\underbrace{0,\dots,0}}{N_\bfp}),
&
\nn
\eea
where we indicated the number of repetitions of the same entry,
and the signs fixed as $\eps_k=(-1)^{k+1}$.
Let us check now the conditions \eqref{condconv}.
First, the norms of the vectors are
\be
\Cv_1^2=(pt^2),
\quad
\Cv_2^2=(p_1t^2)(p_2t^2)(pt^2),
\quad
\Cv_3^2=(p_2t^2),
\quad
\Cv_4^2=0,
\quad
\Cv_5^2=0,
\quad
\Cv_6^2=(pp_1p_2).
\ee
Since the magnetic charges are assumed to belong to the K\"ahler cone \eqref{khcone}, all the norms are non-negative
so that the first condition is satisfied. Then
\be
\begin{split}
&
\Delta_{12}=(p_1t^2)(p_2t^2)(pt^2)^2,
\qquad
\Delta_{23}=(p_1t^2)(p_2t^2)^3,
\\
\Delta_{34}=&\,0,
\qquad
\Delta_{45}=0,
\qquad
\Delta_{56}=0,
\qquad
\Delta_{61}=(pt^2)(pp_1p_2)
\end{split}
\ee
ensures the second condition. Finally, to check the last one, note that
\be
\Cv_1\perp \Cv_2,
\qquad
\Cv_3\perp \Cv_4,
\qquad
\Cv_4\perp \Cv_5,
\qquad
\Cv_5\perp \Cv_6,
\qquad
\Cv_6\perp \Cv_1,
\ee
and
\be
\begin{split}
\Cv_{2\perp 3}=\bigl((p_2t^2)\bft,0,\under{\underbrace{0,\dots,0}}{N_\bfp}\bigr),
\qquad
\Cv_{3\perp 2}=\biggl(\frac{(p_2t^2)}{(pt^2)}\,\bft,\frac{(p_2t^2)}{(pt^2)}\,\bft,\under{\underbrace{0,\dots,0}}{N_\bfp}\biggr).
\end{split}
\ee
This allows to compute
\be
\begin{array}{rclrcl}
\(\Cv_6,\Cv_2\)&=&(pt^2)(p_1p_2t),
& \qquad
\(\Cv_1,\Cv_{3\perp 2}\)&=&(p_2t^2),
\\
\(\Cv_{2\perp 3},\Cv_4\)&=&(p_2t^2)(p_1tt'),
& \qquad
\(\Cv_3,\Cv_5\)&=&(p_2tt'),
\\
\(\Cv_4,\Cv_6\)&=&(p_1p_2t'),
&\qquad
\(\Cv_1,\Cv_5\)&=&(ptt').
\end{array}
\ee
All these expressions are positive which perfectly agrees with the third condition \eqref{condconv} and the signs $\eps_k=(-1)^{k+1}$.
Thus, the conjecture from the previous subsection ensures the convergence of the theta series
with the kernel \eqref{conv-doubletheta2} and thereby of the contact hamiltonian generating the gauge transformation.

\section{Modularity in the presence of D1-D(-1)-instantons}
\label{sec-D1}

In this Appendix we include the contributions of D1-D(-1)-instantons to the geometry of $\cM_H$
and show that their combined effect with D3-instantons is consistent with modular invariance.
As in the main text, we restrict  to the large volume approximation but we take into account
the following instanton contributions:
\begin{enumerate}
\item pure D1-D(-1)-instantons to all orders (computed in \cite{Alexandrov:2009qq,Alexandrov:2012bu});
\item the first order of D3-instantons mixed with all orders of D1-D(-1)-instantons;
\item the second order of pure D3-instantons (computed in the main text).
\end{enumerate}

\subsection{Twistorial description of D1-D(-1)-instantons}

First, we recall the manifestly S-duality invariant description of D1-D(-1)-instantons found in \cite{Alexandrov:2009qq,Alexandrov:2012bu},
which can be achieved by applying a gauge transformation to the twistorial constriction of D-instantons in the type IIA formulation
presented in section \ref{subsec-MH}.
To define this gauge transformation, we introduce an ordering on the charge lattice
according to the phase of the central charge function:
\be
\gamma>\gamma' \quad \mbox{if}\quad 0<\arg\(Z_\gamma Z_{\gamma'}^{-1}\)<\pi.
\label{order-gamma}
\ee
Then for each charge $\gamma$ we define the associated set of D(-1)-brane charges
whose BPS rays lie in the same half-plane as $\ell_\gamma$,\footnote{In this Appendix we will put a tilde on all D1 and D(-1)-brane charges
and their components to distinguish them from those of D3-brane charges.}
\be
\GamD{-1}_\gamma=\left\{\gamD{-1}=(0,0,0,\tilde q_0)\ :\
\tilde q_0\Re Z_\gamma>0
\right\},
\label{latDm1}
\ee
and another set of D1-brane charges
for which the BPS rays are between $\ell_\gamma$ and the imaginary axis,
\be
\GamD{1}_\gamma=\left\{\gamD{1}=(0,0,\tilde q_a,\tilde q_0)\in H_2^+\cup H_2^-\ :\
\Nq(\gamD{1})=\Nq(\gamma) \ {\rm and}\
\begin{array}{c}
\gamD{1}> \gamma \quad \mbox{for}\ \Nq(\gamma)\ {\rm odd}
\\
\gamD{1}\le \gamma \quad \mbox{for}\ \Nq(\gamma)\ {\rm even}
\end{array}
\right\},
\label{latD1}
\ee
where $H_2^+$ is the set of charges corresponding to effective homology classes on $\CYm$,
$H_2^-$ is the set of opposite charges, and $\Nq(\gamma)$ denotes the quadrant which
$\ell_\gamma$ belongs to.\footnote{One can write
$\Nq(\gamma)=\left\lfloor\frac{2}{\pi}\, \arg \(\I Z_\gamma\)\right\rfloor$.}
Note that both the ordering and the two charge sets $\GamD{\pm 1}_\gamma$ may change after crossing a wall of marginal stability.
Given these definitions, we define a holomorphic function which generates the gauge transformation in the patch
$\cU_\gamma$ taken to lie in the counterclockwise direction from the BPS ray $\ell_\gamma$
\be
\gH_{\gamma}=(-1)^{\Nq(\gamma)}\[\frac{1}{2}\sum_{\gamD{-1}\in\GamD{-1}_\gamma} H_{\gamD{-1}}+
\sum_{\gamD{1}\in\GamD{1}_\gamma}H_{\gamD{1}}\].
\label{fungengauge}
\ee
This gauge transformation has a very simple geometric meaning: It simply
rotates the BPS rays corresponding to D1-instantons either to the positive or negative real axis
depending on which one is the closest to the given ray. On the other hand, the D(-1)-BPS rays,
which all go along the imaginary axis, are split into two ``halves" which are also rotated to the two real half-axes.

The gauge transformation \eqref{fungengauge} changes the covering of the twistor space and the associated set of contact hamiltonians.
Now, on top of the non-rotated BPS rays corresponding to D3-instantons, which split $\CP$ into different angular sectors,
one has an infinite set of open patches $\cU_{m,n}$ centered around points $z_+^{m,n}$ where
\be
z_\pm^{m,n} = \mp \frac{m\tau+n}{|m\tau+n|}
\label{poles}
\ee
are the two roots of the equation $m\xi^0(z)+n=0$. Note that $\cU_{km,kn}=\cU_{m,n}$ for $k>0$
and, in particular, $\cU_{0,\pm k}=\cU_\pm$ are the patches around the two poles of $\CP$, $t=0$ and $t=\infty$.
Nevertheless, it is convenient to distinguish all these patches because this allows to make S-duality manifest.
The contact hamiltonians associated with this covering are given by
\be
\begin{array}{c}
H_+=\Fcl(\xi) ,
\qquad
H_-=\bFcl(\xi) ,
\\
\vphantom{\rule{0pt}{1.3cm}}
H_{m,n}=  -\frac{\I}{(2\pi)^3}\!\!\sum\limits_{\tq_a\gamma^a\in H_2^+(\CYm)\cup\{0\}}\!\!\!\! n_{\tq_a}^{(0)}\,
\begin{cases}
\displaystyle\frac{e^{-2\pi \I m \tq_a\xi^a}}{m^2(m\xi^0+n)}, &
\quad  m\ne 0,
\\
\displaystyle (\xi^0)^2 \,\frac{e^{2\pi \I n \tq_a\xi^a/\xi^0}}{n^3}, &\quad  m=0,
\end{cases}
\end{array}
\label{defG1}
\ee
where $\Fcl(X)$ is the classical prepotential \eqref{Flv}. Besides,
we have set $\gfinv_{0}=-\chi_\CYm/2$ and used that
\be\label{instnr}
\begin{split}
\Omega(\gamD{1}) =&\, \gfinv_{q_a} \ \quad\mbox{\rm for}\quad \gamD{1}=(0,0,\pm \tq_a,\tq_0), \quad \{ \tq_a \} \ne 0,
\\
\Omega(\gamD{1})=&\, 2\gfinv_{0}\quad\mbox{\rm for}\quad \gamD{1}=(0,0,0,\tq_0).
\end{split}
\ee
Note that the full holomorphic prepotential is obtained as $F=\Fcl+\Fws$ where $\Fws=\sum_{n>0}H_{0,n}$ and,
given the identifications between the patches, corresponds to the full transition function to $\cU_+$.

Ignoring contributions of D3-instantons for the moment, the effect of D1-D(-1)-instantons on Darboux coordinates
can be described as follows.
First, corrections to the classical mirror map \eqref{symptobd} are captured by
\bea
\delta^{\rm D1}\tzeta_\Lambda & =&
-\half\,  {\sum_{m,n}}' \oint_{C_{m,n}} \frac{\de \varpi'}{2 \pi \I \varpi'} \,
\frac{1/\varpi'-\varpi'}{1/\varpi'+\varpi'}
\, \p_{\xi^\Lambda}H_{m,n}(\varpi')
+\zeta^\Sigma \Re \Fws_{\Lambda\Sigma}(z^\Lambda),
\label{qmirmapb}
\\
\delta^{\rm D1}\sigma & =&-\zeta^\Lambda\tzeta^{\rm D1}_\Lambda
+ {\sum_{m,n}}' \oint_{C_{m,n}} \frac{\de \varpi'}{2 \pi \I \varpi'} \,
\frac{1/\varpi'-\varpi'}{1/\varpi+\varpi}\,
\(1-\xi^\Lambda\p_{\xi^\Lambda}\)H_{m,n}(\varpi')
+\zeta^\Lambda\zeta^\Sigma \Re\Fws_{\Lambda\Sigma}(z^\Lambda),
\nn
\eea
where $C_{m,n}$ are small circles around $\varpi_+^{m,n}$ (pre-images of $z_+^{m,n}$ under \eqref{Cayley})
and the prime on the sum means that the sum goes over all pairs of integers except $m=n=0$.
The meaning of the first integral terms in \eqref{qmirmapb} is to convert the standard kernel
$\frac12\,\frac{t'+t}{t'-t}\, \frac{\de t'}{t'}$ appearing in \eqref{eqTBA} into
\be
K(\varpi,\varpi') \frac{\de t'}{t'}
=\frac{(1+\varpi\varpi')}{(\varpi'-\varpi)(1/\varpi'+\varpi')}\frac{\de \varpi'}{\varpi'}
=\frac12\,\frac{z'+z}{z'-z}\, \frac{\de z'}{z'}\, ,
\ee
which is invariant under S-duality transformations.
Then the instanton contributions to the Darboux coordinates read
\be
\begin{split}
\delta^{\rm D1}\txi_\Lambda&\, =
- {\sum_{m,n}}' \oint_{C_{m,n}} \frac{\de \varpi'}{2 \pi \I \varpi'} \,K(\varpi,\varpi')
\, \p_{\xi^\Lambda}H_{m,n}(\varpi'),
\\
\delta^{\rm D1}\alpha&\, =
-{\sum_{m,n}}' \oint_{C_{m,n}} \frac{\de \varpi'}{2 \pi \I \varpi'} \,K(\varpi,\varpi')
\(1-\xi^\Lambda\p_{\xi^\Lambda}\)H_{m,n}(\varpi'),
\end{split}
\label{twlineIIBb}
\ee
while $\xi^\Lambda$ remain unaffected.
Evaluating the integrals explicitly by residues using that
\be
m\xi^0 + n = \frac{m\btau+n}{z^2-1} \(z-z_+^{m,n}\)\(z-z_-^{m,n}\),
\label{mxi0n}
\ee
keeping only the leading terms in the large volume limit and computing
the combinations \eqref{defdelta}\footnote{Since $\delta^{\rm D1}\xi^a=0$,
this step amounts just to form linear combinations of $\delta^{\rm D1}\txi_0$ and $\delta^{\rm D1}\alpha$.},
the D1-D(-1)-instanton contributions to the Darboux coordinates are found to be
\bea
\hat\delta^{\rm D1}\txi_a &=&  \frac{\I \tau_2}{8\pi^2}\, \sum_{\tq_a\geq 0}\, n_{\tq_a}^{(0)}\, \tq_a\,
{\sum_{m,n}}'\, \frac{e^{-S_{m,n,\tq_a}}}{|m \tau+n|^2},
\nn\\
\hat\delta^{\rm D1}_+\alpha &=&
\frac{\I\tau_2}{8\pi^3}\, \sum_{\tq_a\geq 0}\,\, n_{\tq_a}^{(0)}\,
{\sum_{m,n}}' \Biggl[\pi  \tau_2 \tq_a t^a \,\frac{m\tau+n}{|m\tau+n|}
\Biggr.
+\tau_2\, \frac{m\tau+n}{|m\tau+n|^2}
+\pi  \tq_a\( c^a-\tau b^a- 2\tau_2 z t^a\)\Biggr]\frac{e^{-S_{m,n,\tq_a}}}{|m \tau+n|^2}\, ,
\nn\\
\hat\delta^{\rm D1}_-\alpha &=&
\frac{\I\tau_2}{8\pi^3}\, \sum_{\tq_a\geq 0}\, \, n_{\tq_a}^{(0)}\,
{\sum_{m,n}}' \Biggl[-\pi  \tau_2\, \tq_a t^a\, \frac{m\bar\tau+n}{|m\tau+n|}
\Biggr.
\nn\\
&& \Biggl. \qquad\quad
-\tau_2\,\frac{m\bar\tau+n}{|m\tau+n|^2}
+\pi  \tq_a\( c^a-\bar\tau b^a+2\tau_2\, \frac{m\bar\tau+n}{m\tau+n}\, z t^a\)\Biggr]\frac{e^{-S_{m,n,\tq_a}}}{|m \tau+n|^2}\, ,
\label{D1txia}
\eea
where
\be
S_{m,n,k_a} =2\pi k_a | m \tau+n |\, t^a-2\pi \I k_a (m c^a +n  b^a)
\ee
is the instanton action. Note that all these contributions transform consistently with \eqref{SL2hatdel}.

\subsection{Adding D3-instantons}

The previous subsection gave a description of the first type of the instanton contributions mentioned in the beginning of this Appendix.
The third type is found in the main text. So it remains to understand only the mixed contributions in which D3-instantons are
restricted to the linear order, whereas D1-D(-1)-instantons are taken into account to all orders.
There are three sources for such contributions:
\begin{enumerate}
\item
the effect of the gauge transformation \eqref{fungengauge} on the contact hamiltonians generating D3-instantons;
\item
corrections from D1-D(-1)-instantons to the Darboux coordinates used to calculate the effect of D3-instantons;
\item
corrections from D3-instantons to the Darboux coordinates used to calculate the effect of D1-D(-1)-instantons.
\end{enumerate}

Let us take into account the first effect.
It was found in \cite{Alexandrov:2014rca} that the gauge transformation \eqref{fungengauge}
affects the contact Hamiltonians \eqref{prepHnew} generating D-instantons and maps them to
\be
H^g_\gamma(\xi,\txi)= \frac{\bOm_\gamma}{(2\pi)^2}\,\sigma_\gamma\,
\expe{p^\Lambda \(\txi_\Lambda-\p_{\xi^\Lambda}\gH_{\gamma}(\xi)\)-q_\Lambda \xi^\Lambda} .
\label{Hgam-tr}
\ee
But in our case this result simplifies enormously. Indeed,
in the large volume limit the BPS rays of all D1-instantons are close to the real axis because
\be
|\Im Z_{\tilde\gamma}|=|\tq_a t^a|\gg |\Re Z_{\tilde\gamma}|=|\tq_a b^a+\tq_0|
\qquad \mbox{for }q_a\gamma^a\ne 0.
\ee
On the other hand, the BPS rays of D3-instantons are close to the imaginary axis due to
\be
|\Im Z_\gamma|\approx |(q_a + \kappa_{abc}b^b p^c)t^a|\ll |\Re Z_\gamma|\approx \hf\,(pt^2).
\ee
As a result, for a D3-brane charge $\gamma$, the set $\GamD{1}_\gamma$ \eqref{latDm1} is empty and $\gH_{\gamma}$
appearing in \eqref{Hgam-tr} contains only the contribution of D(-1)-instantons.
Geometrically, this happens because the gauge transformation rotates the D1-BPS rays to the closest real axis,
and in the process this does not require to exchange their relative positions with any of D3-BPS rays once we are in the large volume limit.
Furthermore, the contribution of D(-1)-instantons also drops out because it depends only on $\xi^0$,
but in our case $p^0=0$ so that $H^g_\gamma=H_\gamma$.
Thus, the first effect is actually absent.

The second effect can be easily taken into account by adding the instanton corrections \eqref{D1txia} to $\cXcl_\gamma$
computed in \eqref{cXzero}, which is to be replaced by
\be
\cXclD_\gamma=\cXcl_\gamma \,\expe{p^a \hat\delta^{\rm D1}\txi_a}.
\ee
Since the additional factor does not depend on the electric charge $q_\Lambda$,
it simply modifies the classical action adding to it a modular invariant term. Thus, it amounts to the following replacement
\be
S^{\rm cl}_\bfp\ \mapsto\ S^{\rm D1}_\bfp = \frac{\tau_2}{2}\,p^a\(\kappa_{abc}t^b t^c
+ \frac{1}{4\pi^2}\, \sum_{\tq_a\geq 0}\, n_{\tq_a}^{(0)}\, \tq_a\,
{\sum_{m,n}}'\, \frac{e^{-S_{m,n,\tq_a}}}{|m \tau+n|^2}\)
 - \I p^a \tc_a .
\label{newScl}
\ee

Finally, the last effect appears due to the D3-instanton correction to the Darboux coordinate $\xi^a$, which appears
in the integrands of \eqref{twlineIIBb}.
The problem however is that one cannot use the expression for this correction given in the main text
because the argument $z$ of $\xi^a$, relevant for the computation of this effect,
belongs to the contour $C_{m,n}$ and does not scale towards zero in the large volume limit,
as was assumed in the body of this paper.
Instead, for finite $z$, using the freedom to adjust the mirror map,
the D3-instanton part of $\xi^a$ can be brought to the following form
(see \cite[(B.11)]{Alexandrov:2012au})
\be
\delta^{\rm D3}\xi^a=\frac{1}{4\pi^2} \sum_{\gamma\in\Gamma_+}p^a\sigma_\gamma\bar\Omega(\gamma)
\[ \int_{\ell_\gamma} \frac{\de z'}{z'-z} \,\frac{1-z'^2}{1-z^2}
\, \cX_\gamma
-\int_{\ell_{-\gamma}} \frac{\de z'}{z'-z} \,\frac{1-z'^2}{1-z^2}\, \frac{z^3}{(z')^3}
\, \cX_{-\gamma}\].
\ee
Restricting to the linear order in D3-instantons and taking only the limit $t^a\to\infty$,
which implies $z'\to 0$ or $\infty$ depending on whether $\gamma\in\Gamma_+$ or $\Gamma_-$,
while keeping $z$ fixed, one finds
\be
\begin{split}
\delta^{\rm D3}\xi^a= &\, \frac{1}{4\pi^2(z^2-1)} \sum_{\gamma\in\Gamma_+}p^a\sigma_\gamma\bar\Omega(\gamma)
\[ \frac{1}{z}\int_{\ell_\gamma} \de z'\, \cXclD_\gamma
-z^3\, \int_{\ell_{-\gamma}} \frac{\de z'}{(z')^2} \, \cXclD_{-\gamma}\]
\\
=&\, \frac{1}{z^2-1}\sum_{\bfp} p^a \(z^{-1}\cFq_{\bfp}+z^3 \overline{\cFq_{\bfp}}\),
\end{split}
\label{corrxia}
\ee
where we used the function $\cFq_{\bfp}$ defined in \eqref{defcJ}, \eqref{expcJ}.
Evaluating the correction \eqref{corrxia} and its first derivative at $z_+^{m,n}$, one obtains
\bea
{\xmn^a}&\equiv& m \delta^{\rm D3}\xi^a(z_+^{m,n})=\frac{\I|m\tau+n|^3}{2\tau_2}\sum_{\bfp} p^a
\(\frac{\cFq_{\bfp}}{(m\tau+n)^2}+\frac{\overline{\cFq_{\bfp}}}{(m\bar\tau+n)^2}\),
\label{D3ximn}\\
{\xpmn^a}&\equiv& m^2 \p_z\delta^{\rm D3}\xi^a(z_+^{m,n})
=
\frac{|m\tau+n|^2(m\bar\tau+n)}{\tau_2^2(m\tau_1+n)}\(
\sum_{\bfp}  p^a\Re\cFq_{\bfp} -\frac{\I m^2\tau_2^3}{|m\tau+n|^3}\,\xmn^a\).
\label{D3dximn}
\eea

To find the corresponding instanton contributions to the Darboux coordinates, it is sufficient to add the correction \eqref{corrxia}
to all instances of $\xi^a$ appearing in \eqref{twlineIIBb}, expand the resulting expressions to the first order in $\delta^{\rm D3}\xi^a$,
and evaluate the integrals by residues. This straightforward procedure leads to the following results
\be
\begin{split}
\hat\delta^{\mbox{\scriptsize D3-D1}}\txi_a
=&\, \frac{\tau_2}{4\pi}\, \sum_{\tq_a\geq 0}\, n_{\tq_a}^{(0)}\, \tq_a\,
{\sum_{m,n}}'\, \frac{\tq_b\xmn^b}{|m \tau+n|^2}\,e^{-S_{m,n,\tq_a}},
\\
\hat\delta^{\mbox{\scriptsize D3-D1}}_{+}\alpha
=&\, \frac{\tau_2}{8\pi^2} \sum_{\tq_a \ge 0} n^{(0)}_{\tq_a}\tq_a
{\sum_{m,n}}'\bigg[\frac{2\pi\tq_b\xmn^b}{|m\tau+n|^2} \(c^a - \tau b^a + \frac{\tau_2 t^a (m\tau+n)}{|m\tau+n|}\)
\\
&\, \qquad\qquad
-\frac{\I}{2}\sum_{\bfp}p^a \,\frac{m\tau+n}{|m\tau+n|}\(\frac{\cFq_{\bfp}}{(m\tau+n)^2} -\frac{3\overline{\cFq_{\bfp}}}{(m\btau+n)^2}\)
\bigg]\,e^{-S_{m,n,\tq_a}},
\\
\hat\delta^{\mbox{\scriptsize D3-D1}}_{-}\alpha
=&\,\frac{\tau_2}{8\pi^2} \sum_{\tq_a \ge 0} n^{(0)}_{\tq_a}\tq_a
{\sum_{m,n}}'\bigg[\frac{2\pi\tq_b\xmn^b}{|m\tau+n|^2} \(c^a - \btau b^a - \frac{\tau_2 t^a (m\btau+n)}{|m\tau+n|}\)
\\
&\, \qquad\qquad
-\frac{\I}{2}\sum_{\bfp}p^a \,\frac{m\btau+n}{|m\tau+n|}\(\frac{3\cFq_{\bfp}}{(m\tau+n)^2} -\frac{\overline{\cFq_{\bfp}}}{(m\btau+n)^2}\)
\bigg]\,e^{-S_{m,n,\tq_a}}.
\end{split}
\label{mixedD1D3}
\ee
Since $\xmn^a$ is modular invariant, as follows from \eqref{D3ximn} and the fact that $\cFq_\bfp$
is a modular form of weight $\(-\tfrac32,\tfrac12\)$, it is easy to check that
all contributions transform consistently with \eqref{SL2hatdel}.
Note that we kept terms of different order in the large volume expansion: the terms proportional to $t^a$ in \eqref{mixedD1D3}
are of one order more comparing to all other terms. This is important for verification of the modular symmetry
because it does not mix the terms depending on $t^a$ with other types of terms and hence their invariance must be checked independently.

The full instanton contributions to Darboux coordinates, which comprise all types of corrections we wanted to include,
can be presented in the following form
\be
\hat\delta\Xi^I=\hat\delta^{\rm D1}\Xi^I+\hat\delta^{\mbox{\scriptsize D3-D1}}\Xi^I+ \ff^I[\tcJ_\bfp],
\label{full-dXi}
\ee
where $\ff^I$ are the functionals introduced in \eqref{deffunctional} and $\tcJ_\bfp$ is defined in \eqref{deftcJp}.
The only difference of the last term comparing to the main text is that in the functions $\tcJ_\bfp$ and $\tcF_\bfp$,
when they are expressed in terms of theta series, the factor $e^{-S^{\rm cl}_\bfp}$ is replaced by $e^{-S^{\rm D1}_\bfp}$
(see \eqref{newScl}; equivalently, the factor $\expe{p^a \hat\delta^{\rm D1}\txi_a}$ should be included into
the definition \eqref{Vignerasth-z} of $z$-dependent theta series),
and all differential operators commute with this factor {\it by definition}.\footnote{After redefining
them according to
$\Dop{\bfp}_a\mapsto e^{-S^{\rm D1}_\bfp} \Dop{\bfp}_a e^{S^{\rm D1}_\bfp}$, etc.}
To achieve the modular covariance, one then performs the same gauge transformation as in section \ref{sec-gauge}.
It has exactly the same effect as before by replacing $\tcJ_\bfp$ by $\hcJ_\bfp$ in the last term in \eqref{full-dXi}.
Since the first two terms have the right transformation properties by themselves, the proof of modularity
is identical to the one given in section \ref{sec-modularity}.

\providecommand{\href}[2]{#2}\begingroup\raggedright\endgroup


\begin{thebibliography}{10}

\bibitem{RoblesLlana:2006is}
D.~Robles-Llana, M.~Ro\v{c}ek, F.~Saueressig, U.~Theis, and S.~Vandoren,
  ``{Nonperturbative corrections to 4D string theory effective actions from
  SL(2,Z) duality and supersymmetry},'' {\em Phys. Rev. Lett.} {\bf 98} (2007)
  211602,
\href{http://www.arXiv.org/abs/hep-th/0612027}{{\tt hep-th/0612027}}.

\bibitem{Alexandrov:2011va}
S.~Alexandrov, ``{Twistor Approach to String Compactifications: a Review},''
  {\em Phys.Rept.} {\bf 522} (2013) 1--57,
\href{http://www.arXiv.org/abs/1111.2892}{{\tt 1111.2892}}.

\bibitem{Alexandrov:2013yva}
S.~Alexandrov, J.~Manschot, D.~Persson, and B.~Pioline, ``{Quantum
  hypermultiplet moduli spaces in N=2 string vacua: a review},'' in {\em
  {Proceedings, String-Math 2012, Bonn, Germany, July 16-21, 2012}},
  pp.~181--212.
\newblock 2013.
\newblock
\href{http://www.arXiv.org/abs/1304.0766}{{\tt 1304.0766}}.
\newblock

\bibitem{Alexandrov:2010ca}
S.~Alexandrov, D.~Persson, and B.~Pioline, ``{Fivebrane instantons, topological
  wave functions and hypermultiplet moduli spaces},'' {\em JHEP} {\bf 1103}
  (2011) 111, \href{http://www.arXiv.org/abs/1010.5792}{{\tt 1010.5792}}.

\bibitem{Alexandrov:2014mfa}
S.~Alexandrov and S.~Banerjee, ``{Fivebrane instantons in Calabi-Yau
  compactifications},'' {\em Phys.Rev.} {\bf D90} (2014) 041902,
\href{http://www.arXiv.org/abs/1403.1265}{{\tt 1403.1265}}.

\bibitem{Alexandrov:2014rca}
S.~Alexandrov and S.~Banerjee, ``{Dualities and fivebrane instantons},'' {\em
  JHEP} {\bf 1411} (2014) 040,
\href{http://www.arXiv.org/abs/1405.0291}{{\tt 1405.0291}}.

\bibitem{Alexandrov:2008gh}
S.~Alexandrov, B.~Pioline, F.~Saueressig, and S.~Vandoren, ``{D-instantons and
  twistors},'' {\em JHEP} {\bf 03} (2009) 044,
\href{http://www.arXiv.org/abs/0812.4219}{{\tt 0812.4219}}.

\bibitem{Alexandrov:2009zh}
S.~Alexandrov, ``{D-instantons and twistors: some exact results},'' {\em J.
  Phys.} {\bf A42} (2009) 335402,
\href{http://www.arXiv.org/abs/0902.2761}{{\tt 0902.2761}}.

\bibitem{Gaiotto:2008cd}
D.~Gaiotto, G.~W. Moore, and A.~Neitzke, ``{Four-dimensional wall-crossing via
  three-dimensional field theory},'' {\em Commun.Math.Phys.} {\bf 299} (2010)
  163--224, \href{http://www.arXiv.org/abs/0807.4723}{{\tt 0807.4723}}.

\bibitem{Alexandrov:2009qq}
S.~Alexandrov and F.~Saueressig, ``{Quantum mirror symmetry and twistors},''
  {\em JHEP} {\bf 09} (2009) 108,
\href{http://www.arXiv.org/abs/0906.3743}{{\tt 0906.3743}}.

\bibitem{Alexandrov:2012au}
S.~Alexandrov, J.~Manschot, and B.~Pioline, ``{D3-instantons, Mock Theta Series
  and Twistors},'' {\em JHEP} {\bf 1304} (2013) 002,
\href{http://www.arXiv.org/abs/1207.1109}{{\tt 1207.1109}}.

\bibitem{Maldacena:1997de}
J.~M. Maldacena, A.~Strominger, and E.~Witten, ``{B}lack hole entropy in
  {M}-theory,'' {\em JHEP} {\bf 12} (1997) 002,
\href{http://www.arXiv.org/abs/hep-th/9711053}{{\tt hep-th/9711053}}.

\bibitem{Alexandrov:2008nk}
S.~Alexandrov, B.~Pioline, F.~Saueressig, and S.~Vandoren, ``{Linear
  perturbations of quaternionic metrics},'' {\em Commun. Math. Phys.} {\bf 296}
  (2010) 353--403,
\href{http://www.arXiv.org/abs/0810.1675}{{\tt 0810.1675}}.

\bibitem{Zwegers-thesis}
S.~Zwegers, ``Mock theta functions.'' PhD dissertation, Utrecht University,
  2002.

\bibitem{Alexandrov:2016tnf}
S.~Alexandrov, S.~Banerjee, J.~Manschot, and B.~Pioline, ``{Multiple
  D3-instantons and mock modular forms I},'' {\em Comm. Math. Phys.} (2016)
\href{http://www.arXiv.org/abs/1605.05945}{{\tt 1605.05945}}.

\bibitem{Manschot:2009ia}
J.~Manschot, ``{Stability and duality in N=2 supergravity},'' {\em
  Commun.Math.Phys.} {\bf 299} (2010) 651--676,
  \href{http://www.arXiv.org/abs/0906.1767}{{\tt 0906.1767}}.

\bibitem{Hosono:1993qy}
S.~Hosono, A.~Klemm, S.~Theisen, and S.-T. Yau, ``{Mirror symmetry, mirror map
  and applications to Calabi-Yau hypersurfaces},'' {\em Commun. Math. Phys.}
  {\bf 167} (1995) 301--350,
\href{http://www.arXiv.org/abs/hep-th/9308122}{{\tt hep-th/9308122}}.

\bibitem{Alexandrov:2016enp}
S.~Alexandrov, S.~Banerjee, J.~Manschot, and B.~Pioline, ``{Indefinite theta
  series and generalized error functions},''
\href{http://www.arXiv.org/abs/1606.05495}{{\tt 1606.05495}}.

\bibitem{Vigneras:1977}
M.-F. Vign\'eras, ``{S\'eries th\^eta des formes quadratiques ind\'efinies},''
  {\em Springer Lecture Notes} {\bf 627} (1977) 227 -- 239.

\bibitem{Cecotti:1988qn}
S.~Cecotti, S.~Ferrara, and L.~Girardello, ``{G}eometry of type {II}
  superstrings and the moduli of superconformal field theories,'' {\em Int. J.
  Mod. Phys.} {\bf A4} (1989)
2475.

\bibitem{Ferrara:1989ik}
S.~Ferrara and S.~Sabharwal, ``{Q}uaternionic manifolds for type {II}
  superstring vacua of {C}alabi-{Y}au spaces,'' {\em Nucl. Phys.} {\bf B332}
  (1990)
317.

\bibitem{Antoniadis:1997eg}
I.~Antoniadis, S.~Ferrara, R.~Minasian, and K.~S. Narain, ``{$R^4$ couplings in
  M- and type II theories on Calabi-Yau spaces},'' {\em Nucl. Phys.} {\bf B507}
  (1997) 571--588,
\href{http://www.arXiv.org/abs/hep-th/9707013}{{\tt hep-th/9707013}}.

\bibitem{Gunther:1998sc}
H.~G{\"u}nther, C.~Herrmann, and J.~Louis, ``{Quantum corrections in the
  hypermultiplet moduli space},'' {\em Fortsch. Phys.} {\bf 48} (2000)
  119--123,
\href{http://www.arXiv.org/abs/hep-th/9901137}{{\tt hep-th/9901137}}.

\bibitem{Antoniadis:2003sw}
I.~Antoniadis, R.~Minasian, S.~Theisen, and P.~Vanhove, ``String loop
  corrections to the universal hypermultiplet,'' {\em Class. Quant. Grav.} {\bf
  20} (2003) 5079--5102,
\href{http://www.arXiv.org/abs/hep-th/0307268}{{\tt hep-th/0307268}}.

\bibitem{Robles-Llana:2006ez}
D.~Robles-Llana, F.~Saueressig, and S.~Vandoren, ``String loop corrected
  hypermultiplet moduli spaces,'' {\em JHEP} {\bf 03} (2006) 081,
\href{http://www.arXiv.org/abs/hep-th/0602164}{{\tt hep-th/0602164}}.

\bibitem{Alexandrov:2007ec}
S.~Alexandrov, ``{Quantum covariant c-map},'' {\em JHEP} {\bf 05} (2007) 094,
\href{http://www.arXiv.org/abs/hep-th/0702203}{{\tt hep-th/0702203}}.

\bibitem{MR664330}
S.~M. Salamon, ``Quaternionic {K}\"ahler manifolds,'' {\em Invent. Math.} {\bf
  67} (1982), no.~1, 143--171.

\bibitem{Bohm:1999uk}
R.~B{\"o}hm, H.~G{\"u}nther, C.~Herrmann, and J.~Louis, ``{Compactification of
  type IIB string theory on Calabi-Yau threefolds},'' {\em Nucl. Phys.} {\bf
  B569} (2000) 229--246,
\href{http://www.arXiv.org/abs/hep-th/9908007}{{\tt hep-th/9908007}}.

\bibitem{Alexandrov:2012bu}
S.~Alexandrov and B.~Pioline, ``{S-duality in Twistor Space},'' {\em JHEP} {\bf
  1208} (2012) 112,
\href{http://www.arXiv.org/abs/1206.1341}{{\tt 1206.1341}}.

\bibitem{Gaiotto:2006wm}
D.~Gaiotto, A.~Strominger, and X.~Yin, ``{The M5-brane elliptic genus:
  Modularity and BPS states},'' {\em JHEP} {\bf 08} (2007) 070,
\href{http://www.arXiv.org/abs/hep-th/0607010}{{\tt hep-th/0607010}}.

\bibitem{deBoer:2006vg}
J.~de~Boer, M.~C.~N. Cheng, R.~Dijkgraaf, J.~Manschot, and E.~Verlinde, ``{A
  farey tail for attractor black holes},'' {\em JHEP} {\bf 11} (2006) 024,
\href{http://www.arXiv.org/abs/hep-th/0608059}{{\tt hep-th/0608059}}.

\bibitem{Denef:2007vg}
F.~Denef and G.~W. Moore, ``{Split states, entropy enigmas, holes and halos},''
  {\em JHEP} {\bf 1111} (2011) 129,
\href{http://www.arXiv.org/abs/hep-th/0702146}{{\tt hep-th/0702146}}.

\bibitem{MR781735}
M.~Eichler and D.~Zagier, {\em The theory of {J}acobi forms}, vol.~55 of {\em
  Progress in Mathematics}.
\newblock Birkh\"auser Boston Inc., Boston, MA, 1985.

\bibitem{kudla2016theta}
S.~Kudla, ``Theta integrals and generalized error functions,''
  \href{http://www.arXiv.org/abs/1608.03534}{{\tt 1608.03534}}.

\bibitem{westerholt2016indefinite}
M.~Westerholt-Raum, ``Indefinite theta series on tetrahedral cones,''
  \href{http://www.arXiv.org/abs/1608.08874}{{\tt 1608.08874}}.

\bibitem{MR2605321}
D.~Zagier, ``Ramanujan's mock theta functions and their applications (after
  {Z}wegers and {O}no-{B}ringmann),'' {\em Ast\'erisque} (2009), no.~326, Exp.
  No. 986, vii--viii, 143--164 (2010). S{\'e}minaire Bourbaki. Vol. 2007/2008.

\bibitem{MR1623706}
L.~G{\"o}ttsche and D.~Zagier, ``Jacobi forms and the structure of {D}onaldson
  invariants for {$4$}-manifolds with {$b_+=1$},'' {\em Selecta Math. (N.S.)}
  {\bf 4} (1998), no.~1, 69--115.

\bibitem{0961.14022}
L.~G{\"o}ttsche, ``{Theta functions and Hodge numbers of moduli spaces of
  sheaves on rational surfaces.},'' {\em Commun. Math. Phys.} {\bf 206} (1999),
  no.~1, 105--136.

\bibitem{Manschot:2014cca}
J.~Manschot, ``{Sheaves on $\mathbb{P}^2$ and generalized Appell functions},''
\href{http://www.arXiv.org/abs/1407.7785}{{\tt 1407.7785}}.

\bibitem{Toda:2013}
Y.~Toda, ``{Flops and the S-duality conjecture},'' {\em Duke Math. J.} {\bf
  164} (2015) 2293--2339,
\href{http://www.arXiv.org/abs/1311.7476}{{\tt 1311.7476}}.

\end{thebibliography}

\end{document}